\documentclass[10pt,doublespace]{article}

\usepackage{graphics}
\usepackage{epsfig}

\title{Theory of Current Noise and Photon Noise in \\ Quantum Cascade Lasers}

\author{Farhan Rana, Rajeev J. Ram \\ 
{\em Research Laboratory for Electronics,} \\ 
{\em Massachusetts Institute of Technology, Cambridge, MA 02139}}

\date{\today}

\newcommand{\NarrowMargins}{
\setlength{\oddsidemargin}{-0.00 in}
\setlength{\evensidemargin}{-0.00 in}
\setlength{\textwidth}{6.5 in}
\setlength{\textheight}{9.25 in}
\setlength{\topmargin}{-0.0 in}}

\NarrowMargins

\begin{document}

\maketitle

\begin{abstract}
A comprehensive model for the photon number fluctuations and the current noise in quantum cascade lasers is presented. It is shown that the photon intensity noise in quantum cascade lasers exhibits little amplitude squeezing even when noise in the drive current is suppressed below the shot noise value. This is in contrast to interband semiconductor diode lasers in which the laser intensity noise can be squeezed well below the shot noise limit by high impedance suppression of fluctuations in the drive current. The theoretical model presented in this paper self-consistently accounts for the suppression of current noise in electron transport in multiple quantum well structures due to various electronic correlations. The nature of these electronic correlations is discussed. Mechanisms responsible for the reduced photon number squeezing in intersubband lasers are elucidated. Scaling of the laser intensity noise and the current noise with the number of cascaded gain stages is also described. Direct current modulation response of quantum cascade lasers is also studied, and it is shown that contrary to the predictions in the literature of terahertz modulation bandwidth for these lasers, bandwidth of almost all quantum cascade lasers that have been reported in the literature is limited by the inverse photon lifetime inside the laser cavity to tens of gigahertz.\\ \\
\noindent 78.67.De, 72.70.+m, 73.21.Fg           
\end{abstract}

\newpage

\tableofcontents

\newpage

\listoffigures

\newpage

\section{Introduction}
Unipolar quantum cascade lasers (QCLs) utilizing intersubband transitions to generate photons have become important sources of light in the mid-infrared region (5 $\mu$m - 15 $\mu$m). In this paper a comprehensive model for the photon intensity fluctuations in QCLs is reported for the first time. Current noise associated with electron transport through the active regions is also studied and its effect on the photon intensity noise is evaluated.

QCLs are different from interband semiconductor diode lasers in three important ways which can have a significant impact on their noise properties:
\begin{enumerate}
\item Electron transport in QCLs takes place by tunneling between states in adjacent quantum wells. It is well known that electronic correlations in resonant tunneling in quantum well structures can suppress (or enhance) current noise by providing a negative (or positive) feedback~\cite{buttiker,davis,pelleg}. High impedance suppression of the current noise in semiconductor diode lasers results in light output with squeezed photon number fluctuations~\cite{yama1}. It is therefore intriguing whether suppression of the current noise can also lead to squeezing in QCLs. Any model for the photon noise in QCLs must take into account these electronic correlations self-consistently. 

\item In diode lasers the carrier density in the energy level involved in the lasing action does not increase beyond its threshold value and, therefore, the noise contributed by the non-radiative recombination and generation processes also remains unchanged beyond threshold. In QCLs the electron densities in the upper and lower lasing states do not clamp at threshold, and keep increasing when the bias current is increased beyond threshold. As a result, non-radiative processes contribute significantly to photon noise even at high bias currents. 

\item Since all the gain sections in a QCL are connected electrically and optically, electron density fluctuations and photon emission events in different gain sections become correlated. The effect of these correlations on the photon noise in interband cascade lasers has already been discussed in detail~\cite{farhan1,farhan2}, and it is the aim of this paper to investigate the role of these correlations in QCLs. 
\end{enumerate}

\section{Outline}
In section~\ref{SteadyState} the non-linear rate equations for the electron and photon densities in QCLs are presented. The steady state solution of these rate equations below and above threshold are described. In section~\ref{NoiseModels} the non-linear rate equations are linearized to obtain Langevin rate equations for the fluctuations in the electron and photon densities. Electron transport in the multiple quantum well structure of QCLs is discussed in detail, and a self-consistent model for the fluctuations in the electron charge densities and the electron current density is presented. It is shown that a self-consistent description of the fluctuations in the charge and current densities can be carried out in terms of a few device parameters. Langevin noise sources are also used to model the noise associated with electron transport by tunneling. Section~\ref{NoiseModels} is the main part of this paper. In section~\ref{SolOfEq} the set of coupled linearized Langevin rate equations for the fluctuations in the electron densities in different levels of all the cascaded gain stages and the fluctuations in the photon density are solved under the constraints imposed by the biasing electrical circuit. In addition, the direct current modulation response of QCLs is also evaluated and the maximum possible modulation bandwidth is discussed. The analytical and numerical results on the current noise and the photon noise in QCLs are presented and discussed in section~\ref{DisscussCN} and section~\ref{DisscussPN}, respectively. In these sections the results obtained are compared with the current and photon noise in interband semiconductor diode lasers. Readers not familiar with the results on the current and photon noise in diode lasers are encouraged to read Appendix~\ref{intnoise} in which a detailed model for the noise in diode lasers is presented.

\section{Rate Equations and Steady State Solutions} \label{SteadyState}

\begin{figure}[phtb]
\begin{center}
    \epsfig{file=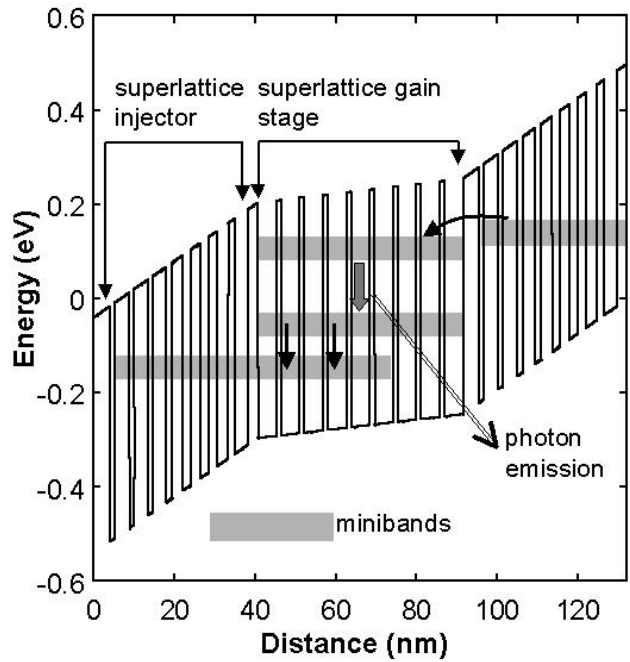,angle=0,width=3.5 in}
    \caption{ Superlattice quantum cascade laser.}
\label{figSQCL}   
\end{center}
\end{figure}

\begin{figure}[phtb]
\begin{center}
    \epsfig{file=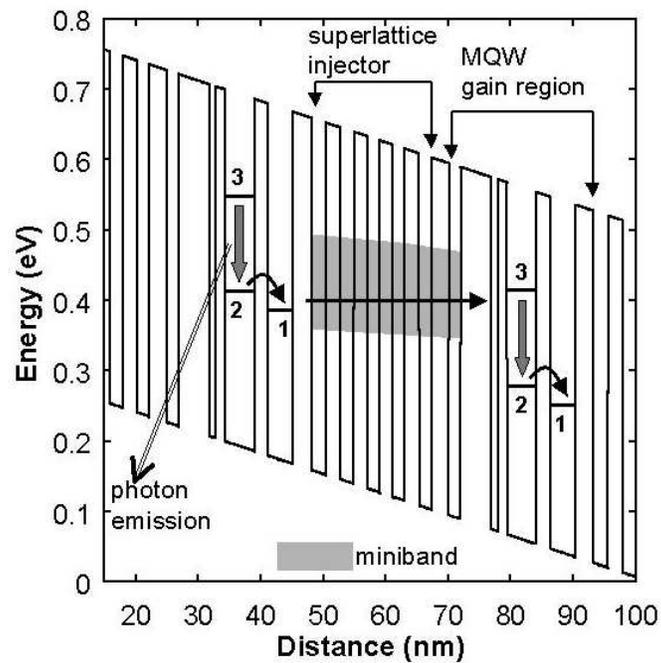,angle=0,width=3.5 in}
    \caption{ Multiple quantum well quantum cascade laser.}
    \label{figMQCL}
\end{center}
\end{figure}

Many different types of QCL structures have been reported in the literature~\cite{faistvtQCL,faistCW,carloCW,carloPandC,faistHP,carloGaAs,cgdiskQCL,cgsingle,cgdfb,atsuperQCL1,atsuperQCL2,atsuperQCL3,gssupQCL}. Almost all of these QCL structures can be classified into two categories:
\begin{enumerate}
\item {\em Superlattice QCLs} in which the gain stage consists of a superlattice structure and the photons are emitted when the electrons make transitions between two minibands of this superlattice. These minibands are actually clusters of closely spaced energy levels (Fig.~\ref{figSQCL})~\cite{atsuperQCL1,atsuperQCL2,atsuperQCL3,gssupQCL}.
\item {\em Multiple quantum well QCLs} in which the gain stage consists of multiple quantum wells (typically two or three) and the radiative electronic transitions occur between two discrete energy levels (Fig.~\ref{figMQCL})~\cite{faistvtQCL,faistCW,carloCW,carloPandC,faistHP,carloGaAs}.
\end{enumerate}
In both types of QCLs, two successive gain stages are separated by a superlattice structure known as the {\em injector}. The superlattice injector has a mini-gap which prevents the electrons from tunneling out into the injector from the upper energy level(s) of the previous gain stage and, therefore, increases the radiative efficiency. Electrons from the lower energy level(s) of a gain stage can tunnel into the injector, and the injector injects these electrons into the upper energy level(s) of the next gain stage.

In this paper, photon noise and current noise in only multiple quantum well QCLs is discussed. However, the methods described here can easily be used to study noise in QCLs with superlattice gain stages. A single stage of a multiple quantum well QCL structure is shown in Fig.~\ref{figMQCL}~\cite{faistHP}. The operation of the QCL can be described as follows. Electrons tunnel from the energy states in the superlattice injector into level 3 of the gain stage. Photons are emitted when electrons make radiative transitions from level 3 to level 2. Transitions from level 2 to level 1 occur primarily by emission of optical phonons. Electrons leave the gain stage from level 1 by tunneling out into the injector of the next stage. In addition, electrons also make non-radiative transitions from level 3 to levels 2 and 1. The non-linear rate equations for the electron and photon densities can be written as,  
\begin{equation}
\frac{d\, n_{3}^{j}}{d\,t} =  \frac{J_{in}^{j}}{q} - R_{32}(n_{3}^{j},n_{2}^{j}) - R_{31}(n_{3}^{j},n_{1}^{j}) - \Gamma^{j}v_{g}\,g(n_{3}^{j},n_{2}^{j})\left( S_{p} + \frac{n_{sp}}{WL} \right) \label{eqN3_a}
\end{equation}
\begin{equation}
\frac{d\, n_{2}^{j}}{d\,t} =  R_{32}(n_{3}^{j},n_{2}^{j}) - R_{21}(n_{2}^{j},n_{1}^{j}) + \Gamma^{j}v_{g}\,g(n_{3}^{j},n_{2}^{j})\left( S_{p} + \frac{n_{sp}}{WL} \right) \label{eqN2_a}
\end{equation}
\begin{equation}
\frac{d\, n_{1}^{j}}{d\,t} =  R_{31}(n_{3}^{j},n_{2}^{j}) + R_{21}(n_{2}^{j},n_{1}^{j}) - \frac{J_{out}^{j}}{q} \label{eqN1_a}
\end{equation}
\begin{equation}
\frac{d\, S_{p}}{d\,t} = \sum_{j=1}^{N}\, \Gamma^{j}v_{g}\,g(n_{3}^{j},n_{2}^{j})(S_{p} + \frac{n_{sp}}{WL}) - \frac{S_{p}}{\tau_{p}} \label{eqSp_a}
\end{equation}
\begin{equation}
P_{out} = \eta_{o}\,h \nu\,\frac{WL\,S_{p}}{\tau_{p}} \label{eqPo_a}
\end{equation}
In the above equations, $n_{k}^{j}$ is the electron density (cm$^{-2}$) in the $k$th energy level of the $j$th gain stage. $J_{in}^{j}$ and $J_{out}^{j}$ are the electron current densities (cm$^{-2}$) tunneling into level 3 and tunneling out of level 1 of the $j$th gain stage, respectively. Only in steady state $J_{in}^{j}$ equals $J_{out}^{j}$. $S_{p}$ is the photon density (cm$^{-2}$) inside the optical cavity. $S_{p}$ is equal to the total number of photons inside the cavity divided by the width $W$ and the length $L$ of the cavity. $v_{g}$ is the group velocity of the lasing mode and $g$ is the optical gain (cm$^{-1}$) contributed by a single gain stage. $\Gamma^{j}$ is the mode confinement factor for the $j$th gain stage. $N$ is the total number of cascaded gain stages. $R_{32}$ is the net transition rate from level 3 to level 2 through non-radiative processes and spontaneous emission into the non-lasing modes. Similarly, $R_{31}$ and $R_{21}$ are the net transition rates from level 3 and level 2 into level 1, respectively. $n_{sp}$ is the spontaneous emission factor~\cite{coldren}. $P_{out}$ is the output power from the laser. $\eta_{o}$ is the power output coupling efficiency and $\tau_{p}$ is the photon lifetime inside the cavity. The expression for $\tau_{p}$ is,
\begin{equation}
\frac{1}{\tau_{p}} = v_{g}(\alpha_{i} + \alpha_{m}) = v_{g}\,\left[ \alpha_{i} + \frac{1}{L}\log{\left( \frac{1}{\sqrt{r_{1}r_{2}}} \right) } \right] \label{eqtaup}
\end{equation}
where $\alpha_{i}$ is the internal loss of the cavity, $\alpha_{m}$ is the loss from the cavity facets,  and $r_{1}$ and $r_{2}$ are the facet reflectivities. The power output coupling efficiency $\eta_{o}$ from the facet with reflectivity $r_{1}$ is,
\begin{equation}
\eta_{o} = \frac{ (1-r_{1})\sqrt{r_{2}} }{ \left[ (1-r_{1})\sqrt{r_{2}} + (1-r_{2})\sqrt{r_{1}} \right] } \; \frac{\alpha_{m}} {(\alpha_{m} + \alpha_{i})} \label{eqetao1}
\end{equation}       

For simplicity it is assumed that all the gain stages have the same mode confinement factor, i.e. $\Gamma^{j}=\Gamma$ for all $j$. This assumption is valid if all the cascaded gain stages are located close to the peak of the transverse profile of the optical mode where the field strength varies slowly. Even for QCLs with large number of gain stages numerical simulations show that corrections to the solution obtained by assuming all $\Gamma^{j}$ to be equal are small. Under this assumption, the steady state electron densities $n_{k}^{j}$ are the same in all the gain stages, and the index $j$ may be suppressed.

\subsection{Steady State Solutions}

\subsubsection{Below Threshold}

\begin{figure}[htb]
\begin{center}
    \epsfig{file=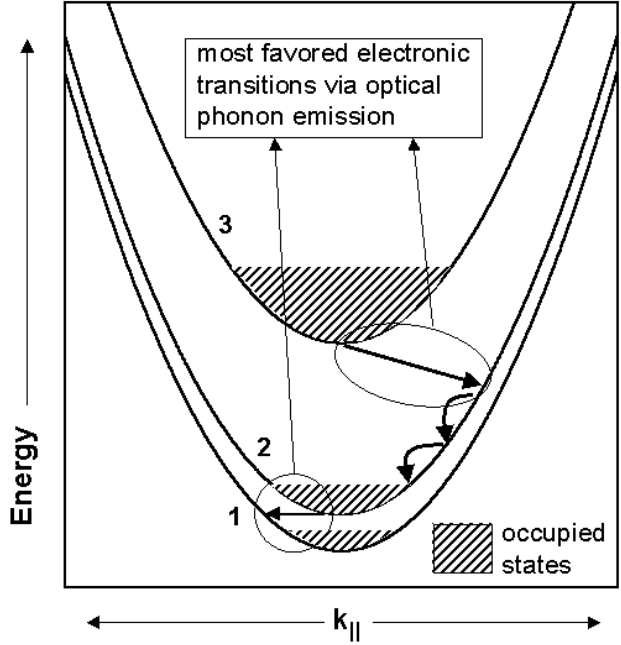,angle=0,width=3.5 in}
    \caption{ Energy subbands of the three levels of the gain stage. Most favored electronic transitions by optical phonon emission are almost horizontal in the $E({\vec k})-{\vec k}$ space.}
    \label{figEngySubband}
\end{center}
\end{figure}

The steady state solution to the rate equations can be found by setting all the time derivatives equal to zero, and putting $J_{in}=J$. Below threshold, steady state carrier densities can be found by putting $S_{p}=0$, and solving the equations (the index $j$ has been suppressed),  
\begin{equation}
R_{32}(n_{3},n_{2}) + R_{31}(n_{3},n_{1}) = \frac{J}{q} \label{dc1}
\end{equation}
\begin{equation}
R_{32}(n_{3},n_{2}) = R_{21}(n_{2},n_{1}) \label{dc2}
\end{equation}
The third equation can be obtained by realizing that $J_{out}$ is also a function of $n_{1}$,
\begin{equation}
J_{out}(n_{1}) = J \label{dc3}
\end{equation}
To proceed further, analytical expressions for the transition rates are required. These transition rates can be approximated as,
\begin{equation}
R_{32}(n_{3},n_{2}) \cong \frac{n_{3}}{\tau_{32}} \label{R32}
\end{equation}
\begin{equation}
R_{31}(n_{3},n_{1}) \cong \frac{n_{3}}{\tau_{31}} \label{R31}
\end{equation}
\begin{equation}
R_{21}(n_{2},n_{1}) \cong \frac{n_{2}}{\tau_{21}} \label{R21}
\end{equation}
\begin{equation}
\frac{J_{out}(n_{1})}{q} \cong \frac{n_{1}}{\tau_{out}} \label{Rout}
\end{equation}
The rationale for the approximations in Equations (\ref{R32})-(\ref{R21}) is that optical phonons are largely responsible for intersubband transitions. As shown in Fig.~\ref{figEngySubband}, optical phonon mediated intersubband transitions that are almost horizontal in $E({\vec k})$-${\vec k}$ plane are more likely to occur~\cite{capasso}. Therefore, the transitions rates from an upper to a lower subband are not much affected by the electron density in the lower subband, as long as the electron density in the lower subband is small. More complicated expressions for these transition rates, such as,
\begin{equation}
R_{qk}(n_{q},n_{k}) = \frac{n_{q}}{\tau_{qk}} - \frac{n_{k}}{\tau_{kq}}
\end{equation}
may be used if necessary.

The expression for $J_{out}$ in Equation (\ref{Rout}) does not depend upon on the electron density in the injector since electrons in the injector states are assumed to relax very quickly into the ground state of the injector which is spatially localized near the next gain stage. Using Equations (\ref{R32})-(\ref{Rout}) in Equations (\ref{dc1})-(\ref{dc3}), expressions for the carrier densities can be obtained as a function of the current density,
\begin{equation}
n_{3} = \frac{J}{q} \, \frac{\tau_{32} \, \tau_{31}}{\tau_{32}+\tau_{31}} \label{n3bth}
\end{equation}
\begin{equation}
n_{2} = \frac{J}{q} \, \frac{\tau_{21} \, \tau_{31}}{\tau_{32}+\tau_{31}} \label{n2bth}
\end{equation}
\begin{equation}
n_{1} = \frac{J}{q} \, \tau_{out} \label{n1bth}
\end{equation}

\subsubsection{Above Threshold}

Above threshold, the gain is clamped to a value determined by equating the gain with the loss,
\begin{equation}
\sum_{j=1}^{N}\,\Gamma^{j}v_{g}\,g(n_{3}^{j},n_{2}^{j}) = N\Gamma v_{g}\,g(n_{3},n_{2}) = \frac{1}{\tau_{p}} \label{lascond}
\end{equation}
For perfectly parabolic subbands, the expression for the gain may be approximated as,
\begin{equation}
g(n_{3},n_{2}) = a \, (n_{3} - n_{2}) \label{gain}
\end{equation}
where $a$ is the differential gain. The carrier and photon densities above threshold can be obtained by solving the equations,
\begin{equation}
R_{31}(n_{3},n_{1}) + R_{21}(n_{2},n_{1}) \cong \frac{n_{3}}{\tau_{31}} + \frac{n_{2}}{\tau_{21}} = \frac{J}{q} \label{dc4}
\end{equation}
\begin{equation}
\frac{J_{out}}{q} \cong \frac{n_{1}}{\tau_{out}} = \frac{J}{q} \label{dc5}
\end{equation}
\begin{equation}
N\Gamma v_{g} \, a (n_{3} - n_{2}) = \frac{1}{\tau_{p}} \label{dc6}
\end{equation}
which results in,
\begin{equation}
n_{3} = \frac{J}{q} \, \frac{\tau_{21}\,\tau_{31}}{\tau_{21}+\tau_{31}} + \left( \frac{1}{N\Gamma v_{g} \, a \tau_{p}}\right)\,\frac{\tau_{31}}{\tau_{21}+\tau_{31}} \label{n3ath}
\end{equation}
\begin{equation}
n_{2} = \frac{J}{q} \, \frac{\tau_{21}\,\tau_{31}}{\tau_{21}+\tau_{31}} - \left( \frac{1}{N\Gamma v_{g} \, a \tau_{p}}\right) \,\frac{\tau_{21}}{\tau_{21}+\tau_{31}} \label{n2ath}
\end{equation}
\begin{equation}
n_{1} = \frac{J}{q} \, \tau_{out} \label{n1ath}
\end{equation}    
\begin{equation}
S_{p} = \eta_{r} \, N \, \frac{(J-J_{th})}{q}\,\tau_{p}  \label{spath}
\end{equation}
\begin{equation}
P_{out} = \eta_{o} \, \eta_{r} \, \frac{h \nu}{q}\,N\,(I-I_{th}) \label{dcpow}
\end{equation}
where the threshold current density $J_{th}$ and the radiative efficiency $\eta_{r}$ are,
\begin{equation}
J_{th} = \frac{q}{N\Gamma v_{g} \, a \tau_{p}} \left( \frac{1}{\tau_{32}} + \frac{1}{\tau_{31}} \right)\,\frac{1}{\left( 1 - \tau_{21} / \tau_{32} \right)}  \label{jth}
\end{equation}
\begin{equation}
\eta_{r} = \left( 1 - \frac{\tau_{21}}{\tau_{32}} \right)\,\frac{\tau_{31}}{(\tau_{21} + \tau_{31})} \label{eqetar}
\end{equation}
The radiative efficiency $\eta_{r}$ for a QCL is defined as that fraction of the total number of electrons injected into each gain stage which contribute to photon emission. Equation~(\ref{dcpow}) shows that the slope efficiency of the laser scales linearly with the number of gain stages $N$.

\begin{figure}[t]
\begin{center}
  \epsfig{file=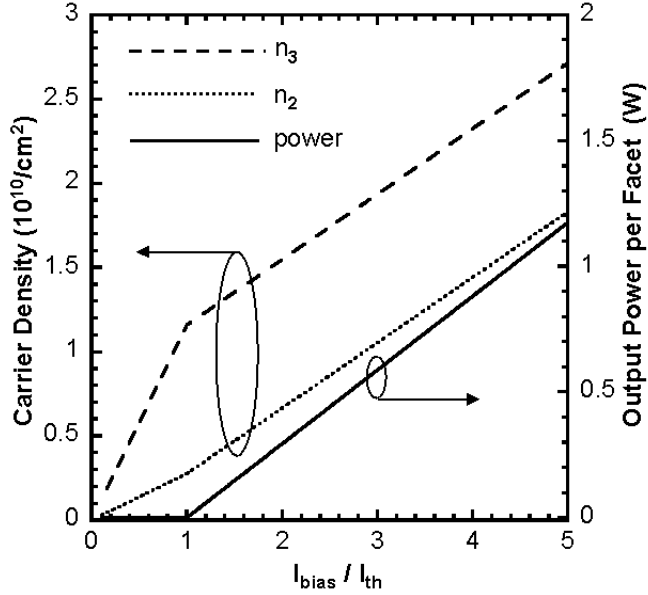,angle=0,width=3.5 in}
    \caption{ Electron densities in level 3 and level 2 of the gain stage, and the output power per facet are plotted as a function of the bias current. There is a discontinuity in the rate of increase of the electron densities with the bias current at threshold.}
    \label{figCarDen}
\end{center}
\end{figure}

Equations (\ref{n3ath}) and (\ref{n2ath}) show that above threshold, even though the gain is clamped to its threshold value, the electron densities keep increasing with the bias current. This is in contrast to semiconductor diode lasers in which the carrier densities do not increase beyond their threshold values. As a result, an increase in the injected current density in QCLs does not only lead to an increase in the photon emission rate but it also leads to an increase in the rate of non-radiative transitions. Therefore, QCLs tend to have radiative efficiencies $\eta_{r}$ significantly smaller than unity. Fig.~\ref{figCarDen} shows the electron densities $n_{3}$ and $n_{2}$ plotted as a function of the bias current. The values of the various device parameters used in generating Fig.~\ref{figCarDen} belong to the QCL reported in~\cite{faistHP}, and these values are given in Table~\ref{table1}. Fig.~\ref{figCarDen} shows that the rate of change of electron densities in levels 3 and 2 with the bias current exhibits discontinuities at threshold. This can be confirmed by comparing Equations (\ref{n3ath}) and (\ref{n2ath}) with Equations (\ref{n3bth}) and (\ref{n2bth}). As will be shown later in this paper, these discontinuities in the rate of increase of electron densities with the bias current result in a discontinuity in the value of the differential resistance of the laser at threshold. 

\section{Theoretical Model for Noise and Fluctuations} \label{NoiseModels}

The model for the noise presented in this paper consists of a set of coupled self-consistent Langevin rate equations for the fluctuations in the electron density in different energy levels of a gain stage. Fluctuations in the electron density are caused by radiative and non-radiative scattering processes, electron tunneling processes and also by fluctuations in the current injected into the gain stage. Fluctuations in the current are a relaxational response to electron scattering and tunneling events occurring inside all the gain stages of the QCL, and they are also caused by sources external to the laser which include thermal noise sources associated with circuit resistances. Photon density fluctuations are also modeled by Langevin rate equations. Electron density fluctuations in different gain stages are all coupled to the photon density fluctuations and also to the fluctuations in the current which flows through all the gain stages connected in series. The system of equations obtained this way can easily be solved analytically or numerically to give the spectral density of the photon number fluctuations and the current fluctuations. The methods described in this paper can be used to study a variety of QCLs that have been reported in the literature. 

\subsection{Linearized Langevin Rate Equations for Electron and Photon densities} 
The non-linear rate equations can be linearized around any bias point to obtain rate equations for the fluctuations. Linearized Langevin rate equations for these fluctuations are,

\begin{eqnarray}
\frac{d\,\delta n_{3}^{j}}{d\,t} & = & \frac{\delta J_{in}^{j}}{q} - \frac{\delta n_{3}^{j}}{\tau_{32}} - \frac{\delta n_{3}^{j}}{\tau_{31}} - \Gamma^{j}v_{g} \left[ a \left( \delta n_{3}^{j} - \delta n_{2}^{j} \right) \left( S_{p} + \frac{n_{sp}}{WL} \right) + g(n_{3}^{j},n_{2}^{j}) \delta S_{p} \right]  \nonumber \\
                                &   & - f_{32}^{j} - f_{31}^{j} - f_{RN}^{j} \label{eqN3}
\end{eqnarray}
\begin{equation}
\frac{d\,\delta n_{2}^{j}}{d\,t} = \frac{\delta n_{3}^{j}}{\tau_{32}} - \frac{\delta n_{2}^{j}}{\tau_{21}} + \Gamma^{j}v_{g} \left[ a \left( \delta n_{3}^{j} - \delta n_{2}^{j} \right) \left( S_{p} + \frac{n_{sp}}{WL} \right) + g(n_{3}^{j},n_{2}^{j}) \delta S_{p} \right] + f_{32}^{j} - f_{21}^{j} + f_{RN}^{j} \label{eqN2}
\end{equation} 

\begin{equation}
\frac{d\,\delta n_{1}^{j}}{d\,t} = \frac{\delta n_{3}^{j}}{\tau_{31}} + \frac{\delta n_{2}^{j}}{\tau_{21}} + f_{31}^{j} + f_{21}^{j} - \frac{\delta J_{out}^{j}}{q} \label{eqN1}
\end{equation} 

\begin{equation}
\frac{d\,\delta S_{p}}{d\,t} = \sum_{j=1}^{N} \, \Gamma^{j}v_{g}\left[ a \left( \delta n_{3}^{j} - \delta n_{2}^{j} \right)\left(  S_{p} + \frac{n_{sp}}{WL} \right) + g(n_{3}^{j},n_{2}^{j}) \delta S_{p} \right] - \frac{\delta S_{p}}{\tau_{p}} - F_{L} + \sum_{j=1}^{N}\,f^{j}_{RS}  \label{eqSp}
\end{equation}

\begin{equation}
\delta P_{out} =  \eta_{o}\,h \nu\,\frac{WL \,\delta S_{p}}{\tau_{p}} + F_{o} \label{eqPo}
\end{equation}
Equations (\ref{R32})-(\ref{R21}) have been used above for approximating the transition rates $R_{qk}$. $f_{32}$, $f_{31}$, and $f_{21}$ are Langevin sources which model the noise associated with the non-radiative intersubband transitions and also the radiative transitions into the non-lasing modes. $f_{RN}$ and $f_{RS}$ are Langevin sources which model the noise in photon emission and absorption from the lasing mode. $F_{L}$ and $F_{o}$ describe the noise associated with photon loss from the cavity~\cite{coldren}. All the Langevin noise sources have a white spectral density and their correlations can be found by the methods described in Ref.~\cite{coldren}. All the non-zero correlations among the noise sources are given in Appendix~\ref{langcor}.

\subsection{Linearized Electron Transport, Coulomb Correlations and Noise} \label{transport}
In order to determine $\delta J_{in}^{j}$ and $J_{out}^{j}$ the electron transport through the active region needs to be looked at in detail. Self-consistent modeling of electron transport in multiple quantum well structures poses a significant challenge, and the steady state current-voltage characteristics of QCLs are difficult to compute accurately. In this paper a slightly different approach has been adopted which is more useful for the problem under consideration. A self-consistent model for the fluctuations in the electron current density and the electron charge density is presented. It is shown that self-consistent analysis of current density and charge density fluctuations can be carried out in terms of only a few device parameters. The values of these parameters can either be determined experimentally or computed theoretically from more detailed self-consistent transport models. The method used in this paper to estimate the value of each parameter will be discussed when we compare the theoretical model with the experimental results. 

The expression for the direct sequential tunneling current density from the injector state into level 3 of the gain stage can be written as~\cite{APK}, 
\begin{eqnarray}
J_{in} & = & J_{in-forward} - J_{in-backward} \nonumber \\
       & = & 2q \,\int \frac{d^{2}{\vec k}}{(2\pi)^{2}} \, \int \frac{d^{2}{\vec k'}}{(2\pi)^{2}} \; \frac{2\pi}{\hbar} \left| T_{{\vec k},{\vec k'}} \right|^{2} \, \int_{-\infty}^{\infty} dE \; A\left( E - E_{inj}({\vec k}) \right) \, A\left(E - E_{3}({\vec k'})\right) \times \nonumber \\
       &   & \hspace{2.5 in} \bigg[ f\left(E - \mu_{inj} \right) - f\left(E - \mu_{3} \right) \bigg] \label{cur}
\end{eqnarray}
where the forward and backward components of the injection current are,
\begin{eqnarray}
J_{in-forward} & = & 2q \,\int \frac{d^{2}{\vec k}}{(2\pi)^{2}} \, \int \frac{d^{2}{\vec k'}}{(2\pi)^{2}} \; \frac{2\pi}{\hbar} \left| T_{{\vec k},{\vec k'}} \right|^{2} \, \int_{-\infty}^{\infty} dE \; A\left( E - E_{inj}({\vec k}) \right) \, A\left(E - E_{3}({\vec k'})\right) \times \nonumber \\
       &   & \hspace{2.0 in} f\left(E - \mu_{inj} \right)\,\bigg[1 - f\left(E - \mu_{3} \right) \bigg] \label{curf}
\end{eqnarray}
\begin{eqnarray}
J_{in-backward} & = & 2q \,\int \frac{d^{2}{\vec k}}{(2\pi)^{2}} \, \int \frac{d^{2}{\vec k'}}{(2\pi)^{2}} \; \frac{2\pi}{\hbar} \left| T_{{\vec k},{\vec k'}} \right|^{2} \, \int_{-\infty}^{\infty} dE \; A\left( E - E_{inj}({\vec k}) \right) \, A\left(E - E_{3}({\vec k'})\right) \times \nonumber \\
       &   & \hspace{2.0 in} f\left(E - \mu_{3} \right)\,\bigg[1 - f\left(E - \mu_{inj} \right) \bigg] \label{curb}
\end{eqnarray}
$T_{{\vec k},{\vec k'}}$ is the coupling constant, and its related to the transmission probability. $E_{inj}({\vec k})$ and $E_{3}(\vec k)$ are the energies of electrons in the injector state and level 3 of the gain stage, respectively. $A(E)$ is a normalized lorentzian function with FWHM equal to the broadening of the energy levels, and $f(E-\mu)$ is the Fermi-Dirac distribution function with a chemical potential $\mu$. Expressions similar to Equation (\ref{cur}) can also be written for the phonon assisted tunneling current density. The analysis presented in this paper is independent of the specific nature of the electron tunneling mechanisms. In what follows, $E_{inj}$ and $E_{3}$ will stand for $E_{inj}({\vec k}=0)$ and $E_{3}({\vec k}=0)$, respectively. The tunneling current in Equation (\ref{cur}) depends upon the following three quantities, 
\begin{itemize}
\item The difference $(\mu_{inj} - E_{inj})$ between the injector chemical potential and the energy of the injector state.
\item The difference $(\mu_{3} - E_{3})$ between the chemical potential and the energy of level 3 of the gain stage.
\item The relative difference $(E_{inj} - E_{3})$ between the energies of the injector state and level 3 of the gain stage.
\end{itemize}
The current can change if the number of electrons in the injector level or in level 3 of the gain stage changes. The current can also change if the energy of the injector level shifts with respect to the energy of level 3. $\delta J_{in}$ can be written as, 
\begin{eqnarray}
\delta J_{in}^{j} & = & \frac{\delta J_{in} / \delta (\mu_{inj}-E_{inj})}{\delta n_{inj} / \delta (\mu_{inj}-E_{inj})} \: \delta n_{inj}^{j} + \frac{\delta J_{in} / \delta (\mu_{3}-E_{3})} {\delta n_{3} / \delta (\mu_{3}-E_{3})} \: \delta n_{3}^{j} \nonumber \\
                  &   & \mbox{} + \frac{\delta J_{inj}}{\delta (E_{inj}-E_{3})} \: (\delta E_{inj}^{j}- \delta E_{3}^{j}) + qf_{in}^{j} \label{eqvarj} \\
& = & \frac{\delta J_{in}}{\delta n_{inj}} \: \delta n_{inj}^{j} + \frac{\delta J_{in}}{\delta n_{3}} \: \delta n_{3}^{j}  + \frac{\delta J_{inj}}{\delta (E_{inj}-E_{3})} \: (\delta E_{inj}^{j}- \delta E_{3}^{j}) + qf_{in}^{j}  \label{eqvarjgen}
\end{eqnarray}
$f_{in}$ is a Langevin noise source which models the noise in electron tunneling. Noise in electron transport by sequential tunneling in multiple quantum well structures can be described with Langevin noise sources. In Refs.~\cite{davis,pelleg} the current noise in double barrier resonant tunneling structures is evaluated using classical discrete master equations. Under suitable conditions a discrete master equation may be converted into a Fokker-Planck equation, and if the fluctuations are relatively small a Fokker-Planck equation can be linearized around a stable steady state solution (see Ref.~\cite{gardiner} for details). Langevin rate equations can be used in place of linearized Fokker-Planck equations since the two formalisms are equivalent.~\cite{gardiner}. It can be shown that Langevin rate equations yield results identical to those presented in Refs.~\cite{davis,pelleg} for the current noise in double barrier resonant tunneling devices~\cite{farhan3}. A linearized analysis based on Langevin rate equations may become invalid for highly non-linear devices. The correlation function for the noise source $f_{in}$ is,
\begin{eqnarray}
WL\,\langle f_{in}^{j}(t)\,f_{in}^{q}(t') \rangle & = & \frac{1}{q}\,(J_{in-forward} + J_{in-backward}) \, \delta_{jq} \, \delta(t-t') \label{FinjFinj1} \\
& \approx & \frac{J_{in}}{q}\,\chi_{in} \, \delta_{jq} \, \delta(t-t') \label{FinjFinj2}
\end{eqnarray}
The factor $\chi_{in}$ relates the sum of the forward and backward tunneling currents to their difference which is the total injection current $J_{in}$. At low temperatures $\chi_{in}$ is expected to be close to unity since Pauli's exclusion would restrict the available phase space for the backward tunneling current~\cite{pelleg}. For the same reason $\chi_{in}$ is expected to be close to unity for large values of the injection current $J_{in}$. At high temperatures and small values of the injection current $\chi_{in}$ can be larger than unity.

Although Equation (\ref{eqvarjgen}) for the change in current density is derived for the case of direct sequential tunneling, it also holds for the case of phonon assisted tunneling. Even if the energy distribution of electrons inside each energy level in steady state were not a Fermi-Dirac distribution, Equation (\ref{eqvarjgen}) would still hold.

\begin{figure}[t]
\begin{center}
    \epsfig{file=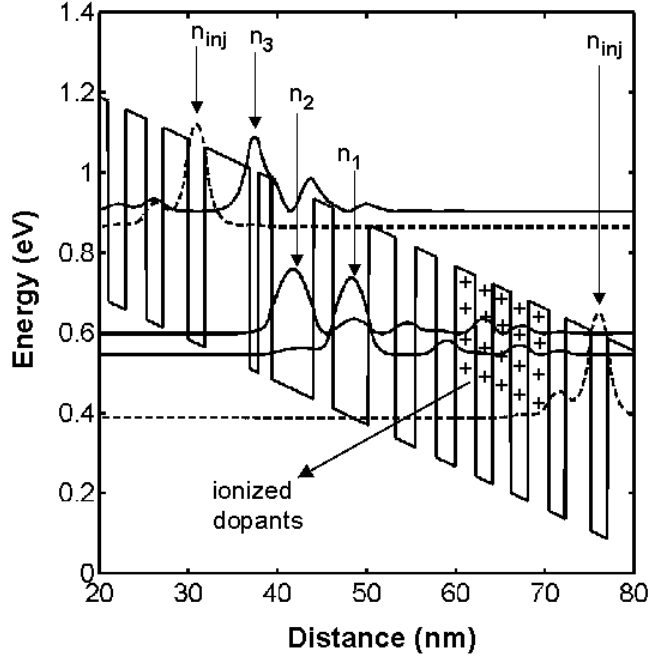,angle=0,width=3.5 in}    
    \caption{ Charge densities associated with the electron densities $\delta n_{inj}$, $\delta n_{3}$, $\delta n_{2}$, and $\delta n_{1}$ are shown. The electron charge densities are imaged on the positively charged ionized dopants present in the superlattice injector of the subsequent stage.} 
    \label{figCapMod}
\end{center}
\end{figure}

It is assumed that the superlattice injector is doped in regions not close to the gain stage. Electric field lines from electron density fluctuations $\delta n_{inj}^{j}$, $\delta n_{3}^{j}$, $\delta n_{2}^{j}$, and $\delta n_{1}^{j}$ are imaged on the ionized dopants in the injector layer of the $(j+1)$th gain stage, as shown in Fig.~\ref{figCapMod}. 
Therefore, the fluctuation $\delta V^{j}$ in the potential difference across the $j$th gain stage can be written as,
\begin{equation}
\delta V^{j} = \frac{q\delta n_{inj}^{j}}{C_{inj}} + \frac{q\delta n_{3}^{j}}{C_{3}} + \frac{q\delta n_{2}^{j}}{C_{2}} + \frac{q\delta n_{1}^{j}}{C_{1}} \label{eqV}
\end{equation}
$C_{inj}$, $C_{3}$, $C_{2}$, and $C_{1}$ are capacitances which determine the incremental change in the potential difference across a gain stage with changes in the electron densities in different energy levels. Using first order quantum mechanical perturbation theory, $\delta E_{inj}^{j} - \delta E_{3}^{j}$ can be related to the fluctuation in the average potential difference between the injector level and level 3 of the gain stage. The fluctuation in the average potential difference between these two levels can also be expressed in terms of capacitances. The expression for $\delta E_{inj}^{j} - \delta E_{3}^{j}$ therefore becomes,
\begin{equation}
\delta E_{inj}^{j} - \delta E_{3}^{j} =   \frac{q^{2}\delta n_{inj}^{j}}{C_{inj}'} - \frac{q^{2}\delta n_{3}^{j}}{C_{3}'} - \frac{q^{2}\delta n_{2}^{j}}{C_{2}'} - \frac{q^{2}\delta n_{1}^{j}}{C_{1}'} \label{eqengydiff}
\end{equation}
Using Equation~(\ref{eqV}) and Equation~(\ref{eqengydiff}), Equation~(\ref{eqvarjgen}) can be cast in the form,
\begin{eqnarray}
\frac{\delta J_{in}^{j}}{q} & = & \left(\frac{1}{t_{in}} + \frac{G_{in}}{C_{inj}'}\right) \delta n_{inj}^{j} - \left(\frac{1}{t_{3}} + \frac{G_{in}}{C_{3}'}\right) \delta n_{3}^{j} - \left(\frac{G_{in}}{C_{2}'}\right)\delta n_{2}^{j} - \left(\frac{G_{in}}{C_{1}'}\right)\delta n_{1}^{j} + f_{in}^{j} \label{eqvarjin1} \\
                  & = & \left(\frac{C_{inj}}{t_{in}} + \frac{C_{inj}}{C_{inj}'}G_{in}\right) \: \frac{\delta V^{j}}{q} - \left[ \left(\frac{1}{t_{in}} + \frac{G_{in}}{C_{inj}'}\right)\frac{C_{inj}}{C_{3}} + \left(\frac{1}{t_{3}} + \frac{G_{in}}{C_{3}'}\right) \right] \delta n_{3}^{j} \nonumber \\
                  &   & \mbox{} - \left[ \left(\frac{1}{t_{in}} + \frac{G_{in}}{C_{inj}'}\right)\frac{C_{inj}}{C_{2}} + \left(\frac{G_{in}}{C_{2}'}\right) \right]\delta n_{2}^{j} - \left[ \left(\frac{1}{t_{in}} + \frac{G_{in}}{C_{inj}'}\right)\frac{C_{inj}}{C_{1}} + \left(\frac{G_{in}}{C_{1}'}\right) \right]\delta n_{1}^{j} \nonumber \\
& & \mbox{} + f_{in}^{j} \label{eqvarjin}
\end{eqnarray}
In Equation~(\ref{eqvarjin}) $G_{in}$, $t_{in}$, and $t_{3}$ are given by,
\begin{equation}
G_{in} = q \frac{\delta J_{in}}{\delta (E_{inj}-E_{3})} \hspace{0.5 cm} , \hspace{0.5 cm} \frac{1}{t_{in}} = \frac{1}{q} \frac{\delta J_{in}}{\delta n_{inj}} \hspace{0.5 cm} , \hspace{0.5 cm} \frac{1}{t_{3}} = - \frac{1}{q} \frac{\delta J_{in}}{\delta n_{3}}  \label{eqGintint3} 
\end{equation}

More generally, there may be more than just one energy level in the injector from which electrons get injected into level 3 of the gain stage. Equation (\ref{eqvarjin1}) can be modified appropriately to take into account the contributions from all the energy levels inside the injector. However, if the values of $t_{in}$ are roughly the same for all such states in the injector then the final form of Equation (\ref{eqvarjin1}) will remain unchanged, but $\delta n_{inj}$ will then represent the total electron density in all the injector states.

Similarly, the fluctuation $\delta J_{out}^{j}$ in the tunneling current density from level 1 of the gain stage into the injector is given by the expression,
\begin{eqnarray}
\delta J_{out}^{j} & = &  \frac{\delta J_{out} / \delta (\mu_{1}-E_{1})}{\delta n_{1} / \delta (\mu_{1}-E_{1})} \: \delta n_{1}^{j} + \frac{\delta J_{out}}{\delta (E_{1}-E_{inj}^{\prime} )} \: (\delta E_{1}^{j}- \delta E_{inj}^{\prime j}) + qf_{out}^{j} \label{eqvarjout} \\
& = & \frac{\delta J_{out}}{\delta n_{1}} \: \delta n_{1}^{j} + \frac{\delta J_{out}}{\delta (E_{1}-E_{inj}^{\prime} )} \: (\delta E_{1}^{j}- \delta E_{inj}^{\prime j}) + qf_{out}^{j} \label{eqvarjoutgen}
\end{eqnarray}
The Langevin noise source $f_{out}^{j}$ has the correlation function,
\begin{eqnarray}
WL\,\langle f_{out}^{j}(t)\,f_{out}^{q}(t') \rangle & = & \frac{1}{q}\,(J_{out-forward} + J_{out-backward}) \, \delta_{jq} \, \delta(t-t') \label{FoutFout1} \\
& \approx & \frac{J_{out}}{q} \, \chi_{out} \, \delta_{jq} \, \delta(t-t') \label{outjFout2}
\end{eqnarray} 
In a well designed QCL the backward tunneling current from the injector of the next stage into level 1 of the gain stage is small, and $\chi_{out}$ is expected to be close to unity. $E_{inj}^{\prime j}$ is the energy of the injector level of the next stage into which electrons tunnel from level 1 of the gain stage. $\delta E_{1}^{j} - \delta E_{inj}^{\prime j}$, as before, can be expressed in terms of capacitances,
\begin{equation}
\delta E_{1}^{j} - \delta E_{inj}^{\prime j} =   \frac{q^{2}\delta n_{inj}^{j}}{C_{inj}''} + \frac{q^{2}\delta n_{3}^{j}}{C_{3}''} + \frac{q^{2}\delta n_{2}^{j}}{C_{2}''} + \frac{q^{2}\delta n_{1}^{j}}{C_{1}''} \label{eqengydiff2}
\end{equation}
Using Equation~(\ref{eqV}) and Equation~(\ref{eqengydiff2}), $\delta J_{out}^{j}$ becomes,
\begin{eqnarray}
\frac{\delta J_{out}^{j}}{q} & = & \left( \frac{C_{inj}}{C_{inj}''}G_{out} \right) \: \frac{\delta V^{j}}{q} - \left( \frac{G_{out}}{C_{inj}''} \frac{C_{inj}}{C_{3}} - \frac{G_{out}}{C_{3}''} \right) \: \delta n_{3}^{j} - \left( \frac{G_{out}}{C_{inj}''} \frac{C_{inj}}{C_{2}} - \frac{G_{out}}{C_{2}''} \right) \: \delta n_{2}^{j} \nonumber \\
                             &   & \mbox{} - \left[ \frac{G_{out}}{C_{inj}''} \frac{C_{inj}}{C_{1}} - \left( \frac{1}{t_{out}} + \frac{G_{out}}{C_{3}''} \right) \right] \: \delta n_{1}^{j} + f_{out}^{j} \label{eqvarjout2}
\end{eqnarray}
where $t_{out}$ and $G_{out}$ are,
\begin{equation}
\frac{1}{t_{out}} = \frac{1}{q} \frac{\delta J_{out}}{\delta n_{1}} \hspace{0.5 cm} , \hspace{0.5 cm} G_{out} = q \frac{\delta J_{out}}{\delta (E_{1}-E_{inj}^{\prime})}  \label{eqtoutGout}
\end{equation}
In Equation~(\ref{eqvarjout}) it has been assumed that electrons in the injector relax into the ground state state of the injector sufficiently fast so that electron occupation in the injector levels do not effect the electron escape rate out of level 1 of the gain stage. 

Note that $G_{in}$ and $G_{out}$ can be positive or negative depending upon the relative alignment of the energy levels $E_{inj}$ and $E_{3}$ in the steady state. The scheme used in deriving Equation~(\ref{eqvarjin}) and Equation~(\ref{eqvarjout2}) is fairly general and can be used to derive self-consistent linearized transport equations for a variety of QCL structures. Approximations can be made to simplify Equation (\ref{eqvarjin}) and Equation (\ref{eqvarjout2}). Expression for $\delta J_{in}^{j}$ can also be written as,
\begin{eqnarray}
\frac{\delta J_{in}^{j}}{q} & = & \frac{1}{\tau_{in}}\, \delta n_{inj}^{j} - \frac{1}{\tau_{3}} \, \delta n_{3}^{j} - \frac{1}{\tau_{2}} \, \delta n_{2}^{j} - \frac{1}{\tau_{1}} \, \delta n_{1}^{j} + f_{in}^{j} \label{eqvarjinnew} \\
& = & \left(\frac{C_{inj}}{\tau_{in}}\right) \: \frac{\delta V^{j}}{q} - \left( \frac{1}{\tau_{in}}\frac{C_{inj}}{C_{3}} + \frac{1}{\tau_{3}} \right) \: \delta n_{3}^{j} \nonumber \\
                  &   & - \mbox{ } \left( \frac{1}{\tau_{in}} \frac{C_{inj}}{C_{2}} + \frac{1}{\tau_{2}} \right) \:  \delta n_{2}^{j} - \left(\frac{1}{\tau_{in}}\frac{C_{inj}}{C_{1}} + \frac{1}{\tau_{1}} \right) \: \delta n_{1}^{j} + f_{in}^{j} \label{eqvarjins}
\end{eqnarray}
For the sake of economy of notation new parameters have been introduced in the above equation,
\begin{equation}
\frac{1}{\tau_{in}} = \frac{1}{t_{in}} + \frac{G_{in}}{C_{inj}'} \hspace{0.5 cm},\hspace{0.5 cm} \frac{1}{\tau_{3}} = \frac{1}{t_{3}} + \frac{G_{in}}{C_{3}'} \hspace{0.5 cm},\hspace{0.5 cm} \frac{1}{\tau_{2}} = \frac{G_{in}}{C_{2}'}  \hspace{0.5 cm},\hspace{0.5 cm} \frac{1}{\tau_{1}} = \frac{G_{in}}{C_{1}'} 
\end{equation}
Simple electrostatic arguments can be used to show that $\tau_{2}$ and $\tau_{1}$ will be large, and can be assumed to be infinite.

The injector is assumed to have a large number of closely spaced energy levels. $J_{out}$ is, therefore, largely insensitive to the relative shifts in $E_{1}$ and $E_{inj}'$. This implies that terms containing $G_{out}$ in the expression for $\delta J_{out}^{j}$ may be neglected, and the simplified expression for $\delta J_{out}^{j}$ becomes,
\begin{equation}
\frac{\delta J_{out}^{j}}{q} = \frac{1}{\tau_{out}} \: \delta n_{1}^{j} + f_{out}^{j} \label{eqvarjout2s}
\end{equation}
where $\tau_{out}$ is just $t_{out}$. Equations (\ref{eqvarjins}) and (\ref{eqvarjout2s}) show that in addition to the parameters given in the electron and photon density rate equations (Equations (\ref{eqN3})-(\ref{eqSp})), the paramters necessary for describing electron transport through the gain stage are $C_{inj}$, $C_{3}$, $C_{2}$, $C_{1}$, $\tau_{in}$, $\tau_{3}$, $\tau_{2}$, and $\tau_{1}$.

\subsection{Displacement Currents}
The noise current $\delta J_{ext}$,  which flows in the external circuit, is not equal to $\delta J_{in}^{j}$ or $\delta J_{out}^{j}$. $\delta J_{ext}$ also includes displacement currents and is given by the expression,
\begin{equation}
\delta J_{ext} = \delta J_{in}^{j} + q\frac{d\:\delta n_{inj}^{j}}{d\,t} \label{eqvarext0}
\end{equation}
Since all the gain stages are connected electrically in series, the same current $\delta J_{ext}$ flows through all the gain stages. The second term on the right hand side of Equation~(\ref{eqvarext0}) is the contribution to $\delta J_{ext}$ from displacement currents. Differentiating both sides of Equation~(\ref{eqV}) with respect to time and rearranging yields,  
\begin{displaymath}
q\frac{d\,n_{inj}^{j}}{d\,t} =  C_{inj}\frac{d \delta \, V^{j}}{d \, t} - \sum_{k=1}^{3} \: q\frac{C_{inj}}{C_{k}}\frac{d \, \delta n_{k}^{j}}{d \, t}
\end{displaymath}
Using the above equation the expression for $\delta J_{ext}$ becomes,
\begin{eqnarray}
\delta J_{ext} & = & \delta J_{in}^{j} + C_{inj}\frac{d \delta \, V^{j}}{d \, t} - \sum_{k=1}^{3} \: q\frac{C_{inj}}{C_{k}}\frac{d \, \delta n_{k}^{j}}{d \, t}  \label{eqvarext1} \\
                   & = & \delta J_{out}^{j} + C_{inj}\frac{d \, \delta V^{j}}{d \, t} + \sum_{k=1}^{3}\: q\left( 1 - \frac{C_{inj}}{C_{k}} \right)\frac{d \, \delta n_{k}^{j}}{dt}  \label{eqvarext2}
\end{eqnarray}
Equation~(\ref{eqvarext2}) follows from Equation~(\ref{eqvarext1}) by using the particle number conservation equation,
\begin{displaymath}
\sum_{k=1}^{3} \, q\, \frac{d \, \delta n_{k}^{j}}{dt} = \delta J_{in}^{j} - \delta J_{out}^{j} 
\end{displaymath} 
Equation~(\ref{eqvarext1}) and Equation~(\ref{eqvarext2}) satisfy the Ramo-Shockley theorem~\cite{ramoshock}.

\subsection{Differential Resistance}

Below threshold, the total differential resistance $R_{d}$ of all the gain stages can be calculated by substituting Equations (\ref{n3bth})-(\ref{n1bth}) in Equation (\ref{eqvarjins}),
\begin{eqnarray}
R_{d} & = & \frac{N}{WL}\frac{\tau_{in}}{C_{inj}}\;\left[1 + \left(\frac{1}{\tau_{in}}\frac{C_{inj}}{C_{3}} + \frac{1}{\tau_{3}} \right)\frac{\tau_{32} \tau_{31}}{\tau_{32} + \tau_{31}} \right. \nonumber \\
      &   & \left. \mbox{} + \left(\frac{1}{\tau_{in}}\frac{C_{inj}}{C_{2}} + \frac{1}{\tau_{2}} \right)\frac{\tau_{21} \tau_{31}}{\tau_{32} + \tau_{31}} + \left(\frac{1}{\tau_{in}}\frac{C_{inj}}{C_{1}} + \frac{1}{\tau_{1}} \right)\tau_{out} \right] \label{rdbth} \\
 & = & \frac{N}{WL}\frac{\tau_{in}}{C_{inj}}\;\left( 1+ \theta'_{3}+ \theta'_{2} + \theta_{1}\right) \label{rdbth2}
\end{eqnarray}
Above threshold, the differential resistance can be computed by using Equations (\ref{n3ath})-(\ref{n1ath}) with Equation (\ref{eqvarjins}),
\begin{eqnarray}
R_{d} & = & \frac{N}{WL}\frac{\tau_{in}}{C_{inj}}\;\left[1 + \left(\frac{1}{\tau_{in}}\frac{C_{inj}}{C_{3}} + \frac{1}{\tau_{3}} \right)\frac{\tau_{21} \tau_{31}}{\tau_{21} + \tau_{31}} \right. \nonumber \\
      &   & \left. \mbox{} + \left(\frac{1}{\tau_{in}}\frac{C_{inj}}{C_{2}} + \frac{1}{\tau_{2}} \right)\frac{\tau_{21} \tau_{31}}{\tau_{21} + \tau_{31}} + \left(\frac{1}{\tau_{in}}\frac{C_{inj}}{C_{1}} + \frac{1}{\tau_{1}} \right)\tau_{out} \right] \label{rdath} \\
 & = & \frac{N}{WL}\frac{\tau_{in}}{C_{inj}}\;\left( 1+ \theta_{3}+ \theta_{2}+ \theta_{1}\right) \label{rdath2}
\end{eqnarray}  
Expressions for the parameters $\theta_{3}$, $\theta'_{3}$, $\theta_{2}$, $\theta'_{2}$, and $\theta_{1}$ are given in Appendix~\ref{DiffRes}. Notice the similarity between Equations (\ref{rdbth2}) and (\ref{rdath2}), and Equation (\ref{rdint}) for the differential resistance of interband semiconductor diode lasers given in Appendix~\ref{intnoise}. Unlike the active regions of diode lasers, the active regions of unipolar QCLs are not charge neutral, and as a result various capacitances appear in the expression for the differential resistance of QCLs.

\begin{figure}[t]
\begin{center}
    \epsfig{file=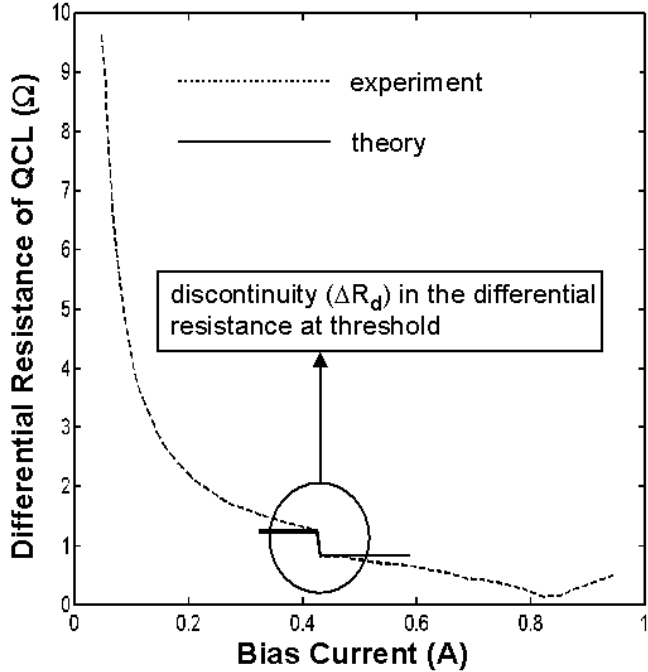,angle=0,width=3.5 in}   
    \caption{ Differential Resistance of a QCL is shown as a function of the bias current. The experimentally measured discontinuity in the differential resistance at threshold is about $0.3 $ $\Omega$. The theoretical model reproduces the discontinuity exactly. The experimental data is taken from Ref.~\cite{faistHP}.}
    \label{figdiffres}
\end{center}
\end{figure}

The discontinuity $\Delta R_{d}$ in the differential resistance at threshold for an $N$ stage QCL is,
\begin{eqnarray}
\Delta R_{d} & = & \frac{N}{WL}\frac{\tau_{in}}{C_{inj}}\;\left[\left(\frac{1}{\tau_{in}}\frac{C_{inj}}{C_{3}} + \frac{1}{\tau_{3}} \right) \left( \frac{\tau_{32} \tau_{31}}{\tau_{32} + \tau_{31}} - \frac{\tau_{21} \tau_{31}}{\tau_{21} + \tau_{31}} \right) \right. \nonumber \\
      &   & \left. \mbox{} + \left(\frac{1}{\tau_{in}}\frac{C_{inj}}{C_{2}} + \frac{1}{\tau_{2}} \right) \left( \frac{\tau_{21} \tau_{31}}{\tau_{32} + \tau_{31}} - \frac{\tau_{21} \tau_{31}}{\tau_{21} + \tau_{31}} \right) \right] \label{delrd} \\
& = & \frac{N}{WL}\frac{\tau_{in}}{C_{inj}}\;\left[ (\theta'_{3}- \theta_{3}) + (\theta'_{2} - \theta_{2}) \right]
\end{eqnarray}
The Incremental change in the potential drop across a gain stage is related to the incremental changes in electron densities through Equation (\ref{eqV}). Therefore, the discontinuity in the differential resistance at threshold results from the discontinuity at threshold in the rate of change of electron densities in levels 3 and 2 of the gain stage with the bias current. Fig.~\ref{figdiffres} shows the calculated and measured differential resistance of a QCL as a function of the bias current. 
The experimental data is taken from Ref.~\cite{faistHP}. The values of the various device parameters are given in Table~\ref{table1}. Values of $\tau_{21}$, $\tau_{31}$ and $\tau_{32}$ are taken from Ref.~\cite{faistHP}. Values of all the capacitances given in Table~\ref{table1} are estimated from the structure of the QCL described in Ref.~\cite{faistHP}. Values of $\tau_{in}$, $\tau_{3}$, and $\tau_{out}$ are estimated from Equations (\ref{eqGintint3}) and (\ref{eqtoutGout}). The total resistance of the ohmic contacts and the superlattice injectors is assumed to be approximately $0.3$ $\Omega$ at threshold. The experimentally observed discontinuity in the differential resistance at threshold is exactly reproduced in the calculated results without the use of any fitting parameters. This agreement suggests that the self-consistent model for the linearized electron transport presented in this paper adequately captures the essential ingredients. 

Diode lasers also exhibit a discontinuity in the differential resistance at threshold. As shown in Appendix~\ref{intnoise}, the discontinuity in the differential resistance of diode lasers at threshold is $K_{_{B}}T/qI_{th}$ times a factor of the order of unity, which can be compared with the more complicated expression given in Equation (\ref{delrd}) for QCLs.

\subsection{Electron Transport in the Superlattice Injector}
In this paper no attention has been given to modeling the electron transport through the superlattice injector. In the absence of any bias current, the energy levels in the injector are not suitably aligned to facilitate electron transport, and the resistance of the injector region is large. As the bias current is gradually increased electrons pile up in different quantum wells until their presence modifies the potential profile and aligns the energy levels such that the electron current can flow. Once the injector has been {\em turned on} in this fashion, the differential resistance of the injector region is negligible, and the only bottleneck for electron transport is the gain stage. As a result of the small differential resistance of the injector region, any current noise originating in the injector region will not couple well into the external circuit. Therefore, electron transport in the injector region may be ignored when modeling noise. If necessary, the impedance of the superlattice injectors can be modeled with a lumped element, and the current noise generated inside the injector regions can be modeled with a voltage source in series (or a current source in parallel) with that element, as shown below. A detailed discussion of the current noise in superlattice structures is beyond the scope of this paper.

\subsection{Biasing Electrical Circuit}
\begin{figure}[t]
\begin{center}
   \epsfig{file=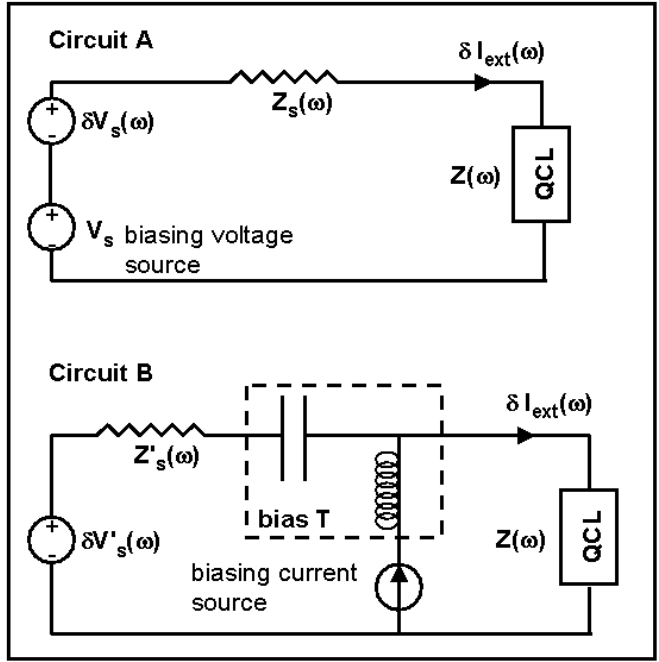,angle=0,width=3.5in} 
    \caption{ Circuits used for biasing QCLs.}
    \label{figBiasCir}
\end{center}
\end{figure}

Two electrical circuits for biasing QCLs are shown in Fig.~\ref{figBiasCir}. 
\begin{figure}[htb]
\begin{center}
   \epsfig{file=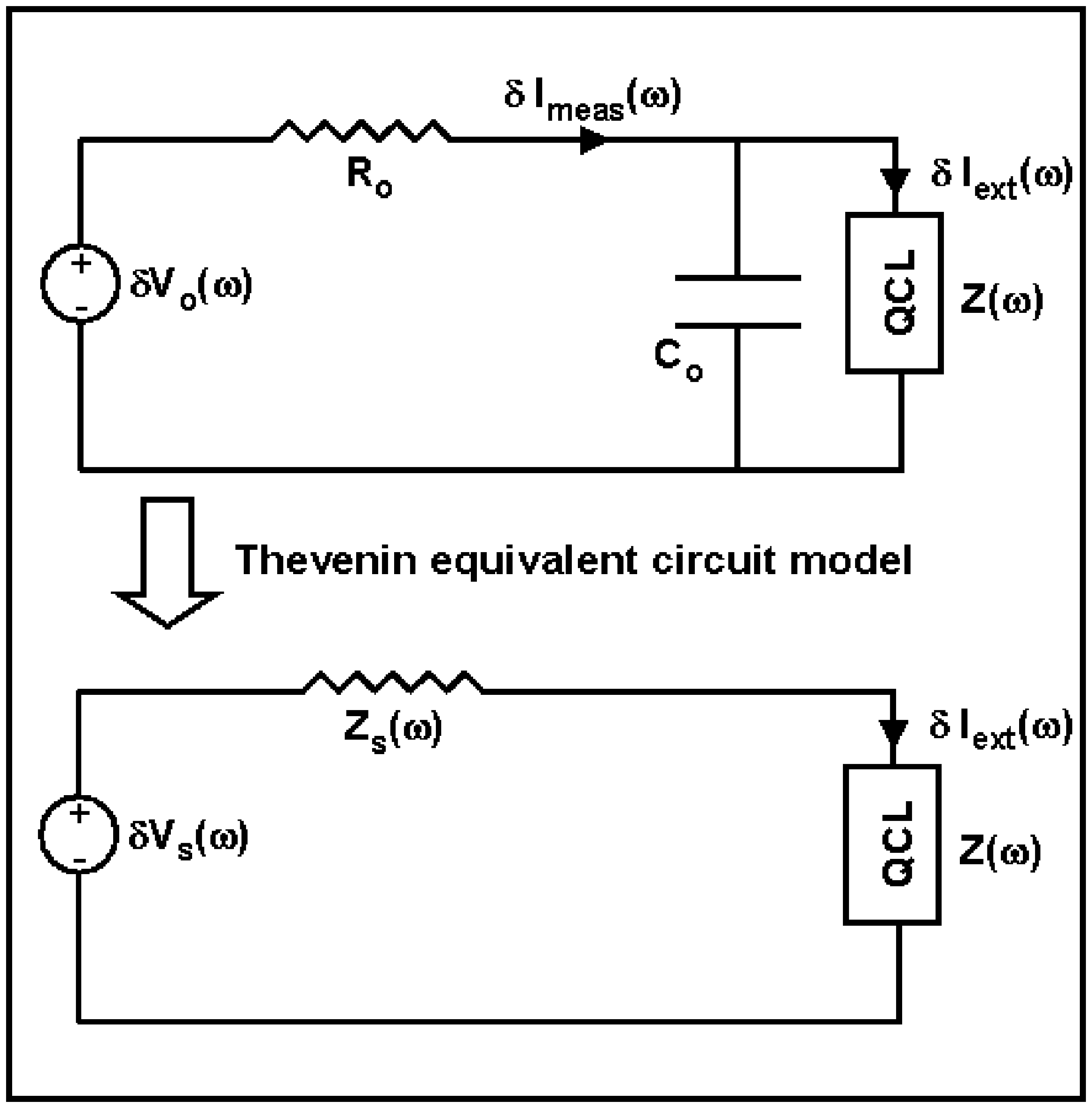,angle=0,width=3.5in}
    \caption{ Thevenin equivalent circuit model indicating the distinction between $\delta I_{ext}(\omega)$ and $\delta I_{meas}(\omega)$.}
    \label{figCapR}
\end{center}
\end{figure}
In circuit A the QCL, with an impedance $Z(\omega)$, is biased with a voltage source $V_{s}$ in series with an impedance $Z_{s}(\omega)$. The thermal noise originating in the impedance $Z_{s}(\omega)$ is modeled by adding a voltage noise source $\delta V_{s}$. The differential impedance of the superlattice injector and the current noise generated by the injector can also be modeled with an impedance and a voltage noise source in series (or a current noise source in parallel) with that impedance. For the sake of economy of notation it will be assumed that the impedance $Z_{s}(\omega)$ represents not just an external circuit impedance but the Thevenin equivalent impedance of the superlattice injectors, device ohmic contacts, external circuit resistances, and device and circuit parasitics, and the voltage noise source $\delta V_{s}$ represents the Thevenin equivalent of their individual noise sources. Only the gain stages inside the QCL are not included within $Z_{s}(\omega)$ and they are represented by the impedance $Z(\omega)$. However, $Z(\omega)$ will be loosely referred to as the impedance of the QCL. The current noise generated by the gain stages can also be modeled by adding a current noise source in parallel with $Z(\omega)$, as shown in later sections.

Direct current modulation of the QCL can be achieved by adding an RF voltage source in series with $V_{s}$, and this RF voltage source can also be represented by the voltage source $\delta V_{s}$. From the context it will be clear whether $\delta V_{s}$ represents a RF signal source or a noise source. 

In circuit B the QCL is biased with a current source in series with an ideal inductor, and it is also capacitively coupled to a voltage source $\delta V'_{s}$ with a series impedance $Z'_{s}(\omega)$. If at frequencies of interest the inductor and the coupling capacitor are almost open and short, respectively, then this circuit is also equivalent to circuit A. Therefore, in this paper only circuit A will be considered. In circuit A the current $\delta I_{ext}$ can be expressed as,
\begin{equation}
\delta I_{ext}(\omega) =  \delta J_{ext}(\omega) \, WL = \frac{\delta V_{s}(\omega) - {\displaystyle \sum_{j=1}^{N}}\,\delta V^{j}(\omega)}{Z_{s}(\omega)} \label{eqVs}
\end{equation}
It is important to note here that $\delta I_{ext}(\omega)$ may not be the noise current which would be measured in an experiment. For example, suppose that the QCL has a parasitic capacitance $C_{o}$ in parallel with the actual device, as shown in Fig.~\ref{figCapR}. The QCL is driven with a series resistor $R_{o}$ and a noise voltage source $\delta V_{o}(\omega)$ representing the thermal noise in the resistor $R_{o}$. Fig.~\ref{figCapR} shows the distinction between the noise current $\delta I_{ext}(\omega)$ defined in Equation (\ref{eqVs}), and the noise current $\delta I_{meas}(\omega)$ that would be measured in an experiment. Notice that the Thevenin equivalent impedance $Z_{s}(\omega)$ is a parallel combination of the resistance $R_{o}$ and the capacitance $C_{o}$. $Z_{s}(\omega)$ and $\delta V_{s}(\omega)$ are,
\begin{equation}
Z_{s}(\omega) = \frac{R_{o}}{\left(1 + j\omega \, R_{o}C_{o} \right)} \hspace{1 cm} \delta V_{s}(\omega) = \frac{\delta V_{o}(\omega)}{\left(1 + j\omega \, R_{o}C_{o} \right)}
\end{equation}
and the relation between $\delta I_{ext}(\omega)$ and $\delta I_{meas}(\omega)$ is,
\begin{equation}
\delta I_{ext}(\omega) = \frac{\delta I_{meas}(\omega)}{\left( 1 + j\omega \, Z(\omega)C_{o} \right)} 
\end{equation}
Choosing to define $Z_{s}(\omega)$ this way helps in formulating a noise model that is independent of the specific nature of the device parasitics.

\section{Solution of the Coupled Equations} \label{SolOfEq}

\subsection{Current Modulation Response}

In this section the response $\delta P_{out}(\omega)/\delta I_{ext}(\omega)$ of QCLs to external sinusoidal current modulation $\delta I_{ext}(\omega)$ is determined~\footnote{It is assumed that $P_{out}(t) = P_{out} + {\rm Real} \{ \delta P_{out}(\omega) \: e^{j\,\omega t} \} $ and $I_{ext}(t) = I + {\rm Real} \{ \delta I_{ext}(\omega)\: e^{j\,\omega t} \}$ }. The frequency dependence of the photon noise spectral density of semiconductor lasers is directly related to the frequency dependence of the current modulation response. It is therefore instructive to look at the modulation response of QCLs. The modulation response can be found by solving Equations (\ref{eqN3})-(\ref{eqPo}), together with Equations (\ref{eqvarjins}) and (\ref{eqvarjout2s}), and setting all the noise sources equal to zero. The external circuit constraints expressed in Equations (\ref{eqvarext1}) and (\ref{eqVs}) must also be enforced. Equations (\ref{eqN3})-(\ref{eqN1}) for each gain stage are coupled to the same set of equations for all the other gain stages through Equations (\ref{eqSp}) and (\ref{eqvarext1}). Such a large system of coupled equations can be solved only numerically. A numerical approach, although simple to implement, is not very instructive. With the approximation that all gain stages have the same confinement factor $\Gamma$, a significant portion of the work can be done analytically. This approach will be followed in this paper. All equations, unless stated otherwise, will be expressed in the frequency domain.

The relationship between the current density $\delta J_{ext}(\omega)$, which flows in the external circuit, and the total potential drop $\delta V(\omega)$ across all the gain section can be obtained by using Equation (\ref{eqvarjins}) in Equation (\ref{eqvarext1}), and summing over the index $j$, 
\begin{equation}
\frac{C_{inj}}{\tau_{in}} \, \frac{\delta V(\omega)}{q} = \frac{N}{(1 + j\omega \, \tau_{in})} \: \frac{\delta J_{ext}(\omega)}{q} + \sum_{k=1}^{3} \, \left[ \frac{1}{\tau_{in}}\,\frac{C_{inj}}{C_{k}} + \frac{1}{\tau_{k}\,(1 + j\omega \, \tau_{in})} \right] \, \delta N_{k}(\omega) \label{eqall}
\end{equation}
The following new symbols have been introduced in Equation (\ref{eqall}),
\begin{displaymath}
\delta N_{k}(\omega) = \sum_{j=1}^{N}\,\delta n_{k}^{j}(\omega) 
\: \: \: \: \: \: \: \: 
\{{\rm where} \: k=1,2,3 \}
\: \: \: \: \: \: \: \:
{\rm and}
\: \: \: \: \: \: \: \:
\delta V(\omega) = \sum_{j=1}^{N}\,\delta V^{j}(\omega)
\end{displaymath}
Using Equations (\ref{eqvarjins}), (\ref{eqvarjout2s}) and (\ref{eqall}) in Equations (\ref{eqN3})-(\ref{eqSp}), summing over the index $j$, and arranging the resulting equations in a matrix form gives,
\begin{equation}
\left[
\begin{array}{lccr}
{\bf D}_{11} & {\bf D}_{12} & {\bf D}_{13} & 0 \\
0 & {\bf D}_{22} & {\bf D}_{23} & {\bf D}_{24} \\
{\bf D}_{31} & {\bf D}_{32} & {\bf D}_{33} & {\bf D}_{34} \\
0 & {\bf D}_{42} & {\bf D}_{43} & {\bf D}_{44} \\
\end{array}
\right]
\: 
\left[ 
\begin{array}{c} 
\delta N_{1}(\omega) \\ 
\delta N_{2}(\omega) \\ 
\delta N_{3}(\omega) \\ 
\delta S_{p}(\omega) 
\end{array} 
\right] = 
\frac{N}{(1 + j\omega \, \tau_{in})} \: \frac{\delta J_{ext}(\omega)}{q}
\:
\left[ 
\begin{array}{c}  
0 \\
0 \\
1 \\ 
0  
\end{array} 
\right] 
\label{eqD}
\end{equation}
The coefficients of the matrix {\bf D} can be found from Equations (\ref{eqN3})-(\ref{eqSp}), and they are given in Appendix~\ref{dmatrix}. The solution of Equation (\ref{eqD}) can be written as,
\begin{equation}
\left[ 
\begin{array}{c} 
\delta N_{1}(\omega) \\ 
\delta N_{2}(\omega) \\ 
\delta N_{3}(\omega) \\ 
\delta S_{p}(\omega) 
\end{array} 
\right] = 
\left[ 
\begin{array}{c}  
{\bf D}_{13}^{-1}(\omega)\\ 
{\bf D}_{23}^{-1}(\omega)\\ 
{\bf D}_{33}^{-1}(\omega)\\ 
{\bf D}_{43}^{-1}(\omega)
\end{array} 
\right] 
\:
\frac{N}{(1 + j\omega \, \tau_{in})} \: \frac{\delta J_{ext}(\omega)}{q}
\label{eqinvD}
\end{equation}
The coefficients of the matrix ${\bf D}^{-1}$ are given in Appendix~\ref{idmatrix}. Equation (\ref{eqinvD}) can be used in Equation (\ref{eqall}) to calculate the total impedance $Z(\omega)$ of all the gain stages,
\begin{equation}
Z(\omega) =  \frac{N}{WL} \, \frac{\tau_{inj}}{C_{inj}} \, \frac{1}{(1 + j\omega \, \tau_{in})} \left[1 + \sum_{k=1}^{3} \, \left( \frac{1}{\tau_{in}}\,\frac{C_{inj}}{C_{k}} + \frac{1}{\tau_{k}\,(1 + j\omega \, \tau_{in})} \right) \, {\bf D}_{k3}^{-1}(\omega) \right] \label{eqZ}  
\end{equation} 
$Z(\omega = 0)$ is just the differential resistance $R_{d}$ of the QCL given earlier in Equations (\ref{rdbth}) and (\ref{rdath}). Finally, from Equation (\ref{eqPo}) and Equation (\ref{eqinvD}), the current modulation response can be written as,
\begin{equation}
\frac{\delta P_{out}(\omega)}{\delta I_{ext}(\omega)} =  \eta_{o}\,\frac{h \nu}{q}\,\frac{N}{\tau_{p}} \,\frac{\displaystyle {\bf D}_{43}^{-1}(\omega)}{(1 + j\omega\,\tau_{in})} \label{eqmodres} 
\end{equation}
Above threshold, an analytical approximation for the modulation response can be found in the limit $\omega \tau_{in} \ll 1$ and $\{\tau_{2}\, , \, \tau_{1} \} \rightarrow \infty$ (Appendix~\ref{idmatrix}),
\begin{eqnarray}
\frac{\delta P_{out}(\omega)}{\delta I_{ext}(\omega)} & = & \eta_{o}\,\frac{h \nu}{q}\,\frac{N}{\tau_{p}\,\tau_{st}} \,\left[ j\omega + \left( \frac{1}{\tau_{21}} - \frac{1}{\tau_{32}} \right) \right]\,\left\{ (j\omega)^{3}\,\left(1 + \frac{\tau_{in}}{\tau_{3}} \right) \right. \nonumber \\
& & \left. \mbox{} + (j\omega)^{2}\,\left[ \frac{1}{\tau_{21}}\left( 1 + \frac{\tau_{in}}{\tau_{3}} \right) + \frac{1}{\tau_{31}} + \frac{1}{\tau_{32}} + \frac{1}{\tau_{st}}\left( 2 + \frac{\tau_{in}}{\tau_{3}} \right)  \right] \right. \nonumber \\
& & \left. \mbox{} + j\omega \,\left[ \frac{1}{\tau_{st}}\left( \frac{1}{\tau_{21}} + \frac{1}{\tau_{31}} \right) + \frac{1}{\tau_{21}\,\tau_{31}} + \frac{1}{\tau_{21}\,\tau_{32}} + \frac{1}{\tau_{p}\,\tau_{st}}\left( 2 + \frac{\tau_{in}}{\tau_{3}} \right) \right] \right. \nonumber \\
& & \left. \mbox{} + \frac{1}{\tau_{p}\,\tau_{st}}\,\left(\frac{1}{\tau_{21}} + \frac{1}{\tau_{31}} \right) \right\}^{-1} \label{eqmodres2} \\ 
& = & \eta_{o}\,\frac{h \nu}{q}\,N\,\left[ j\omega \frac{\tau_{21}\,\tau_{31}}{\left( \tau_{21} + \tau_{31} \right)} + \eta_{r} \right]\,H(\omega)
\end{eqnarray}
Expression for the current modulation response function $H(\omega)$ is given in Appendix~\ref{idmatrix}. In Equation (\ref{eqmodres2}) $\tau_{st}$ is the differential lifetime associated with stimulated and spontaneous photon emission into the lasing mode, and is given by the relation,
\begin{equation}
\frac{1}{\tau_{st}} = \Gamma v_{g} \, a \, \left( S_{p} + \frac{n_{sp}}{WL} \right) \label{eqtaust}
\end{equation}   
For frequencies less than the inverse of the smallest time constant of the QCL (which is usually either $\tau_{21}$ or $\tau_{st}$), Equation (\ref{eqmodres2}) can be further simplified, and put in the standard form used for semiconductor diode lasers (see Appendix~\ref{intnoise}, and Ref.~\cite{coldren}),
\begin{equation}
\frac{\delta P_{out}(\omega)}{\delta I_{ext}(\omega)} = \eta_{o}\,\eta_{r}\,\frac{h \nu}{q} \, N \: \frac{\omega_{R}^{2}}{(\omega_{R}^{2} - \omega^{2} + j\omega \gamma)} \label{eqmodres3}
\end{equation}
where $\eta_{r}$ is the radiative efficiency defined in Equation (\ref{eqetar}), and the relaxation oscillation frequency $\omega_{R}$ and the damping constant $\gamma$ are,
\begin{equation}
\omega_{R}^{2} = \frac{\displaystyle \frac{1}{\tau_{p}\,\tau_{st}}\,\left(1 + \frac{\tau_{21}}{\tau_{31}} \right)}{\displaystyle \left[ 1 + \frac{\tau_{21}}{\tau_{31}} + \frac{\tau_{21}}{\tau_{32}} + \frac{\tau_{in}}{\tau_{3}} + \frac{\tau_{21}}{\tau_{st}} \, \left( 2 + \frac{\tau_{in}}{\tau_{3}} \right) \right]}  \label{eqwr}
\end{equation}
\begin{equation}
\gamma = \frac{\displaystyle \left[ \frac{1}{\tau_{st}}\,\left(1 + \frac{\tau_{21}}{\tau_{31}} \right) + \frac{1}{\tau_{31}} + \frac{1}{\tau_{32}} + \frac{\tau_{21}}{\tau_{p}\,\tau_{st}}\,\left(2 + \frac{\tau_{in}}{\tau_{3}} \right) \right] }{\displaystyle \left[ 1 + \frac{\tau_{21}}{\tau_{31}} + \frac{\tau_{21}}{\tau_{32}} + \frac{\tau_{in}}{\tau_{3}} + \frac{\tau_{21}}{\tau_{st}} \, \left( 2 + \frac{\tau_{in}}{\tau_{3}} \right) \right]} \label{eqdamp}
\end{equation}
The damping constant $\gamma$ can be related to $\omega_{R}$,
\begin{equation}
\gamma = K\;\omega_{R}^{2} + \gamma_{o} \label{eqK}
\end{equation}
where, 
\begin{equation}
K = \tau_{p} 
\end{equation} 
\begin{equation}
\gamma_{o}  =  \frac{\displaystyle \left[ \frac{1}{\tau_{31}} + \frac{1}{\tau_{32}} + \frac{\tau_{21}}{\tau_{p}\,\tau_{st}}\,\left(2 + \frac{\tau_{in}}{\tau_{3}} \right) \right] }{\displaystyle \left[ 1 + \frac{\tau_{21}}{\tau_{31}} + \frac{\tau_{21}}{\tau_{32}} + \frac{\tau_{in}}{\tau_{3}} + \frac{\tau_{21}}{\tau_{st}} \, \left( 2 + \frac{\tau_{in}}{\tau_{3}} \right) \right]} \label{eqKg}
\end{equation}
The $K$-factor describes the damping of the QCL modulation response at high photon densities. $\gamma_{o}$ has a weak dependence on the photon density through $\tau_{st}$, and it approaches $1/\tau_{p}$ at large photon densities.

If the condition $\omega_{R} < \gamma/\sqrt{2}$ is satisfied then Equation (\ref{eqmodres3}) describes a second order over-damped system. For many QCLs that have been reported in the literature this condition holds true above threshold. Using the values of device parameters from Table~\ref{table1}, $\omega_{R}$ and $\gamma$ can be calculated. If we assume that the output power of the laser is around 150 mW, then from Equations (\ref{eqtaup}) and (\ref{eqtaust}) $\tau_{p}$ and $\tau_{st}$ are approximately 7 ps and 2.8 ps, respectively. The resulting value of $\gamma$ is more than three times larger than that of $\omega_{R}$. The internal time constants in QCLs are usually smaller than the photon lifetime $\tau_{p}$, and therefore the modulation response of QCLs is over-damped. In contrast, the current modulation response of semiconductor diode lasers is under-damped, and becomes over-damped only at very large bias currents when $\tau_{st}$ becomes small~\cite{coldren}.

For QCLs the 3 dB frequency, which is defined to be the frequency at which the square modulus of the laser modulation response becomes one half of its value at zero frequency, can be found from the simplified expression for the modulation response in Equation (\ref{eqmodres3}),
\begin{equation}
\omega^{2}_{3\:{\rm dB}} = \sqrt{(\frac{\gamma^{2}}{2} - \omega_{R}^{2})^{2} + \omega_{R}^{4}} \: \: \: - \: (\frac{\gamma^{2}}{2} - \omega_{R}^{2}) \label{eqwr3db}
\end{equation}
As the photon density inside the laser cavity increases the 3 dB frequency also increases but it asymptotically approaches an upper limit $\omega_{\rm 3\: dB\:|\: max}$. This maximum attainable 3 dB bandwidth can be calculated from Equation (\ref{eqwr3db}), and it comes out to be,
\begin{equation}
\omega_{\rm 3\: dB\:|\: max} \approx \frac{1}{\tau_{p}} \label{eqwr3dbmax}
\end{equation}
As long as the computed value of $\omega_{\rm 3\: dB\:|\: max}$ is less than $1/\tau_{in}$, $1/\tau_{21}$, and $1/\tau_{st}$, the approximations made in deriving Equation (\ref{eqmodres3}) are justified. Equation (\ref{eqwr3dbmax}) confirms the intuition that a laser cannot be modulated much faster than the inverse of the photon lifetime inside the laser cavity. As shown in Appendix~\ref{intnoise}, in diode lasers the value of $\omega_{\rm 3\: dB\:|\: max}$ equals $\sqrt{2}/\tau_{p}$. The difference of a factor of $\sqrt{2}$ comes from the fact that in diode lasers the modulation response is under-damped (see Appendix~\ref{intnoise}).

As in diode lasers, the photon lifetime imposes a fundamental limit on how fast QCLs can be modulated. It is not uncommon to find predictions of THz modulation bandwidths for QCLs in literature~\cite{shore1}. However, for all the QCLs reported in the literature so far, the photon lifetime is the longest of all the time constants and it is the dominant factor that would limit the modulation bandwidth of these QCLs to tens of GHz instead of THz.

\begin{figure}[t]
\begin{center}
   \epsfig{file=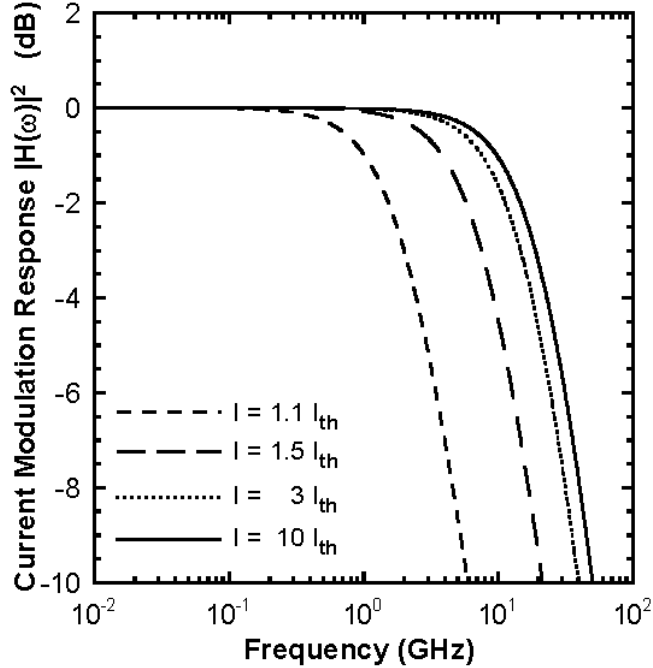,angle=0,width=3.5in}
    \caption{ Direct current modulation response ($20\,\log _{10}[|H(\omega)|/|H(0)|]$) of QCLs is plotted as a function of the frequency for different bias currents. Modulation response shown in the figure has been normalized w.r.t its value at zero frequency. For the device parameters see Table~\ref{table1}.}
    \label{figmodband}
\end{center}
\end{figure}

Fig.~\ref{figmodband} shows the calculated modulation response of a QCL as a function of the frequency for different values of the bias current. 
The values of the different parameters of the QCL are taken from Ref.~\cite{faistHP}, and are given in Table~\ref{table1}. In the numerical calculations values of all the device time constants (except $\tau_{st}$) were assumed to be independent of the bias. Fig.~\ref{figmodband} shows that at low bias currents the 3 dB frequency increases with the bias current, and at high bias currents the 3 dB frequency saturates to a value which is well approximated by $1/(2\pi\tau_{p}) = 21$ GHz. The analysis carried out in this paper does not take into account device heating which may also be important in limiting the modulation bandwidth of QCLs at large biases.     

\begin{figure}[t]
\begin{center}
   \epsfig{file=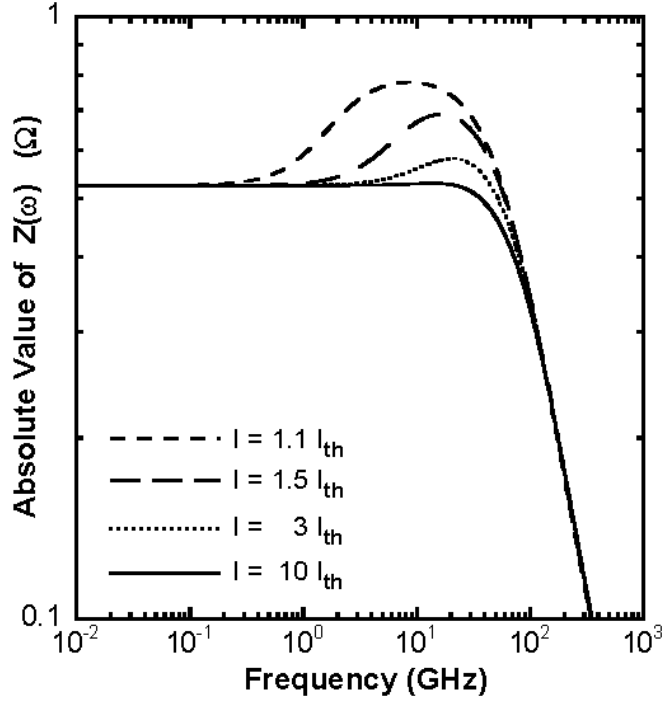,angle=0,width=3.5in}
    \caption{ Absolute value of the QCL impedance $Z(\omega)$ is plotted as a function of the frequency for different bias currents. The peaks in the values of $Z(\omega)$ are not because of relaxation oscillations, since the modulation response of the QCL is over-damped, but because the smallest zero of $Z(\omega)$ is smaller than its smallest pole.}
    \label{figzvsfreq}
\end{center}
\end{figure}

Fig.~\ref{figzvsfreq} shows the impedance $Z(\omega)$ of the QCL plotted as a function of the frequency for different bias currents. 
The peaks in the values of $Z(\omega)$ are not due to relaxation oscillations since, as already pointed out earlier, the modulation response of the QCL is over-damped. The peaks are due to the fact that the smallest zero of $Z(\omega)$ is at a frequency which is smaller than the frequency of its smallest pole. Impedance measurements can therefore provide valuable information about the time scales associated with electron dynamics in QCLs.

\subsection{Laser Intensity Noise and Current Noise}
In this section the current noise in the external circuit and the intensity noise in the output power from QCLs is calculated. In the Langevin equation formalism noise is added through the Langevin noise sources which were introduced in Equations (\ref{eqN3})-(\ref{eqPo}) and also in Equations (\ref{eqvarjins}) and (\ref{eqvarjout2s}). In addition, any noise originating in the external circuit and in the superlattice injectors can also contribute to the current noise and the photon noise, and as already explained earlier, this noise can be represented by the voltage source $\delta V_{s}$. In this paper it is assumed that $\delta V_{s}$ represents the thermal noise originating in the series impedance $Z_{s}(\omega)$, and its correlation function is,
\begin{equation}
\langle \delta V_{s}(\omega)\; \delta V_{s}(\omega') \rangle = 2K_{B}T\,\,{\rm Real} \{ Z_{s}(\omega) \}  \: 2\pi \, \delta(\omega-\omega') \label{VsVs}
\end{equation}
By assuming the above correlation function for the noise source $\delta V_{s}$ we are ignoring any noise which may be contributed by the superlattice injectors.

Including the Langevin noise sources Equation (\ref{eqD}) can be written as,
\begin{equation}
\left[
\begin{array}{l c c r}
{\bf D}_{11} & {\bf D}_{12} & {\bf D}_{13} & 0 \\
0 & {\bf D}_{22} & {\bf D}_{23} & {\bf D}_{24} \\
{\bf D}_{31} & {\bf D}_{32} & {\bf D}_{33} & {\bf D}_{34} \\
0 & {\bf D}_{42} & {\bf D}_{43} & {\bf D}_{44} \\
\end{array}
\right]
\: 
\left[ 
\begin{array}{c} 
\delta N_{1}(\omega) \\ 
\delta N_{2}(\omega) \\ 
\delta N_{3}(\omega) \\ 
\delta S_{p}(\omega) 
\end{array} 
\right] = 
\frac{N}{(1 + j\omega \, \tau_{in})} \: \frac{\delta J_{ext}(\omega)}{q}
\:
\left[ 
\begin{array}{c}  
0 \\
0 \\
1 \\ 
0  
\end{array} 
\right] 
+ 
\left[
\begin{array}{c}
F_{1}(\omega) \\
F_{2}(\omega) \\
F_{3}(\omega) \\
F_{4}(\omega) 
\end{array}
\right] \label{eqDF}
\end{equation}
The expressions for the noise sources $F_{1}$, $F_{2}$, $F_{3}$, and $F_{4}$ are,
\begin{equation}
F_{1}(\omega) =  \sum_{j=1}^{N}\,F^{j}_{1} = \sum_{j=1}^{N} \, \left[ f_{31}^{j}(\omega) + f_{21}^{j}(\omega) - f_{out}^{j}(\omega) \right] \label{eqF1}
\end{equation}
\begin{equation}
F_{2}(\omega) = \sum_{j=1}^{N}\,F^{j}_{2} =  \sum_{j=1}^{N} \, \left[ f_{32}^{j}(\omega) - f_{21}^{j}(\omega) + f_{RN}^{j}(\omega) \right]
\end{equation}
\begin{equation}
F_{3}(\omega) = \sum_{j=1}^{N}\,F^{j}_{3} = \sum_{j=1}^{N} \, \left[ f_{in}^{j}(\omega)\,\frac{j\omega \, \tau_{in}}{(1 + j\omega \, \tau_{in})} - f_{32}^{j}(\omega) - f_{31}^{j}(\omega) - f_{RN}^{j}(\omega) \right]
\end{equation}
\begin{equation}  
F_{4}(\omega) = \sum_{j=1}^{N}\,F^{j}_{4} = \sum_{j=1}^{N} \, \left[ f^{j}_{RS}(\omega) - \frac{F_{L}(\omega)}{N} \right] \label{eqF4}
\end{equation}
The solution of Equation (\ref{eqDF}) can be written as,
\begin{equation}
\left[ 
\begin{array}{c} 
\delta N_{1}(\omega) \\ 
\delta N_{2}(\omega) \\ 
\delta N_{3}(\omega) \\ 
\delta S_{p}(\omega) 
\end{array} 
\right] = 
\left[ 
\begin{array}{c}  
{\bf D}_{13}^{-1}(\omega)\\ 
{\bf D}_{23}^{-1}(\omega)\\ 
{\bf D}_{33}^{-1}(\omega)\\ 
{\bf D}_{43}^{-1}(\omega)
\end{array} 
\right] 
\:
\frac{N}{(1 + j\omega \, \tau_{in})} \: \frac{\delta J_{ext}(\omega)}{q}
\:
+
\:
\left[ 
\begin{array}{c}
\sum_{l=1}^{4}\,{\bf D}_{1l}^{-1}(\omega)\,F_{l}(\omega)\\ 
\sum_{l=1}^{4}\,{\bf D}_{2l}^{-1}(\omega)\,F_{l}(\omega)\\ 
\sum_{l=1}^{4}\,{\bf D}_{3l}^{-1}(\omega)\,F_{l}(\omega)\\ 
\sum_{l=1}^{4}\,{\bf D}_{4l}^{-1}(\omega)\,F_{l}(\omega)
\end{array} 
\right] 
\label{eqinvDF}
\end{equation}
$\delta J_{ext}(\omega)$ in Equation (\ref{eqinvDF}) is an unknown. Using Equation (\ref{eqvarjins}) in Equation(\ref{eqvarext1}), summing over the index $j$, and making use of Equation (\ref{eqinvDF}) yields,
\begin{eqnarray} 
\delta V(\omega) & = & Z(\omega)\,\delta I_{ext}(\omega) \nonumber \\
& & \mbox{} - q\,\frac{\tau_{in}}{C_{inj}}\,\left[ \frac{F_{in}(\omega)}{(1 + j\omega \, \tau_{in})} - \sum_{k=1}^{3} \sum_{l=1}^{4} \, \left( \frac{1}{\tau_{in}} \, \frac{C_{inj}}{C_{k}} + \frac{1}{\tau_{k}(1 + j\omega \, \tau_{in})} \right) {\bf D}_{kl}^{-1}(\omega)\,F_{l}(\omega) \right] \label{eqdeltaV}
\end{eqnarray}
where $F_{in} = \sum_{j=1}^{N}\,f_{in}^{j}$. 
Substituting the value of $\delta V(\omega)$ from Equation (\ref{eqVs}) in Equation (\ref{eqdeltaV}), we get the final expression for the current fluctuations $\delta I_{ext}(\omega)$ in the external circuit,
\begin{eqnarray}
\delta I_{ext}(\omega) & = & \frac{\delta V_{s}(\omega)}{(Z(\omega) + Z_{s}(\omega))} + \frac{q}{(Z(\omega) + Z_{s}(\omega))} \, \frac{\tau_{in}}{C_{inj}}\,\left[ \frac{F_{in}(\omega)}{(1 + j\omega \, \tau_{in})} \right. \nonumber \\ 
& & \hspace{3 cm} \left. \mbox{} - \sum_{k=1}^{3} \sum_{l=1}^{4} \, \left( \frac{1}{\tau_{in}} \, \frac{C_{inj}}{C_{k}} + \frac{1}{\tau_{k}(1 + j\omega \, \tau_{in})} \right) {\bf D}_{kl}^{-1}(\omega)\,F_{l}(\omega) \right] \label{eqdeltaIext}
\end{eqnarray} 
The fluctuation $\delta P_{out}(\omega)$ in the output power can be obtained by substituting Equation (\ref{eqdeltaIext}) in Equation (\ref{eqinvDF}), and using Equation (\ref{eqPo}),
\begin{eqnarray}
\delta P_{out}(\omega) & = & \eta_{o}\,\frac{h \nu}{q}\,\frac{N}{\tau_{p}} \, \frac{{\bf D}_{43}^{-1}(\omega)}{(1 + j\omega\,\tau_{in})}\,\delta I_{ext}(\omega) + \eta_{o}\,\frac{h \nu}{q}\,\frac{WL}{\tau_{p}} \,\left[ q\,\sum_{l=1}^{4}\,{\bf D}_{4l}^{-1}(\omega)\,F_{l}(\omega) \right] + F_{o}(\omega) \label{eqdeltaPo1}
%\delta P_{out}(\omega) & = & \eta_{o}\,\frac{h \nu}{q}\,\frac{N}{\tau_{p}} \, \frac{{\bf D}_{43}^{-1}(\omega)}{(1 + j\omega\,\tau_{in})} \, \left[ \frac{\delta V_{s}(\omega)}{(Z(\omega) + Z_{s}(\omega))} \right. \nonumber \\
%& & \left. \mbox{} + \frac{q}{(Z(\omega) + Z_{s}(\omega))} \, \frac{\tau_{in}}{C_{inj}}\,\left\{ \frac{F_{in}(\omega)}{(1 + j\omega \, \tau_{in})} - \sum_{k=1}^{3} \sum_{l=1}^{4} \, \left( \frac{1}{\tau_{in}} \, \frac{C_{inj}}{C_{k}} + \frac{1}{\tau_{k}(1 + j\omega \, \tau_{in})} \right) {\bf D}_{kl}^{-1}(\omega)\,F_{l}(\omega) \right\} \right] \nonumber \\
%& & \mbox{} + \eta_{o}\,\frac{h \nu}{q}\,\frac{WL}{\tau_{p}} \,\left[ q\,\sum_{l=1}^{4}\,{\bf D}_{4l}^{-1}(\omega)\,F_{l}(\omega) \right] + F_{o}(\omega) \label{eqdeltaPo}
\end{eqnarray}

\section{Current Noise: Results and Discussion} \label{DisscussCN}

\subsection{Circuit Models for the Current Noise} 

\begin{figure}[pht]
\begin{center}
   \epsfig{file=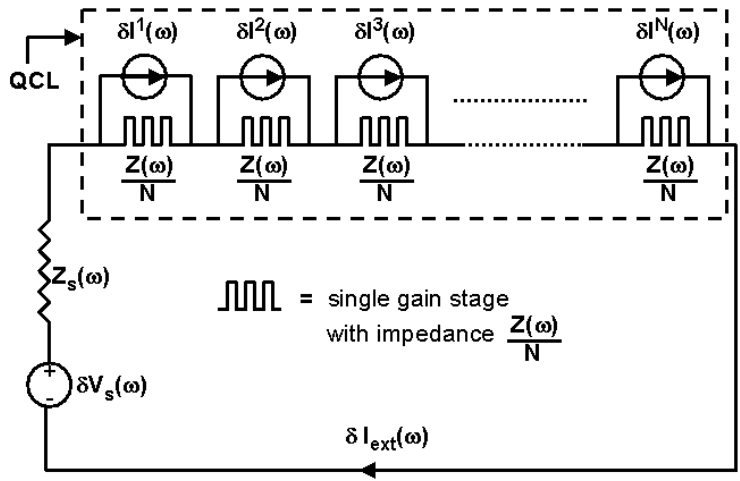,angle=0,width=4.0in}  
    \caption{ Circuit model for the current fluctuations.}
    \label{figCirMod}
\end{center}
\end{figure}

Equation (\ref{eqdeltaIext}) for the current fluctuations in the external circuit can be written as,
\begin{equation}
\delta I_{ext}(\omega) = \frac{\delta V_{s}(\omega)}{(Z(\omega) + Z_{s}(\omega))} + \frac{Z(\omega)/N}{(Z(\omega) + Z_{s}(\omega))} \, \sum_{j=1}^{N} \: \delta I^{j}(\omega) \label{eqCirMod}
\end{equation}
Expression for $\delta I^{j}(\omega)$ is given in Appendix~\ref{CirMod}. Observing that $Z(\omega)/N$ is the differential impedance of a single gain stage, a circuit model for the current fluctuations can be constructed by attaching a current noise source $\delta I^{j}(\omega)$ in parallel with the $j$th gain stage, as shown in Fig.~\ref{figCirMod}. Current noise sources belonging to two different gain stages are not independent but are positively correlated, as shown in Appendix~\ref{CirMod}. This is because carrier density fluctuations in all the gain stages interact with the same optical field.

\begin{figure}[phb]
\begin{center}
   \epsfig{file=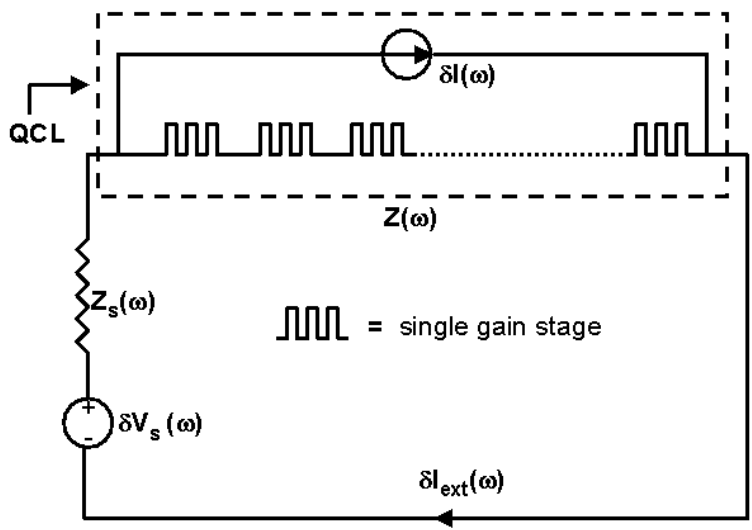,angle=0,width=4.0in} 
    \caption{ A simplified circuit model for the current fluctuations.}
    \label{figsimpleCirMod}
\end{center}
\end{figure}

A simplified circuit model for the current noise, more relevant from the point of view of experiments, is shown in Fig.~\ref{figsimpleCirMod}, where a single current noise source $\delta I(\omega)$ has been added in parallel with all the gain stages of the QCL. Expression for $\delta I(\omega)$ is,
\begin{eqnarray}
\delta I(\omega) & = & \frac{1}{N}\;\sum_{j=1}^{N}\,\delta I^{j}(\omega) \nonumber \\
                 & = & \frac{q}{Z(\omega)}\,\frac{\tau_{in}}{C_{inj}}\,\left[ \frac{F_{in}(\omega)}{(1 + j\omega \, \tau_{in})} - \sum_{k=1}^{3} \sum_{l=1}^{4} \, \left( \frac{1}{\tau_{in}} \, \frac{C_{inj}}{C_{k}} + \frac{1}{\tau_{k}(1 + j\omega \, \tau_{in})} \right) {\bf D}_{kl}^{-1}(\omega)\,F_{l}(\omega) \right] \label{eqdeltaI}
\end{eqnarray}
Equation (\ref{eqCirMod}) shows that the current noise $\delta I(\omega)$ is equal to the current noise $\delta I_{ext}(\omega)$ in the external circuit if $\delta V_{s}(\omega)$ and $Z_{s}(\omega)$ are both zero. This is also obvious from Fig.~\ref{figsimpleCirMod}. The characteristics of the noise source $\delta I(\omega)$ are explored next.

\subsection{Spectral Density and Fano Factor of the Current Noise}
\begin{figure}[t]
\begin{center}
   \epsfig{file=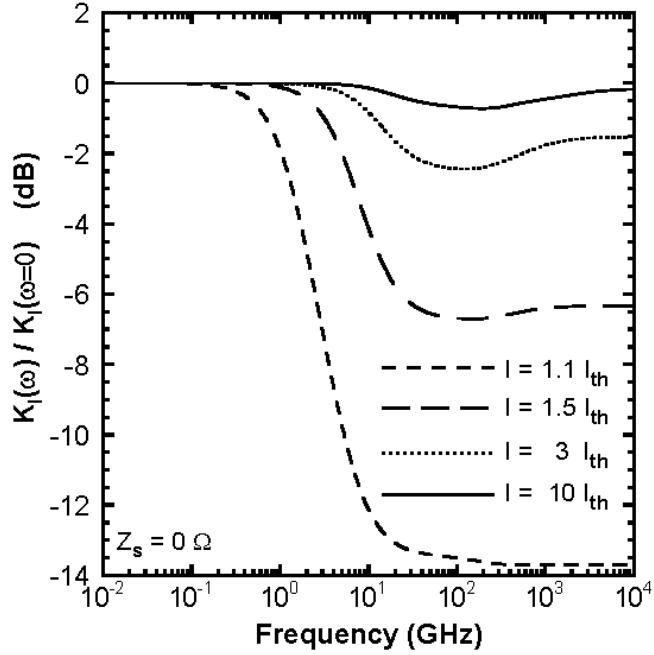,angle=0,width=3.5in}  
    \caption{ Spectral density $K_{I}(\omega)$ of the current noise is plotted as a function of the frequency. The noise spectral density has been normalized w.r.t. its value at zero frequency.} 
    \label{figSpecDenCurFreq}
\end{center}
\end{figure}

\begin{figure}[phb]
\begin{center}
   \epsfig{file=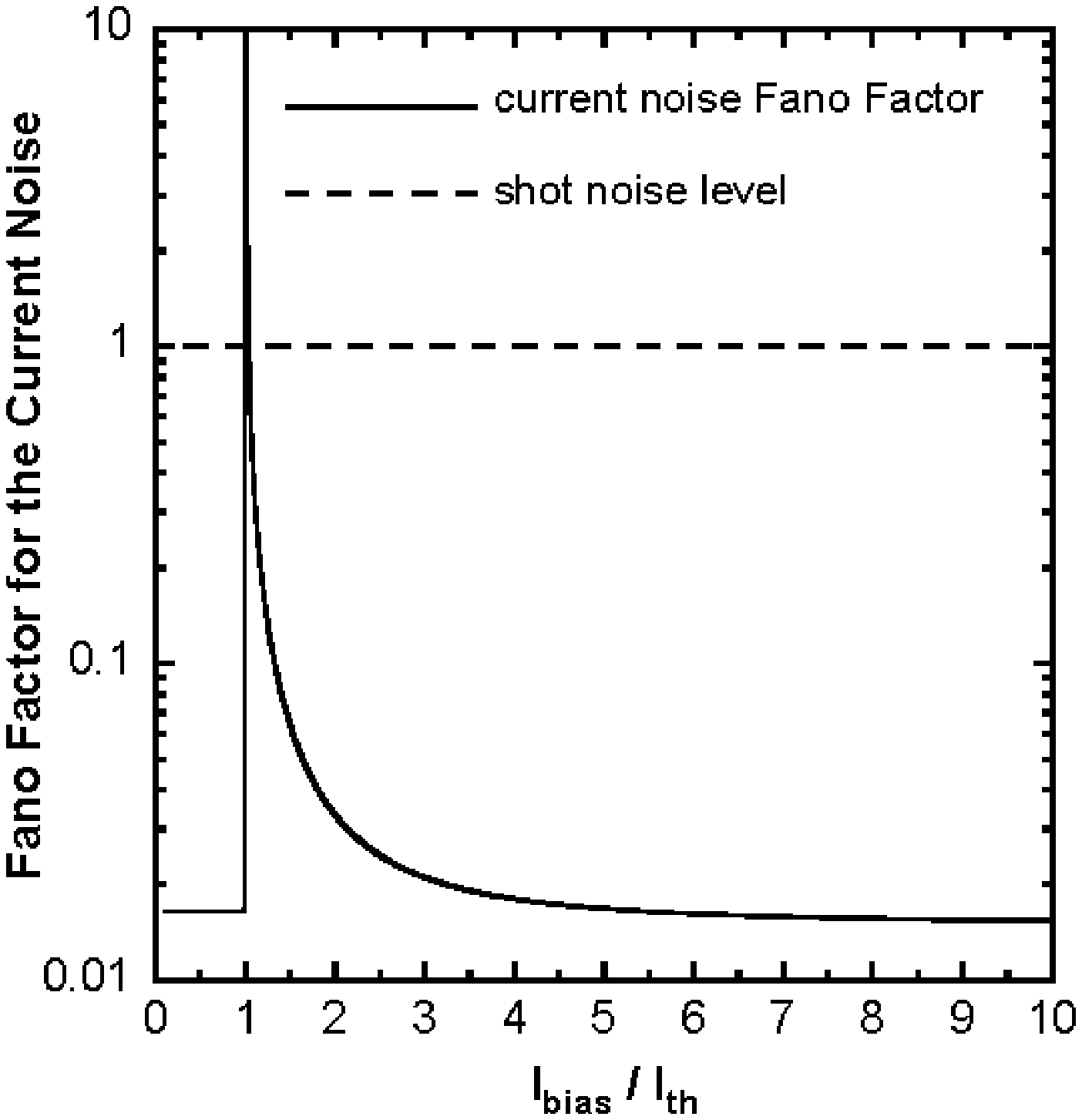,angle=0,width=3.5in} 
    \caption{ Fano Factor for the low frequency current fluctuations is plotted as a function of the bias current.}
    \label{figFanoCurBias}
\end{center}
\end{figure}

The spectral density $K_{I}(\omega)$ of the noise source $\delta I(\omega)$ can be calculated from Equation (\ref{eqdeltaIext}). Most of the numerical results presented in this paper, unless stated otherwise, are for the QCL described in Ref.~\cite{faistHP}. The device parameters for this QCL are given in the Table~\ref{table1}. In the numerical calculations values of all the device time constants (except $\tau_{st}$) were assumed to be independent of bias. The values of $\chi_{in}$ and $\chi_{out}$ were assumed to be unity (see the discussion in section~\ref{transport}). Fig.~\ref{figSpecDenCurFreq} shows the frequency dependence of $K_{I}(\omega)$ for different values of the bias current. As expected, $K_{I}(\omega)$ rolls over near the 3 dB frequency ($\omega_{3\:{\rm dB}}$) for the laser modulation response. Fig.~\ref{figFanoCurBias} shows the Fano Factor (Appendix~\ref{NoiseSpecDen}) for the low frequency fluctuations of the current noise source $\delta I(\omega)$ as a function of the bias current. Near the laser threshold the fluctuations in the electron densities inside the gain stages become large, and consequently, the current noise is also large. Away from the laser threshold the current noise is suppressed far below the shot noise value. 
\begin{figure}[pht]
\begin{center}
   \epsfig{file=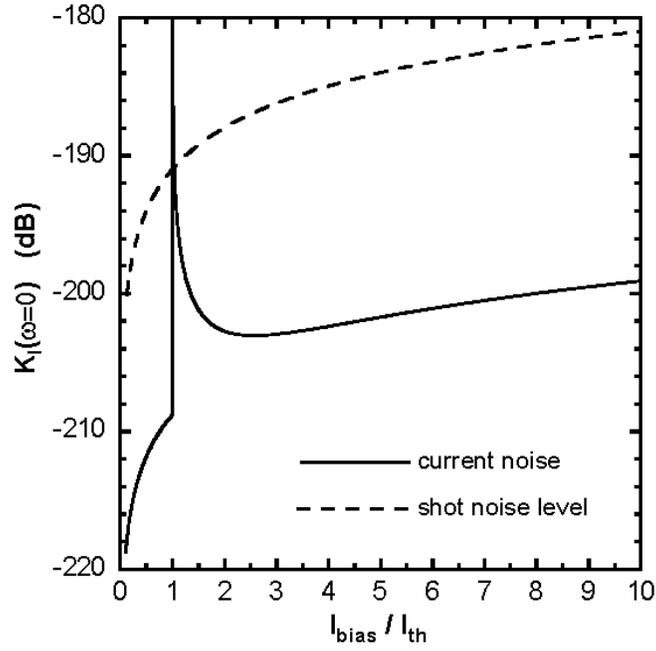,angle=0,width=3.5in}  
  \caption{ Low frequency spectral density $K_{I}(\omega=0)$ of the current noise is plotted as a function of the bias current.}
  \label{figSpecDenCurBias}
\end{center}
\end{figure}

For frequencies less than $\omega_{3\:{\rm dB}}$, analytical expression for $K_{I}(\omega)$ can be found using the expressions for the elements of the matrix ${\bf D}^{-1}$ given in Appendix~\ref{idmatrix},
\begin{equation}
K_{I}(\omega) \Big\vert_{\displaystyle \omega < \omega_{3\:{\rm dB}}} = \left\{ 
\begin{array}{ll}
{\displaystyle
q\frac{I}{N}\;\frac{\left( \chi_{in}^{2} + (\theta'_{3})^{2} + (\theta'_{2})^{2} \, \left(1 + 2\,\frac{\displaystyle \tau_{32}}{\displaystyle \tau_{31}} \right) + (\theta_{1}\,\chi_{out})^{2} \right) }{\left( 1 + \theta'_{3} + \theta'_{2} + \theta_{1} \right)^{2}} \hspace{1 cm} (I<I_{th}) } & \\
{\displaystyle
q\frac{I}{N}\;\frac{\left( \chi_{in}^{2} + ( \theta_{3} + \theta_{2} )^{2} + (\theta_{1}\,\chi_{out})^{2} \right) }{\left( 1 + \theta_{3} + \theta_{2} + \theta_{1}\right)^{2}} \; + \; 2qn_{sp}\;\eta_{r}\frac{(I-I_{th})}{N} \; \frac{\left( \theta_{3} \; \frac{\displaystyle \tau_{st}}{\displaystyle \tau_{21}}  - \theta_{2} \; \frac{\displaystyle \tau_{st}}{\displaystyle \tau_{31}}  \right)^{2}}{(1 + \theta_{3} + \theta_{2} + \theta_{1} )^{2}} \hspace{1 cm} (I>I_{th}) }
\end{array} \right.
\label{eqKI}
\end{equation}
Expressions for the parameters $\theta_{3}$, $\theta'_{3}$, $\theta_{2}$, $\theta'_{2}$ and $\theta_{1}$ are given in Appendix~\ref{DiffRes}. The expression for $K_{I}(\omega)$ above threshold is valid provided, 
\begin{equation}
N\Gamma v_{g}\,g - \frac{1}{\tau_{p}} \approx 0  \hspace{0.5 cm} {\rm and} \hspace{0.5 cm} S_{p} \gg \frac{n_{sp}}{WL} \label{eqcondition} 
\end{equation}
It is insightful to compare the expression for the current noise in Equation (\ref{eqKI}) to the current noise in interband semiconductor diode lasers. Using the model presented in Appendix~\ref{intnoise} one gets for diode lasers (see Appendix~\ref{intnoise} for details),
\begin{equation}
K_{I}(\omega) \Big\vert_{\displaystyle \omega < \omega_{3\:{\rm dB}}} = \left\{ 
\begin{array}{ll}
{\displaystyle
q\,I \hspace{1 cm} (I<I_{th}) } & \\
{\displaystyle
q\,I + 
2q\; I_{th} \;\frac{ \left( \theta' - \theta \right) }{\left( 1 + \theta \right)} +               
\; 2qn_{sp} \; \eta_{i} \; (I-I_{th})\,\frac{\left( \theta \, \frac{\displaystyle \tau_{st}}{\displaystyle \tau_{e}}    \right)^{2}}{\left( 1 + \theta \right)^{2}} \hspace{1 cm} (I>I_{th}) }
\end{array} \right.
\label{eqKIint}
\end{equation} 
where $\eta_{i}$ is the current injection efficiency, $\theta'$ is a number of the order of unity, and $\theta$ is much less than unity (see Appendix~\ref{intnoise}). Comparing Equations (\ref{eqKI}) and (\ref{eqKIint}) one can see that below threshold and also much above threshold (when $\tau_{st} \rightarrow 0$ and $I \gg I_{th}$) the current noise approaches the shot noise value in diode lasers, whereas in QCLs the current noise can be suppressed much below the shot noise value. The mechanisms responsible for the suppression of the current noise in QCLs are discussed below.

\subsubsection{\em Effect of Small Differential Impedance of a Single Gain Stage} 
The total differential impedance of all the gain stages in a $N$-stage QCL is larger than the differential impedance of a single gain stage by a factor of $N$. This reduces the total noise power of the current fluctuations by a factor of $N$, and therefore $K_{I}(\omega)$ has an explicit $1/N$ dependence in Equation (\ref{eqKI}).

\subsubsection{\em Effect of Electronic Correlations} 
The expression for the current fluctuations $\delta I(\omega)$ given in Equation (\ref{eqdeltaI}), for frequencies less than $\omega_{3\:{\rm dB}}$, can also be written as, 
\begin{equation}
\frac{N\, \delta I(\omega)}{q\,WL} = \sum_{j=1}^{N} \left[ f_{in}^{j}(\omega) - \sum_{k=1}^{3} \, \left( \frac{1}{\tau_{in}}\frac{C_{inj}}{C_{k}} + \frac{1}{\tau_{k}} \right) \, \delta n_{k}^{j}(\omega) \right] \label{eqcolcor1}
\end{equation}
Equation (\ref{eqcolcor1}), which is almost identical to Equation (\ref{eqdeltaIint0}) given in Appendix~\ref{intnoise} for semiconductor diode lasers, shows that fluctuations in the electron density in different levels of the gain stage causes fluctuations in the current. The sign of the current fluctuations is such as to restore the electron density to its average value, thus providing a negative feedback. The physical mechanisms responsible for this negative feedback are discussed below. On one hand these {\em electronic correlations} suppress the current noise associated with electron injection into the gain stage by providing negative feedback, and on the other hand they are also responsible for generating current noise in response to electron density fluctuations caused by noise sources internal to the gain stage. Various physical mechanisms included in our model which contribute to these electronic correlations are described below:

\begin{enumerate}
\item {\em Coulomb Correlations}: If the electron density changes in any level of the gain stage then the electrostatic potential energy of level 3 also changes because of coulomb interactions. As a result, the energy level separation $\delta E_{inj} - \delta E_{3}$ also changes, and consequently the total electron current from the injector into the gain stage also changes. Usually QCLs are not biased in the negative differential regime and the value of the conductance $G_{in}$, given by Equation (\ref{eqGintint3}), is positive. Therefore, the change in the current will be such as to restore the electron density in the levels of the gain stage to its steady state value. Coulomb correlations provide negative feedback to regulate electron density fluctuations. If a QCL is biased in the negative differential regime, in which the coulomb correlations provide positive feedback (negative $G_{in}$), the fluctuations may increase substantially and the linearized noise analysis presented in this paper may not be applicable. In our model the effect of coulomb correlations was introduced through the parameters $G_{in}/C'_{k}$ in Equation (\ref{eqvarjin}). 

\item {\em Pauli's Exclusion and Backward Tunneling Current}: If the electron density increases in level 3 of the gain stage then this reduces the phase space available for additional electrons to tunnel into level 3 from the injector due to Pauli's exclusion, and consequently the forward tunneling current from the injector into level 3 decreases from its average value. In addition, an increase in electron density in level 3 also increases the backward tunneling current from level 3 into the injector and this also reduces the net current from the injector into level 3 (recall from Equation (\ref{cur}) that the net current is the difference of the forward and backward tunneling currents). In our model both these effects were introduced through the parameter $t_{3}^{-1}$ in Equation (\ref{eqvarjin}). We remind the readers that later in Equations (\ref{eqvarjinnew}) and (\ref{eqvarjins}) $t_{3}^{-1}$ and $G_{in}/C'_{3}$ were absorbed in the definition of $\tau_{3}^{-1}$, and $G_{in}/C'_{2}$ and $G_{in}/C'_{3}$ were relabeled as $\tau_{2}^{-1}$ and $\tau_{1}^{-1}$, respectively. Therefore, coulomb correlations, Pauli's exclusion and backward tunneling current account for the presence of the terms $\delta n_{k}^{j}(\omega)/\tau_{k}$ in Equation (\ref{eqcolcor1}).

\item {\em Injector Electron Density Response}: 
Here we explain the presence of the terms $\left( C_{inj}/ \tau_{in}\,C_{k} \right)\delta n_{k}^{j}(\omega)$ in Equation (\ref{eqcolcor1}). Recall that the current fluctuations $\delta I(\omega)$ can be evaluated by looking at the current fluctuations $\delta I_{ext}(\omega)$ in the external circuit when $Z_{s}(\omega)$ is zero, and all external voltage sources are incrementally shorted, and the sum of the fluctuations in voltage  across all the gain stages (i.e. $\sum_{j=1}^{N} \delta V^{j}$) is, therefore, also zero. Under these conditions the relationship between the fluctuations in the carrier densities, expressed earlier in Equation (\ref{eqV}), becomes,
\begin{equation}
\sum_{j=1}^{N} \delta n_{inj}^{j}(\omega) =  - \sum_{j=1}^{N} \: \sum_{k=1}^{3} \, \frac{C_{inj}}{C_{k}} \, \delta n_{k}^{j}(\omega) \label{eqNinj}
\end{equation}
Equation (\ref{eqNinj}) can be used to write Equation (\ref{eqcolcor1}) as, 
\begin{equation} 
\frac{N \, \delta I(\omega) }{q \, WL} = \sum_{j=1}^{N} \, \left[ \frac{1}{\tau_{in}} \, \delta n_{inj}^{j}(\omega) - \sum_{k=1}^{3} \, \left( \frac{1}{\tau_{k}}  \, \delta n_{k}^{j}(\omega) \right) + f_{in}^{j}(\omega) \right]  \label{eqvarjinnew2}
\end{equation}
Equation (\ref{eqvarjinnew2}) shows that the current fluctuations are proportional to the total fluctuations in the electron density in the injector states of all the stages. Since $\sum_{j=1}^{N} \delta V^{j}(\omega) = 0$, a net increase in the electron density in different levels of all the gain stages must result in a net decrease of the electron density in all the injector states, and consequently, the current being injected into the gain stages must also decrease. This effect is captured through the terms $\left( C_{inj}/\tau_{in}\,C_{k} \right) \delta n_{k}^{j}(\omega)$ appearing in Equation (\ref{eqcolcor1}).
\end{enumerate}

As a result of the electronic correlations described above the current noise associated with electron injection into the gain stages, which is represented in our model through the noise sources $f_{in}^{j}(\omega)$, is suppressed. Electron density fluctuations caused by sources internal to the gain stage contribute more strongly towards the current fluctuations because of the same correlations. To see this in a more transparent fashion it is best to write Equation (\ref{eqcolcor1}) in terms of all the noise sources. Below threshold Equation (\ref{eqcolcor1}) becomes,
\begin{eqnarray}
\frac{N \, \delta I(\omega)}{q\,WL} % & = &  \sum_{j=1}^{N} \, \left[ f_{in}^{j}(\omega) - \left( \frac{1}{\tau_{32}} + \frac{1}{\tau_{31}} \right)\left( \theta'_{3}\, \delta n_{3}^{j}(\omega) + \theta'_{2} \, \frac{\tau_{32}}{\tau_{21}} \, \delta n_{2}^{j} (\omega) \right) - \left( \frac{1}{\tau_{out}} \right) \, \theta_{1} \, \delta n_{1}^{j}(\omega) \right] \label{eqcolcor2} \\
& = & {\displaystyle \frac{1}{\left( 1 + \theta'_{3}+ \theta'_{2} + \theta_{1} \right) } \, \sum_{j=1}^{N} \left[ f_{in}^{j}(\omega) + \theta_{1} f_{out}^{j}(\omega) + \left( \theta'_{3}+ \theta'_{2} \right) f_{31}^{j}(\omega) \right. } \nonumber \\
&  & \hspace{4 cm} {\displaystyle \left. \mbox{} + \theta'_{2} \left( 1 + \frac{\displaystyle \tau_{32}}{\displaystyle \tau_{31}} \right) f_{21}^{j}(\omega) + \left( \theta'_{3} - \theta'_{2} \frac{\displaystyle \tau_{32}}{\displaystyle \tau_{31}} \right) f_{32}^{j}(\omega) \right] }  \label{eqcolcor3}
\end{eqnarray}
Above threshold $\delta I(\omega)$ is,
\begin{eqnarray} 
\frac{N \, \delta I(\omega)}{q\,WL} % & = & \sum_{j=1}^{N} \, \left[ f_{in}^{j}(\omega) - \left( \frac{1}{\tau_{21}} + \frac{1}{\tau_{31}} \right) \left( \theta_{3}\, \delta n_{3}^{j}(\omega) + \theta_{2}\, \delta n_{2}^{j}(\omega) \right) - \left( \frac{1}{\tau_{out}} \right) \, \theta_{1}\, \delta n_{1}^{j}(\omega) \right]  \label{eqcolcor4} \\
& = & {\displaystyle \frac{1}{\left( 1 + \theta_{3}+ \theta_{2}+ \theta_{1} \right) } \, \sum_{j=1}^{N} \left[ f_{in}^{j}(\omega) + \theta_{1}f_{out}^{j}(\omega) + \left( \theta_{3}+ \theta_{2}\right) \left( f_{31}^{j}(\omega) + f_{21}^{j}(\omega)\right)  \right. } \nonumber   \\
&  & \hspace{4 cm} {\displaystyle \left. \mbox{} + \left( \theta_{3}\frac{\displaystyle \tau_{st}}{\displaystyle \tau_{21}} - \theta_{2}\frac{\displaystyle \tau_{st}}{\displaystyle \tau_{31}} \right) \left( f_{RS}^{j}(\omega) - \frac{\displaystyle F_{L}(\omega)}{\displaystyle N} \right) \right]}  \label{eqcolcor5} 
\end{eqnarray} 
Note that the strength of the electronic correlations depends on the values of the parameters $\theta_{3}$, $\theta'_{3}$, $\theta_{2}$, $\theta'_{2}$ and $\theta_{1}$ (Appendix~\ref{DiffRes}). From Equations (\ref{eqcolcor3}) and (\ref{eqcolcor5}) it is clear that larger values of these parameters will result in stronger electronic correlations, larger suppression of the current noise associated with electron injection into the gain stage, and also larger contribution to the current noise from the noise sources internal to the gain stage. The reader is encouraged to compare Equations (\ref{eqcolcor3}) and (\ref{eqcolcor5}) with the corresponding expressions for semiconductor diode lasers given in Equations (\ref{eqdeltaIint1}) and (\ref{eqdeltaIint2}) in Appendix~\ref{intnoise}.

\begin{figure}[t]
\begin{center}
   \epsfig{file=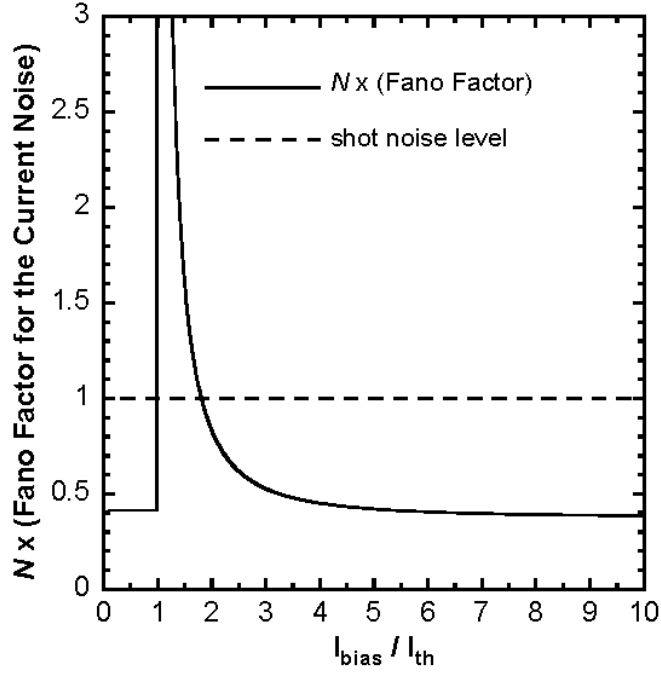,angle=0,width=3.5in}
    \caption{ $N$ times the Fano Factor for the low frequency current fluctuations is plotted as a function of the bias current. }
    \label{figFanoCurBias2}
\end{center}
\end{figure}

A quantitative measure of the role played by the electronic correlations in suppressing the current noise can be obtained by multiplying the Fano Factor of the current noise by $N$. It has been mentioned earlier that a factor of $1/N$ appears in Equation (\ref{eqKI}) because the total differential impedance of all the gain stages is larger than the differential impedance of a single gain stage by a factor of $N$. Therefore, multiplying the current noise Fano Factor by $N$ removes this explicit $1/N$ dependence in the current noise, and the resulting expression can only be less than unity because of electronic correlations. Fig.~\ref{figFanoCurBias2} shows the current noise Fano Factor from Fig.~\ref{figFanoCurBias} multiplied by $N$. 
Below threshold and much above threshold $N$ times the current noise Fano Factor is less than 0.5. This implies that electronic correlations are responsible for suppressing the current noise by a factor greater than 2. From Equation (\ref{eqKI}), expression for the current noise Fano Factor $F_{I}$ can be written as,
\begin{equation}
N \times F_{I}(\omega) \Big\vert_{\displaystyle \omega < \omega_{3\:{\rm dB}}} = \left\{ 
\begin{array}{lll}
{\displaystyle
\frac{\left( \chi_{in}^{2} + (\theta'_{3})^{2} + (\theta'_{2})^{2}\,\left( 1 + 2\,\frac{\displaystyle \tau_{32}}{\displaystyle \tau_{31}} \right) + (\theta_{1}\,\chi_{out})^{2} \right) }{\left( 1 + \theta'_{3}+ \theta'_{2} + \theta_{1} \right)^{2}} \hspace{1 cm} (I < I_{th}) } & \\
& \\
{\displaystyle
\frac{\left( \chi_{in}^{2} + (\theta_{3}+ \theta_{2})^{2} + (\theta_{1}\,\chi_{out})^{2} \right) }{\left( 1 + \theta_{3}+ \theta_{2}+ \theta_{1} \right)^{2}} \hspace{1 cm} (I \gg I_{th}) }
\end{array} \right.
\label{eqFI}
\end{equation}
For semiconductor diode lasers, using Equation (\ref{eqKIint}), one gets,
\begin{equation}
F_{I}(\omega) \Big\vert_{\displaystyle \omega < \omega_{3\:{\rm dB}}} = 1 \hspace{1 cm} \left( I < I_{th} \hspace{0.5 cm} {\rm and} \hspace{0.5 cm} I \gg I_{th} \right) \label{eqFIint}
\end{equation}

At frequencies much higher than the inverse of the smallest time constant of the QCL the current noise $\delta I(\omega)$ is just the capacitive response to the various electronic transitions which occur inside the gain stages. In the limit $\omega \rightarrow \infty$, $K_{I}(\omega)$ is given by the expression,
\begin{eqnarray}
K_{I}(\omega) \Big\vert_{\displaystyle \omega \rightarrow \infty } & = & q\,\frac{I}{N}\,\left[ \left(1 - \frac{C_{inj}}{C_{3}} \right)^{2}\,\chi_{in}^{2} + \left( \frac{C_{inj}}{C_{1}} \right)^{2}\,\chi_{out}^{2} \right]  \nonumber \\
& & \mbox{} + \frac{q^{2}WL}{N}\,\left[ R_{31}\left( \frac{C_{inj}}{C_{3}} - \frac{C_{inj}}{C_{1}} \right)^{2}
+ R_{21}\left( \frac{C_{inj}}{C_{2}} - \frac{C_{inj}}{C_{1}} \right)^{2} \right. \nonumber \\
& & \mbox{} + \left. R_{32}\left( \frac{C_{inj}}{C_{3}} - \frac{C_{inj}}{C_{2}} \right)^{2} \right] \hspace{ 1 cm} \left( I < I_{th} \right)
\end{eqnarray}
Above threshold an extra term,
\begin{equation}
q(2n_{sp} - 1)n_{r}\frac{(I-I_{th})}{N}\left( \frac{C_{inj}}{C_{3}} - \frac{C_{inj}}{C_{2}} \right)^{2} \nonumber
\end{equation}
is added to the above equation to account for the stimulated transitions. Semiconductor diode lasers on the other hand are charge neutral. Therefore, in the limit $\omega \rightarrow \infty$, the current noise in diode lasers is just the noise associated with carrier injection into the active region (see Appendix~\ref{intnoise}),
\begin{equation}
K_{I}(\omega) \Big\vert_{\displaystyle \omega \rightarrow \infty} = \left\{ 
\begin{array}{ll}
{\displaystyle
q\,I\,\left( 1 + 2\,\theta' \right) \hspace{1 cm} (I<I_{th}) } & \\
{\displaystyle
q\,I\,\left( 1 + 2\,\theta \right) + 2q\;I_{th}\,\left( \theta' - \theta \right) \hspace{1 cm} (I>I_{th}) }
\end{array} \right.
\end{equation}

\subsection{Scaling of the Current Noise with the Number of Cascade Stages}

In QCLs spectral density $K_{I}(\omega)$ of the current noise obeys a simple scaling relation with respect to the number of cascaded gain stages $N$, and this relation can be determined from Equation (\ref{eqdeltaI}),
\begin{equation}
N^{2}K_{I}(\omega,I/I_{th},N)  = {N'}^{2}K_{I}(\omega,I/I_{th},N') \label{eqscaleKI}
\end{equation}
According to the above equation, the spectral density of the current noise, when expressed as a function of $I/I_{th}$, scales as $1/{N^{2}}$. This scaling relation for $K_{I}(\omega)$ holds for all frequencies provided that the transition rates $R_{jk}(n_{j},n_{k})$ and the material gain $g(n_{3},n_{2})$ are linear functions of the electron densities and the total mode confinement factor also scales linearly with the number of cascade stages $N$. 

\subsection{Spectral density of the Current Noise in the External Circuit}

The quantity that can be measured experimentally is the spectral density of the current noise $\delta I_{ext}(\omega)$ which flows in the external circuit. Spectral density of the current noise in the external circuit provides a valuable tool for studying the high speed dynamics of QCLs. When $Z_{s}(\omega) \ne 0 $ $\Omega$, which is usually the case, then $K_{I_{ext}}(\omega)$ is not the same as $K_{I}(\omega)$. Expression for $K_{I_{ext}}(\omega)$ follows from Equation (\ref{eqCirMod}),
\begin{eqnarray}
K_{I_{ext}}(\omega) & = & \frac{K_{V_{s}}(\omega)}{\left\vert Z(\omega) + Z_{s}(\omega) \right\vert^{2}} + \left\vert \frac{Z(\omega)}{Z(\omega) + Z_{s}(\omega)} \right\vert^{2} K_{I}(\omega) \nonumber \\
& = & \frac{2K_{B}T\,{\rm Real} \{ Z_{s}(\omega) \}}{\left\vert Z(\omega) + Z_{s}(\omega) \right\vert^{2}} + \left\vert \frac{Z(\omega)}{Z(\omega) + Z_{s}(\omega)} \right\vert^{2} K_{I}(\omega) \label{eqKIext} 
\end{eqnarray}
Equation (\ref{eqKIext}) shows that in the presence of a large impedance $Z_{s}(\omega)$ the current fluctuations in the external circuit are suppressed. The total differential impedance of a QCL is usually less than $1 $ $\Omega$. Therefore, for even a moderately large impedance $Z_{s}(\omega)$ the current noise in the external circuit is dominated by the thermal noise from the impedance $Z_{s}(\omega)$. Experimental measurement of the current noise would, therefore, require a relatively sensitive measurement scheme. High impedance suppression of the current noise in the external circuit can influence the laser intensity noise, as shown in Equation (\ref{eqdeltaPo1}) and in more detail in the following section.

\section{Photon Noise: Results and Discussion} \label{DisscussPN}

\subsection{Spectral Density and Fano Factor of the Laser Intensity Noise}

\begin{figure}[pt]
\begin{center}
   \epsfig{file=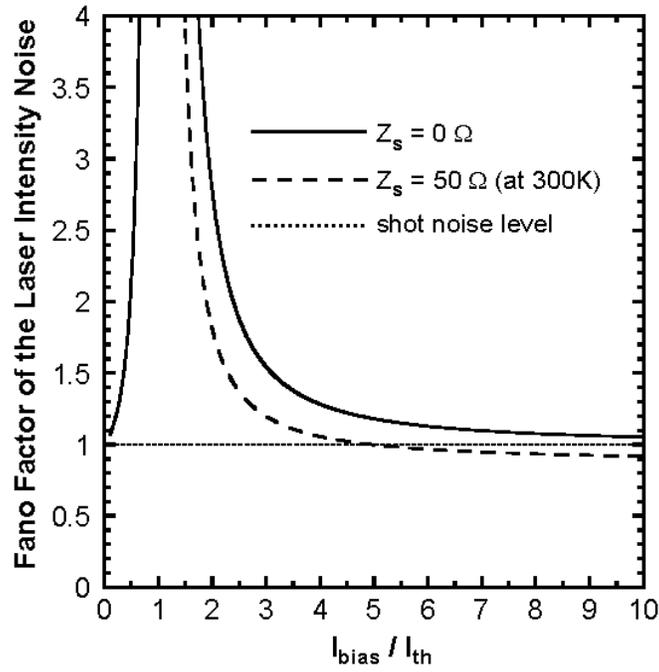,angle=0,width=3.5in}  
    \caption{ Fano Factor for the (low frequency) noise in the laser intensity is plotted as a function of the bias current.}  
    \label{figFanoPowBias}
\end{center}
\end{figure}

\begin{figure}[pb]
\begin{center}
   \epsfig{file=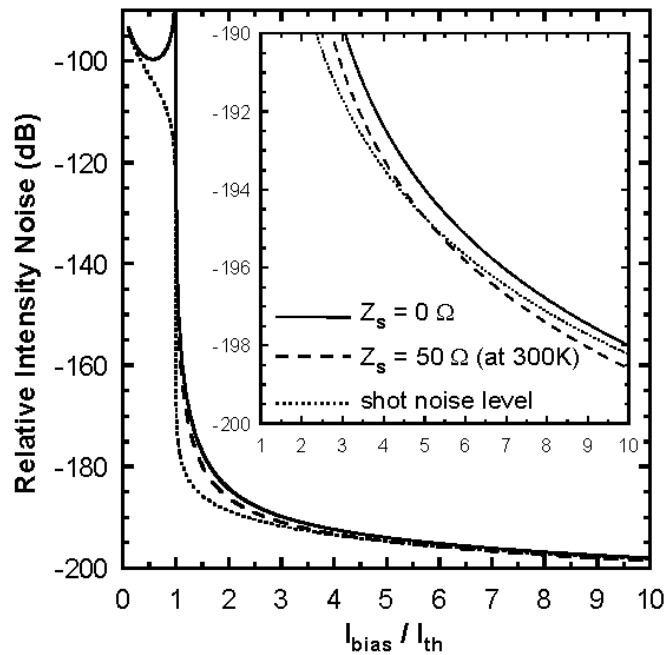,angle=0,width=3.5in}  
    \caption{ Low frequency Relative Intensity Noise (RIN) is plotted as a function of the bias current. Very small amount of squeezing (less than 0.4 dB) is exhibited at high bias levels even when the circuit current fluctuations are suppressed with a $50 $ $\Omega$ impedance.}
    \label{figRINBias}
\end{center}
\end{figure}

\begin{figure}[t]
\begin{center}
   \epsfig{file=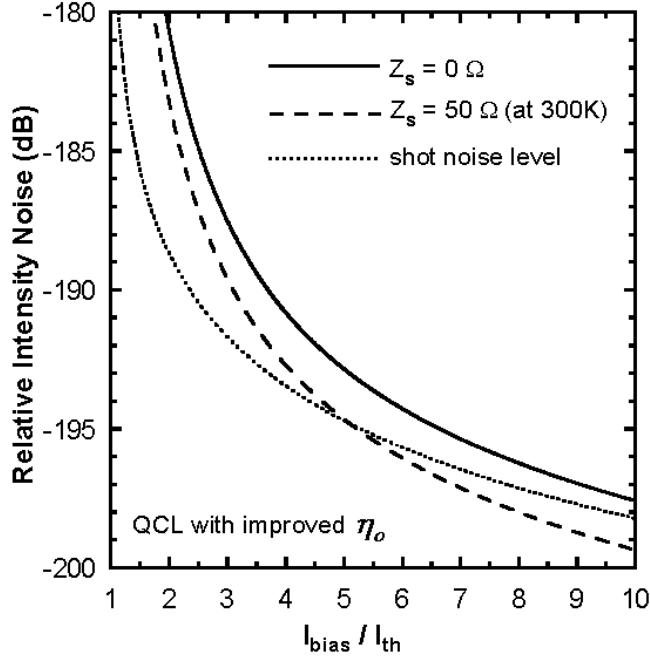,angle=0,width=3.5in}
    \caption{ Relative Intensity Noise (RIN) for the QCL with improved $\eta_{o}$ is shown. Only around 1 dB of squeezing is possible at high bias levels and when $Z_{s} = 50 $ $\Omega$.} 
    \label{figRINBias2}
\end{center}
\end{figure}

The spectral density $K_{P}(\omega)$ of the laser intensity noise can be calculated from Equation (\ref{eqdeltaPo1}). The Fano Factor for the low frequency fluctuations in the laser output power is plotted as a function of the bias current in Fig.~\ref{figFanoPowBias}. The Relative Intensity Noise (RIN) is plotted in Fig.~\ref{figRINBias}. 
\begin{figure}[htb]
\begin{center}
   \epsfig{file=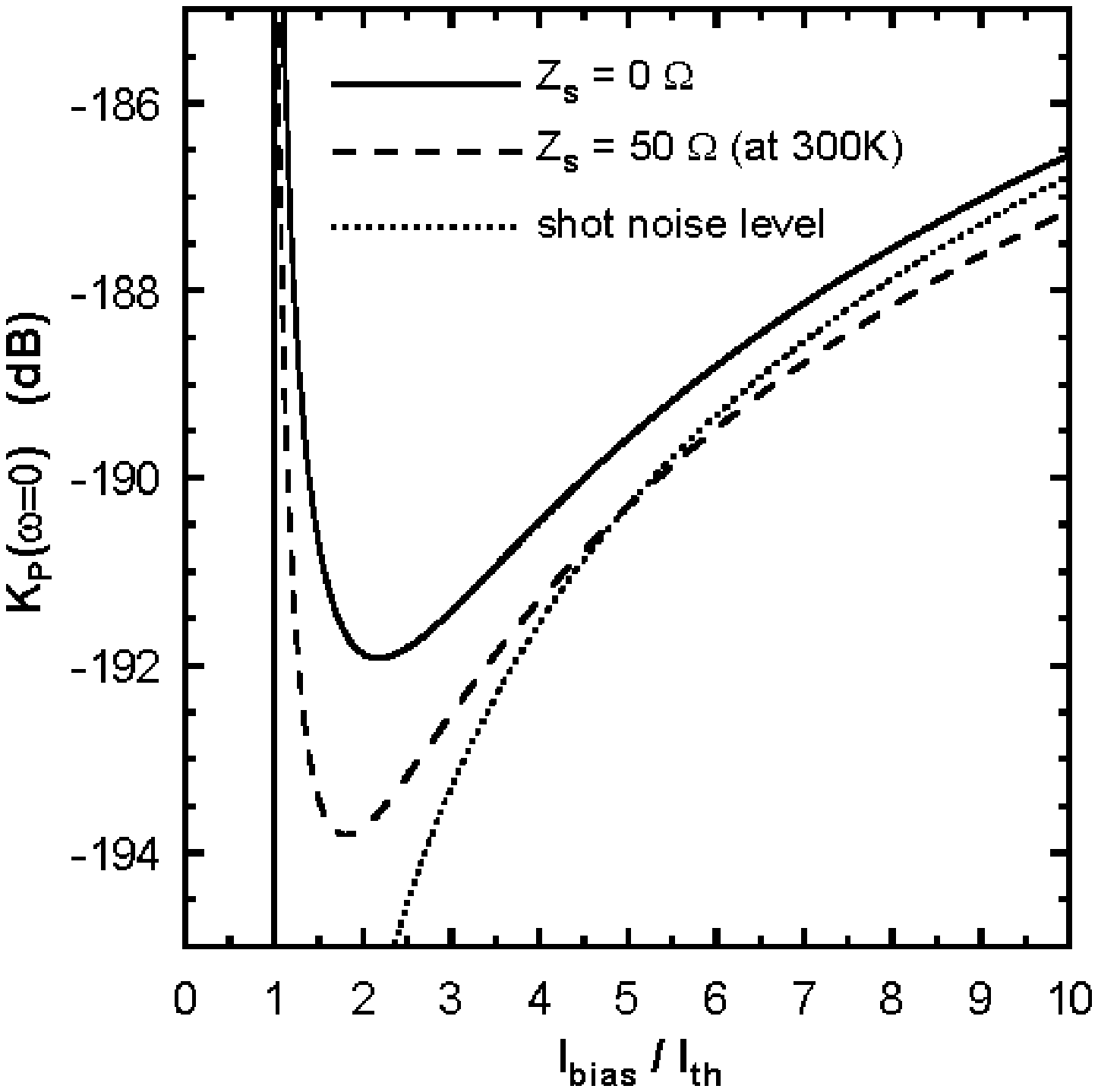,angle=0,width=3.5in}
  \caption{ Low frequency Spectral density $K_{P}(\omega=0)$ of the laser intensity fluctuations is plotted as a function of the bias current. When $Z_{s}$ is large ($50 $ $\Omega$) small amounts of squeezing is possible at high bias levels.}
  \label{figSpecDenCur}
\end{center}
\end{figure}
In each figure the respective shot noise limit is also shown. It is assumed that the light coming out from both the facets of the laser is collected before the noise is evaluated. This is equivalent to assuming that the output coupling efficiency $\eta_{o}$, defined earlier in Equation (\ref{eqetao1}), is,
\begin{equation}
\eta_{o} = \frac{\alpha_{m}} {(\alpha_{m} + \alpha_{i})} \label{eqetao2}
\end{equation}
In practice this can be achieved by HR/AR coating the laser facets so that that most of the light comes out from only one facet of the laser. When the value of the external impedance $Z_{s}$ is $0$ $\Omega$, the photon noise remains above the shot noise limit. Even at high bias levels, no amplitude squeezing is observed despite the fact that the current noise is suppressed much below the shot noise value as shown earlier in Fig.~\ref{figFanoCurBias}. When $Z_{s}=50$ $\Omega$, and the current noise in the circuit is further suppressed, a very small amount of squeezing is observed at high bias levels (less than 0.2 dB at $I = 10\,I_{TH}$). This is in sharp contrast to the characteristics of interband semiconductor diode lasers which exhibit substantial amount of amplitude squeezing (around 3-5 dB) when the noise in the circuit current is suppressed far below the shot noise value~\cite{interbandsq}. The absence of any significant amount of squeezing in QCLs can be attributed to the following factors,
\vspace{0.25 cm}
\begin{enumerate}
\item {\em Small Output Coupling Efficiency}: Most of the QCLs reported in literature have relatively long cavity lengths and large values for internal waveguide losses. The output coupling efficiency $\eta_{o}$ is thus small. The small output coupling efficiency significantly reduces the amount of squeezing that can be achieved. This, however, does not impose a fundamental limit on the amount of squeezing achievable in QCLs, but only a practical limit. 
\vspace{0.25 cm}     
\item {\em Non-Radiative Transitions}: As shown earlier, in QCLs above threshold the electron densities in different energy levels of a gain stage do not remain fixed at their threshold values. The electron densities keep increasing when the bias current is increased beyond threshold. As a result, the contribution of non-radiative electronic transitions to the photon noise also keeps increasing with the bias current. This is very different from what happens in semiconductor diode lasers in which above threshold the carrier densities in the quantum wells, and the non-radiative recombination rates, remain clamped to their threshold values. Amplitude squeezing is therefore expected to be less in QCLs than in semiconductor diode lasers, even if the output coupling efficiency of QCLs is improved. The QCL, whose characteristics are shown in Fig.~\ref{figFanoPowBias} and Fig.~\ref{figRINBias}, has a 3 mm long cavity, a waveguide loss of 11 cm$^{-1}$, and an output coupling efficiency of only 28\%. Consider a QCL with a 500 $\mu$m long cavity, a waveguide loss of 5 cm$^{-1}$, and an output coupling efficiency of 84\%. The values of all the other parameters of this QCL are identical to those given in Table~\ref{table1}. Fig.~\ref{figRINBias2} shows the relative intensity noise when this QCL is driven with a $50$ $\Omega$ resistor in series. Only about 1 dB of squeezing is observed even at very high bias levels ($I \approx 10\,I_{TH}$). The amount of squeezing achievable in QCLs can be increased by decreasing the lifetime $\tau_{21}$ of electrons in level 2 of the gain stage. This will reduce the rate of increase of the electron density in levels 3 and 2 with the bias current above threshold, inrease the radiative efficiency $\eta_{r}$, and reduce the contribution of non-radiative electronic transitions to the photon noise.      
\end{enumerate}
\vspace{0.25 cm}

\begin{figure}[pt]
\begin{center}
   \epsfig{file=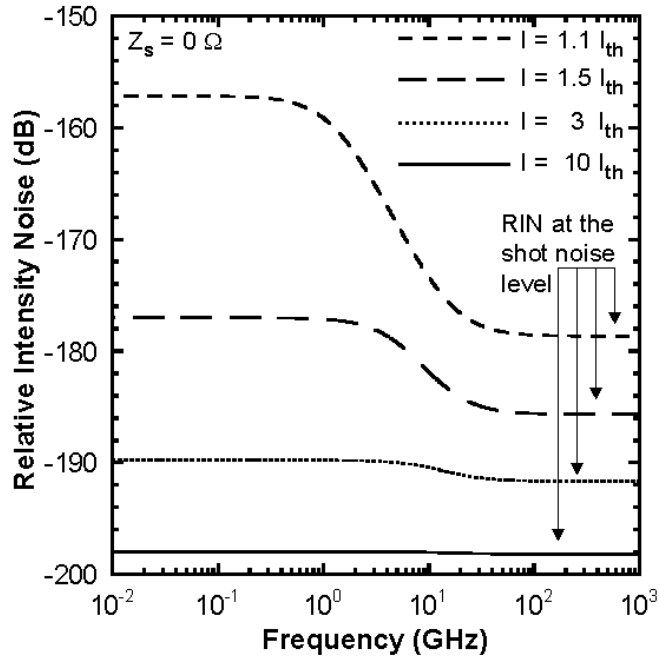,angle=0,width=3.5in}  
    \caption{ Relative Intensity Noise (RIN) is plotted as a function of the frequency for different bias currents ($Z_{s} = 0$ $\Omega$). At high frequencies the RIN reaches the shot noise limit.}
    \label{figRINFreq}
\end{center}
\end{figure}

\begin{figure}[pb]
\begin{center}
   \epsfig{file=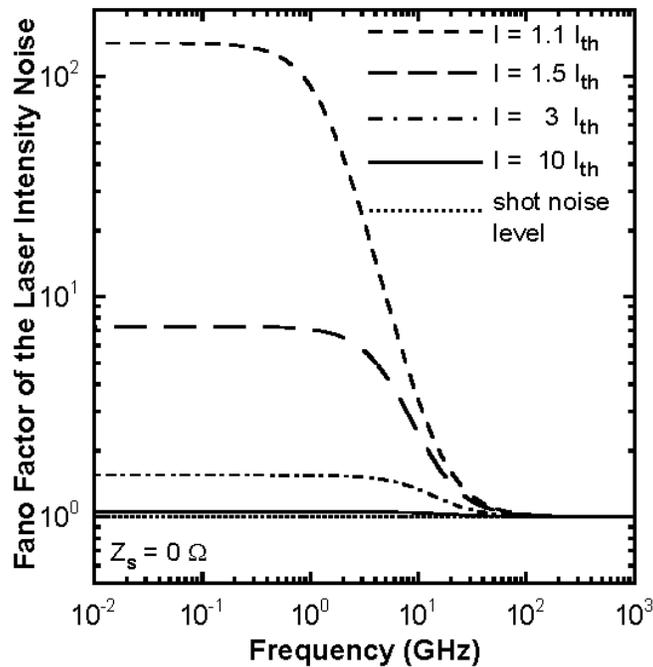,angle=0,width=3.5in}
    \caption{ Fano Factor for the laser intensity noise is plotted as a function of the frequency ($Z_{s}=0$ $\Omega$). The intensity noise at frequencies higher than the inverse of the photon lifetime in the cavity is dominated by the photon partition noise at the output facet.}
    \label{figFanoFreq}
\end{center}
\end{figure}

Fig.~\ref{figRINFreq} shows the RIN as a function of the frequency for different values of the bias current assuming $Z_{s}(\omega)=0$. The RIN also rolls over at the frequency $\omega_{3\:{\rm dB}}$. Fig.~\ref{figFanoFreq} shows that the Fano Factor for the laser intensity noise as a function of the frequency. As in all other lasers, at frequencies much higher than the inverse of the photon lifetime inside the cavity, the RIN is dominated by the noise from photon partition at the output facet. Therefore, 
\begin{equation}
K_{P}(\omega) \Big\vert_{\displaystyle \omega > 1/\tau_{p}} = h\nu \,P_{out}
\end{equation}

\begin{figure}[htb]    
\begin{center}
   \epsfig{file=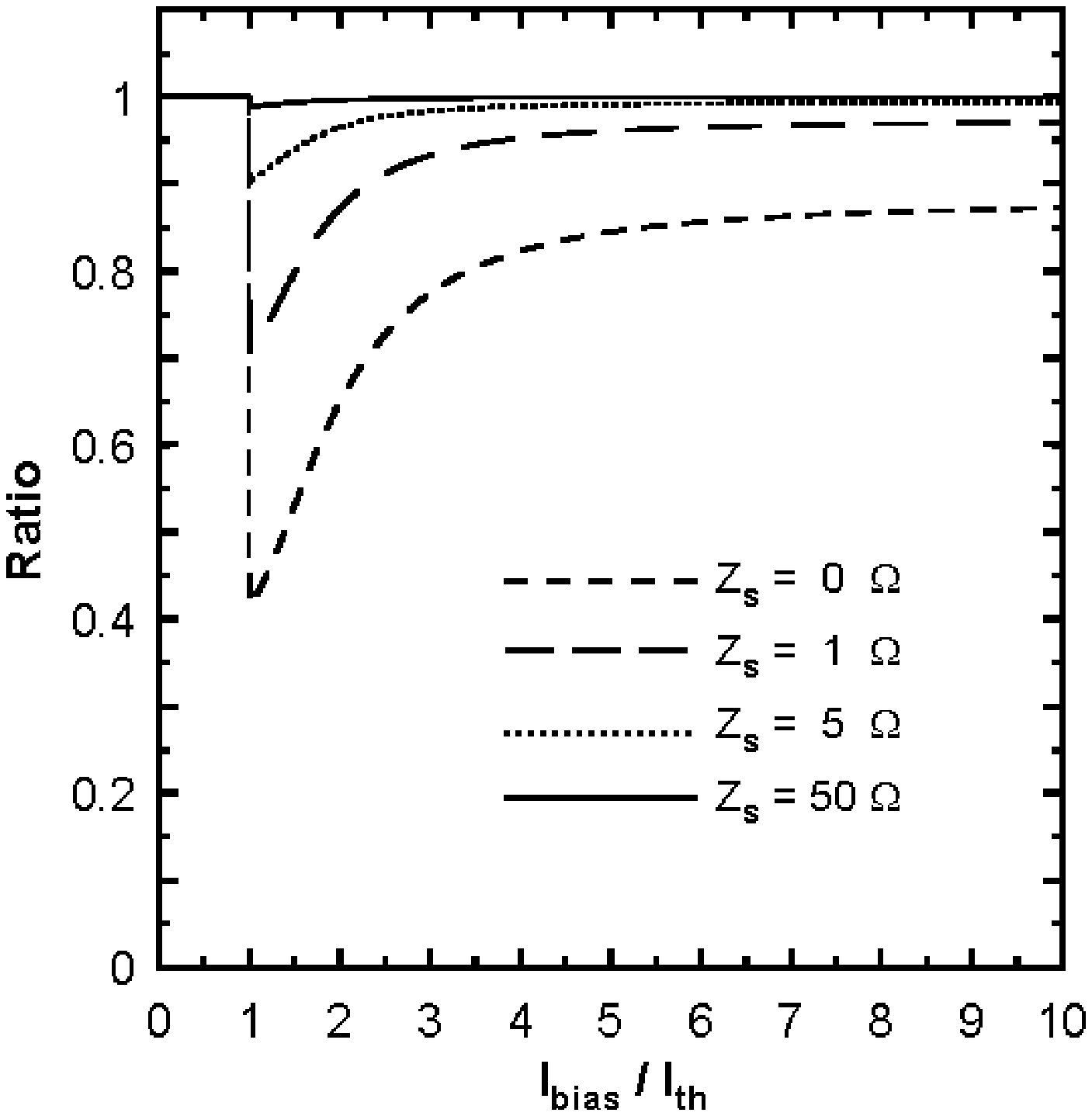,angle=0,width=3.5in}
    \caption{ Ratio of the RIN obtained by ignoring the term containing the current fluctuations in Equation (\ref{eqdeltaPo1}) to the actual RIN is plotted as a function of the bias current for different values of the impedance $Z_{s}$. The current fluctuations are suppressed when $Z_{s}$ is large and the error incurred in calculating RIN is, therefore, small. }  
    \label{figContribCur}
\end{center}
\end{figure}

In this paper careful attention has been given to modeling the current fluctuations in the external circuit. The question arises if such detailed modeling of the current fluctuations is necessary for calculating the photon intensity noise. In Equation (\ref{eqdeltaPo1}) the current noise $\delta I_{ext}(\omega)$ is included in the first term on the right hand side. It should be noted here that the first and the second term on the right hand side in Equation (\ref{eqdeltaPo1}) are correlated, and the spectral density of the photon intensity noise cannot be obtained by a simple addition of the spectral densities of these two terms. In Fig.~\ref{figContribCur} the ratio of the spectral density of the photon intensity noise obtained by ignoring the term containing $\delta I_{ext}(\omega)$ in Equation (\ref{eqdeltaPo1}) to the actual spectral density of the photon intensity noise is plotted as a function of the bias current for different values of the impedance $Z_{s}$. When the laser is biased a little above threshold the fluctuations in the current are large, and the error incurred by ignoring the term containing $\delta I_{ext}(\omega)$ in Equation (\ref{eqdeltaPo1}) is also large. Also, when $Z_{s}$ is much larger than the total differential impedance of the QCL, the current fluctuations in the circuit are suppressed, and the term containing $\delta I_{ext}(\omega)$ can be ignored in Equation (\ref{eqdeltaPo1}). 

For large $Z_{s}(\omega)$, using the expressions for the elements of the matrix ${\bf D}^{-1}$ in Appendix~\ref{idmatrix}, analytical expression for the spectral density of the low frequency fluctuations in the laser intensity noise can be obtained, 
\begin{eqnarray}
K_{P}(\omega) \Big\vert_{\displaystyle \omega < \omega_{3\:{\rm dB}}} 
% & = & h \nu \, P_{out} \, \left[  1 - \eta_{o} + 2\eta_{o}\,n_{sp}\,\left( \frac{\tau_{31}+\tau_{32}}{\tau_{31}+\tau_{21}} \right)^{2}\left(\frac{\tau_{st}}{\tau_{32}}\right)^{2} \right] \nonumber \\
% & & \mbox{} + (\eta_{o}\,h\,\nu)^{2}\,N\,WL\,\left[ \frac{n_{3}}{\tau_{32}} + \left( \frac{\tau_{32}-\tau_{21} }{\tau_{31}+\tau_{21}}\right)^{2}\left(\frac{\tau_{31}}{\tau_{32}}\right)^{2} \,\frac{n_{3}}{\tau_{31}} \right. \nonumber \\
% & & \left. \mbox{} + \left( \frac{\tau_{31}+\tau_{32} }{\tau_{31}+\tau_{21}}\right)^{2}\left(\frac{\tau_{21}}{\tau_{32}}\right)^{2}\,\frac{n_{2}}{\tau_{21}} \right] \label{eqKp1} \\
& = & h \nu \, P_{out} \, \left[  1 - \eta_{o} + 2\eta_{o}\,n_{sp}\,\left( \frac{\tau_{31}}{\tau_{21} + \tau_{31}} \right)^{2} \left( \frac{1}{\tau_{31}} + \frac{1}{\tau_{32}} \right)^{2}\;\tau_{st}^{2} \right] \nonumber \\
 & & \mbox{} + (\eta_{o}\,h\,\nu)^{2}\,N\,WL\,\left[ R_{32} + \eta_{r}^{2}\,R_{31} + (1-\eta_{r})^{2}\,R_{21} \right] \label{eqKp2} \\
& = & h \nu \, P_{out} \, \left[  1 - \eta_{o} + 2\eta_{o}\,n_{sp}\,\left( \frac{\tau_{31}}{\tau_{21} + \tau_{31}} \right)^{2} \left( \frac{1}{\tau_{31}} + \frac{1}{\tau_{32}} \right)^{2}\;\tau_{st}^{2} \right] \nonumber \\
 & & \mbox{} + (\eta_{o}\,h\,\nu)^{2}\,N\,\left[  \alpha \, \frac{I}{q} + \beta \, \frac{I_{th}}{q} \right]
\label{eqKp4}
\end{eqnarray} 
$\eta_{r}$ in the above Equation is the radiative efficiency defined in Equation (\ref{eqetar}). The constants $\alpha$ and $\beta$ are,
\begin{equation}
%\alpha = \frac{\tau_{21} \, \tau_{31}}{ \left( \tau_{21} + \tau_{31} \right)} \left[ \frac{1}{\tau_{32}} + \frac{\eta_{r}^{2}}{\tau_{31}} + \frac{(1 - \eta_{r})^{2}}{\tau_{21}} \right] 
\alpha = \eta_{r}\left( 1 - \eta_{r} \right) + 2\left( 1 - \eta_{r} \right) \frac{\tau_{31}}{\tau_{21} + \tau_{31}}\left( \frac{\displaystyle \tau_{21}}{\displaystyle \tau_{32}} \right)
\end{equation}
\begin{equation}
%\beta = \frac{\tau_{31} \, \tau_{32}}{\left( \tau_{31} + \tau_{32} \right)} \left[ \frac{1}{\tau_{32}} + \frac{\eta_{r}^{2}}{\tau_{31}} - \frac{(1 - \eta_{r})^{2}}{\tau_{31}} \right] 
\beta = \eta_{r} - 2 \eta_{r}\left( 1 - \eta_{r} \right) \frac{\tau_{32}}{\tau_{31} + \tau_{32}}
\end{equation}
The above expression for $K_{P}(\omega)$ is valid for frequencies smaller than $\omega_{3\:{\rm dB}}$, and when the laser is biased above threshold when the conditions given by Equation (\ref{eqcondition}) are satisfied. The expression given in Equation (\ref{eqKp4}) is almost identical to the expression for $K_{P}(\omega)$ for semiconductor diode lasers (when the later are also biased with a high impedance current source). Using the model presented in Appendix~\ref{intnoise} one gets for diode lasers (see Appendix~\ref{intnoise} for details),
\begin{eqnarray}
K_{P}(\omega) \Big\vert_{\displaystyle \omega < \omega_{3\:{\rm dB}}} & = & h \nu \, P_{out} \, \left[  1 - \eta_{o} + 2\eta_{o}\,n_{sp}\,\left( \frac{1}{\tau_{nr}} + \frac{1-\eta_{i}}{\tau_{e}}  \right)^{2}\,\tau_{st}^{2} \right]  \nonumber \\ 
& & \mbox{} + (\eta_{o} \, h \nu)^{2}\left[ \eta_{i}(1 - \eta_{i}) \frac{I}{q} + \eta_{i}\frac{I_{th}}{q} \right] \label{eqKp3}
\end{eqnarray} 
where $\eta_{i}$ is the current injection efficiency into the quantum wells~\cite{coldren}. Comparing Equations (\ref{eqKp2}) and (\ref{eqKp3}), it can be seen that the noise associated with the non-radiative transitions appears in a slightly more complex form in Equation (\ref{eqKp2}). The noise contribution from the non-radiative transitions in diode lasers is constant above threshold and it has been expressed in terms of the threshold current in Equation (\ref{eqKp3}). In QCLs, as already pointed out earlier, the noise contribution from the non-radiative transitions keeps increasing with bias beyond the laser threshold. Since only a fraction $\eta_{r}$ of the electrons injected in level 3 of the gain stage end up producing photons, a multiplicative factor $\eta_{r}^{2}$ appears with the noise associated with the transitions from level 3 to level 1. A fraction $1-\eta_{r}$ of the vacancies left by removing electrons from level 2 get filled by radiative transitions from level 3 to level 2, and therefore a factor $(1-\eta_{r})^{2}$ appears with the noise associated with the transitions from level 2 to level 1. All the electrons taken out of level 2 and injected into level 3 will end up producing photons since, 
\begin{equation}
(1-\eta_{r}) + \eta_{r}= 1
\end{equation}  
and, therefore, the noise contribution from the non-radiative transitions from level 3 to level 2 has no multiplicative factor in Equation (\ref{eqKp2}). The noise associated with the electron transitions from level 1 into the injector of the next stage does not directly contribute to photon noise at low frequencies. These transitions contribute to the current noise in the external circuit, which can in turn contribute to the photon noise. But we have assumed in Equation (\ref{eqKp2}) that $Z_{s}(\omega)$ is large and the current fluctuations are suppressed. Similarly the noise associated with the electron transitions from the injector into level 3 of the gain stage is also suppressed at low frequencies when $Z_{s}(\omega)$ is large. 

In diode lasers, since the current injection efficiency $\eta_{i}$ is less than unity, the partition noise associated with carrier leakage from the SCH region contributes a term to the intensity noise which increases linearly with the bias current even beyond the laser threshold, as shown in Equation (\ref{eqKp3}). Since $\eta_{i}$ is usually close to unity in well designed diode lasers~\cite{coldren}, the contribution of this term to the intensity noise is small. 

The Fano factors for the laser intensity noise much above threshold (when $\tau_{st} \rightarrow 0$ and $I \gg I_{th}$) in QCLs and diode lasers can be calculated from Equations (\ref{eqKp4}) and (\ref{eqKp3}),
\begin{equation}
F_{P}(\omega) \Big\vert_{\displaystyle \left( \omega < \omega_{3\:{\rm dB}} \: , \: I \gg I_{th} \right) } = \left\{
\begin{array}{ll}
{\displaystyle 1 - \eta_{o} \, \eta_{r} + 2 \eta_{o} \, \frac{\displaystyle \left( 1 - \eta_{r} \right) }{\displaystyle \eta_{r}}\frac{\displaystyle \tau_{31}}{\displaystyle \tau_{21} + \tau_{31}}  \left( \frac{\displaystyle \tau_{21}}{\displaystyle \tau_{32}} \right)} \hspace{1 cm} {\rm QCLs}  & \\
{\displaystyle 1 - \eta_{o}\,\eta_{i}}  \hspace{1 cm} {\rm Diode \: Lasers} 
\end{array} \right. \label{maxsq}
\end{equation}      
Equation (\ref{maxsq}) gives the maximum photon number squeezing possible in QCLs and in diode lasers. In diode lasers both $\eta_{i}$ and $\eta_{o}$ can be larger than 0.85~\cite{coldren}. Therefore, the intensity noise in diode lasers can be suppressed more than 5 dB below the shot noise value. For the QCL whose parameters are listed in Table~\ref{table1}, $\eta_{r}$ and $\eta_{o}$ have the values 0.66 and 0.28, respectively, and, consequently, the maximum possible squeezing is only 0.6 dB.

\subsection{Scaling of the Laser Intensity Noise with the Number of Cascade Stages}

In QCLs $K_{p}(\omega)$ obeys a simple scaling relation with respect to the number of cascaded stages $N$, and this relation can easily be deduced from Equation (\ref{eqdeltaPo1}), 
\begin{equation}
K_{P}(\omega,I/I_{th},N) = K_{P}(\omega,I/I_{th},N') \label{eqKpScale}
\end{equation}
According to Equation (\ref{eqKpScale}), the spectral density of the photon noise, when expressed as a function of $I/I_{th}$, is independent of the value of $N$. The scaling relation for $K_{p}(\omega)$ holds for all frequencies and when $Z_{s}(\omega)$ is very large or when $Z_{s}(\omega)=0$ $\Omega$, provided that the transition rates $R(n_{j},n_{q})$ and the material gain $g(n_{3},n_{2})$ are linear functions of electron densities and the total mode confinement factor also scales linearly with the number of cascaded stages $N$. In Ref.~\cite{cgscale} it is shown that the total mode confinement factor scales with the number of cascaded stages according to the expression $\sim \, erf(0.019N)$, which is almost linear in $N$ for $N < 40$.

\subsection{Effect of Multiple Longitudinal Modes on the Measured Intensity Noise}

Most QCLs reported in literature lase with multiple longitudinal modes. Although the intensity noise of each longitudinal mode can be large, the intensity noise of all the modes taken together is expected to be adequately described by the single mode analysis carried out in this paper. This is because the intensity noise in different lasing modes is negatively correlated, as it is in the case of semiconductor diode lasers~\cite{yama3}. However, this demands that in experiments designed to measure the intensity noise attention must also be paid to optimizing the light collection efficiency such that photons are collected from all the lasing modes, otherwise intensity noise in excess of that described by Equation (\ref{eqKp2}) can be introduced.

\section{Conclusion}

A comprehensive model for treating noise and fluctuations in intersubband quantum cascade lasers has been presented. The current noise exhibited by QCLs is much below the shot noise value. Suppression of the current noise in QCLs is largely due to the small differential resistance of individual gain stages compared to the total differential resistance of all the cascaded gain stages. In addition, electronic correlations also tend to suppress the current noise. However, unlike semiconductor diode lasers, current noise suppression does not lead to significant photon number squeezing in QCLs. In QCLs the contribution to the photon intensity noise coming from the non-radiative electronic transitions keeps increasing with bias beyond the laser threshold, and this reduces the amount of photon number squeezing achievable in QCLs compared to semiconductor diode lasers. In addition, poor laser output coupling efficiencies and the lack of availability of photon detectors with good quantum efficiencies in the mid-infrared wavelength region make the experimental observation of photon number squeezing in QCLs difficult.    

The current modulation response of QCLs has also been investigated. It has been found that the direct current modulation response of many QCLs that have been reported in the literature is over-damped since, in contrast to diode lasers, the photon lifetime inside the optical cavity in QCLs is usually the longest time constant. The modulation bandwidth is also limited by the inverse photon lifetime. At present, only quantum well infrared photodetectors (QWIPs) have a bandwidth wide enough that they could be used to study the modulation response of QCLs. However, the current noise provides an alternate way of studying the high speed dynamics of QCLs, and as shown in this paper, the modulation bandwidth of QCLs can be found by looking at the spectral density of the current noise in the external circuit. 

Although in this paper the emphasis has been on a specific multiple quantum well QCL structure, the theoretical methods and techniques presented in this paper can be used to study a variety of QCLs that have been reported in the literature. 

\section*{Acknowledgments}
This work was supported by DARPA, Rome Laboratories, and ONR.

\appendix

\section{Correlations Among the Langevin Noise Sources} \label{langcor}

\begin{eqnarray}
WL\,\langle f_{32}^{j}(t)\,f_{32}^{q}(t') \rangle & = & (R_{3 \rightarrow 2} + R_{2 \rightarrow 3}) \, \delta_{jq} \, \delta(t-t') \nonumber \\
                                                  & \approx & R_{32}(n_{3}^{j},n_{2}^{j}) \, \delta_{jq} \, \delta(t-t') \label{F32F32}
\end{eqnarray}
\begin{eqnarray}
WL\,\langle f_{31}^{j}(t)\,f_{31}^{q}(t') \rangle & = & (R_{3 \rightarrow 1} + R_{1 \rightarrow 3}) \, \delta_{jq} \, \delta(t-t') \nonumber \\
                                                  & \approx & R_{31}(n_{3}^{j},n_{1}^{j}) \, \delta_{jq} \, \delta(t-t') \label{F31F31}
\end{eqnarray}
\begin{eqnarray}
WL\,\langle f_{21}^{j}(t)\,f_{21}^{q}(t') \rangle & = & (R_{2 \rightarrow 1} + R_{1 \rightarrow 2}) \, \delta_{jq} \, \delta(t-t') \nonumber \\
                                                  & \approx & R_{21}(n_{2}^{j},n_{1}^{j}) \, \delta_{jq} \, \delta(t-t') \label{F21F21}
\end{eqnarray}
\begin{equation}
WL\,\langle f_{RN}^{j}(t)\,f_{RN}^{q}(t') \rangle = \Gamma^{j}v_{g}\,g(n_{3}^{j},n_{2}^{j})\left[ (2n_{sp} - 1)S_{p} + \frac{n_{sp}}{WL} \right]\,\delta_{jq}\,\delta(t-t') \label{FrnFrn}
\end{equation}
\begin{equation}
WL\,\langle f^{j}_{RS}(t)\,f^{q}_{RS}(t') \rangle =  \Gamma^{j}v_{g}\,g(n_{3}^{j},n_{2}^{j})\left[ (2n_{sp} - 1)S_{p} + \frac{n_{sp}}{WL} \right]\,\delta_{jq}\,\delta(t-t') \label{FrsFrs}
\end{equation}
\begin{equation}
WL\,\langle f^{j}_{RS}(t)f_{RN}^{q}(t') \rangle = \Gamma^{j}v_{g}\,g(n_{3}^{j},n_{2}^{j})\left[ (2n_{sp} - 1)S_{p} + \frac{n_{sp}}{WL} \right]\,\delta_{jq}\,\delta(t-t') \label{FrsFrn}
\end{equation}
\begin{equation}
WL\,\langle F_{L}(t)\,F_{L}(t') \rangle = \frac{S_{p}}{\tau_{p}} \, \delta(t-t') \label{FlFl}
\end{equation}
\begin{equation}
\langle F_{o}(t)\,F_{o}(t') \rangle = \eta_{o}\,(h \nu)^{2}\,\frac{WL\,S_{p}}{\tau_{p}} \, \delta(t-t') \label{FoFo}
\end{equation}
\begin{equation}
\langle F_{o}(t)\,F_{L}(t') \rangle = \eta_{o}\,h \nu\,\frac{S_{p}}{\tau_{p}} \, \delta(t-t') \label{FoFl}
\end{equation}

\section{Differential Resistance of a QCL} \label{DiffRes}
The expression in Equations (\ref{rdath}) and (\ref{rdbth}) for the differential resistance $R_{d}$ of a QCL can be put in the form,
\begin{equation}
R_{d} = \left\{
\begin{array}{ll}
{\displaystyle \frac{N}{WL}\,\frac{\tau_{in}}{C_{inj}}\,\left( 1+ \theta'_{3}+ \theta'_{2} + \theta_{1}\right) \hspace{1 cm} (I<I_{th})  } & \\
{\displaystyle \frac{N}{WL}\,\frac{\tau_{in}}{C_{inj}}\,\left( 1+ \theta_{3}+ \theta_{2}+ \theta_{1}\right) \hspace{1 cm} (I>I_{th}) }
\end{array} \right.
\end{equation}
The dimensionless parameters $\theta_{3}$, $\theta_{3}'$, $\theta_{2}$, $\theta_{2}'$, and $\theta_{1}$ that have been used in the above Equation are as follows,
\begin{equation}
\theta_{3}= \left( \frac{1}{\tau_{in}}\frac{C_{inj}}{C_{3}} + \frac{1}{\tau_{3}} \right) \; \frac{\tau_{31} \tau_{21}}{(\tau_{21} + \tau_{31})} \hspace{0.25 cm} , \hspace{0.25 cm} \theta'_{3}= \left( \frac{1}{\tau_{in}}\frac{C_{inj}}{C_{3}} + \frac{1}{\tau_{3}} \right) \; \frac{\tau_{32} \tau_{31}}{(\tau_{32} + \tau_{31})}
\end{equation}
\begin{equation}
\theta_{2}= \left( \frac{1}{\tau_{in}}\frac{C_{inj}}{C_{2}} + \frac{1}{\tau_{2}} \right) \; \frac{\tau_{31} \tau_{21}}{(\tau_{21} + \tau_{31})} \hspace{0.25 cm} , \hspace{0.25 cm} \theta'_{2} = \left( \frac{1}{\tau_{in}}\frac{C_{inj}}{C_{2}} + \frac{1}{\tau_{2}} \right) \; \frac{\tau_{31} \tau_{21}}{(\tau_{32} + \tau_{31})}
\end{equation}
\begin{equation}
\theta_{1}= \left( \frac{1}{\tau_{in}}\frac{C_{inj}}{C_{1}} + \frac{1}{\tau_{1}} \right)\;\tau_{out}
\end{equation}

\section{Elements of Matrix {\bf D}}  \label{dmatrix}
The non-zero elements of the matrix ${\bf D}$ are,
\begin{equation}
{\bf D}_{11} =  j\omega + \frac{1}{\tau_{out}} \hspace{0.25 cm} , \hspace{0.25 cm} {\bf D}_{12} = - \frac{1}{\tau_{21}} \hspace{0.25 cm} , \hspace{0.25 cm} {\bf D}_{13} = - \frac{1}{\tau_{31}} 
\end{equation}
\begin{equation}
{\bf D}_{22} = j\omega + \frac{1}{\tau_{21}} + \Gamma v_{g} \, a \, \left( S_{p} + \frac{n_{sp}}{WL} \right) 
\end{equation}
\begin{equation}
{\bf D}_{23} = -\frac{1}{\tau_{32}} - \Gamma v_{g} \, a \, \left( S_{p} + \frac{n_{sp}}{WL} \right)
\end{equation}
\begin{equation}
{\bf D}_{24} = -{\bf D}_{34} = -N\,\Gamma v_{g} \, g(n_{3},n_{2})
\end{equation}
\begin{equation}
{\bf D}_{31} = \frac{1}{\tau_{1}} \frac{j\omega \, \tau_{in}}{(1 + j\omega\,\tau_{in})}
\end{equation}
\begin{equation}
{\bf D}_{32} = \frac{1}{\tau_{2}} \frac{j\omega \, \tau_{in}}{(1 + j\omega\,\tau_{in})} - \Gamma v_{g} \, a \, \left( S_{p} + \frac{n_{sp}}{WL} \right)
\end{equation}
\begin{equation}
{\bf D}_{33} = j\omega + \frac{1}{\tau_{3}} \frac{j\omega \, \tau_{in}}{(1 + j\omega\,\tau_{in})} + \frac{1}{\tau_{32}} + \frac{1}{\tau_{31}} + \Gamma v_{g} \, a \, \left( S_{p} + \frac{n_{sp}}{WL} \right)
\end{equation}  
\begin{equation}
{\bf D}_{42} = - {\bf D}_{43} = \Gamma v_{g} \, a \, \left( S_{p} + \frac{n_{sp}}{WL} \right)
\end{equation}
\begin{equation}
{\bf D}_{44} = j\omega + \frac{1}{\tau_{p}} - N\,\Gamma v_{g} \, g(n_{3},n_{2}) 
\end{equation}

\section{Important Elements of Matrix {\bf D}$^{-1}$} \label{idmatrix}
Above threshold, elements of the matrix ${\bf D}^{-1}$ in the limit $\omega \tau_{in} \ll 1$, and $\{ \tau_{2} \, , \, \tau_{1} \} \rightarrow \infty$ are as follows:

\begin{equation}
{\bf D}^{-1}_{11} = \frac{\tau_{out}}{(j\omega\,\tau_{out} + 1)}
\end{equation}
\begin{eqnarray}
{\bf D}^{-1}_{12} &  = & \tau_{out}\,\tau_{p}\,\tau_{st}\,\left\{ (j\omega)^{2}\,\left( 1 + \frac{\tau_{in}}{\tau_{3}} \right) \frac{1}{\tau_{21}} + j\omega\,\left[ \frac{1}{\tau_{st}}\left( \frac{1}{\tau_{21}} + \frac{1}{\tau_{31}} \right) + \frac{1}{\tau_{21}\,\tau_{31}} + \frac{1}{\tau_{21}\,\tau_{32}}  \right] \right. \nonumber \\
 & & \hspace{3 cm} \left. \mbox{} + \frac{1}{\tau_{p}\,\tau_{st}}\,\left(\frac{1}{\tau_{21}} + \frac{1}{\tau_{31}} \right) \right\} \frac{\tau_{21}\,\tau_{31}}{(\tau_{21} + \tau_{31})} \: \frac{\displaystyle H(\omega)}{\displaystyle (j\omega\,\tau_{out} + 1) }
\end{eqnarray}
\begin{eqnarray}
{\bf D}^{-1}_{13} &  = & \tau_{out}\,\tau_{p}\,\tau_{st}\,\left\{ (j\omega)^{2}\,\left( \frac{1}{\tau_{31}} \right) + j\omega\,\left[ \frac{1}{\tau_{st}}\left( \frac{1}{\tau_{21}} + \frac{1}{\tau_{31}} \right) + \frac{1}{\tau_{21}\,\tau_{31}} + \frac{1}{\tau_{21}\,\tau_{32}} \right] \right. \nonumber \\
 & & \hspace{3 cm} \left. \mbox{} + \frac{1}{\tau_{p}\,\tau_{st}}\,\left(\frac{1}{\tau_{21}} + \frac{1}{\tau_{31}} \right) \right\} \frac{\tau_{21}\,\tau_{31}}{(\tau_{21} + \tau_{31})} \: \frac{\displaystyle H(\omega)}{\displaystyle (j\omega\,\tau_{out} + 1) }
\end{eqnarray}
\begin{equation}
{\bf D}^{-1}_{14} = \tau_{out}\,\tau_{st}\,\left\{ j\omega\,\left[ \frac{1}{\tau_{21}}\left( 1 + \frac{\tau_{in}}{\tau_{3}} \right) - \frac{1}{\tau_{31}} \right] \right\} \frac{\tau_{21}\,\tau_{31}}{(\tau_{21} + \tau_{31})} \: \frac{\displaystyle H(\omega)}{\displaystyle (j\omega\,\tau_{out} + 1) }
\end{equation}

\begin{equation}
{\bf D}^{-1}_{21} \approx 0;
\end{equation}
\begin{equation}
{\bf D}^{-1}_{22} =\tau_{p}\tau_{st}\,\left[ (j\omega)^{2}\,\left( 1 + \frac{\tau_{in}}{\tau_{3}} \right) + j\omega\,\left(\frac{1}{\tau_{32}} + \frac{1}{\tau_{31}} + \frac{1}{\tau_{st}} \right) + \frac{1}{\tau_{p}\tau_{st}} \right]\frac{\tau_{21}\,\tau_{31}}{(\tau_{21} + \tau_{31})} \: H(\omega)
\end{equation}
\begin{equation}
{\bf D}^{-1}_{23} = \tau_{p}\,\left[ j\omega\,\left( \frac{\tau_{st}}{\tau_{32}} + 1 \right) + \frac{1}{\tau_{p}} \right]\frac{\tau_{21}\,\tau_{31}}{(\tau_{21} + \tau_{31})} \: H(\omega)
\end{equation}
\begin{equation}
{\bf D}^{-1}_{24} = \tau_{st}\,\left[ j\omega \,\left( 1 + \frac{\tau_{in}}{\tau_{3}} \right) + \frac{1}{\tau_{31}} \right] \frac{\tau_{21}\,\tau_{31}}{(\tau_{21} + \tau_{31})} \: H(\omega)
\end{equation}

\begin{equation}
{\bf D}^{-1}_{31} \approx 0;
\end{equation}
\begin{equation}
{\bf D}^{-1}_{32} = \tau_{p}\,\left( j\omega + \frac{1}{\tau_{p}} \right)\frac{\tau_{21}\,\tau_{31}}{(\tau_{21} + \tau_{31})} \: H(\omega)
\end{equation}
\begin{equation}
{\bf D}^{-1}_{33} = \tau_{p}\tau_{st}\,\left[ (j\omega)^{2} + j\omega\,\left( \frac{1}{\tau_{21}} + \frac{1}{\tau_{st}} \right) + \frac{1}{\tau_{p}\tau_{st}} \right] \frac{\tau_{21}\,\tau_{31}}{(\tau_{21} + \tau_{31})} \: H(\omega)
\end{equation}
\begin{equation}
{\bf D}^{-1}_{34} = - \tau_{st}\,\left( j\omega + \frac{1}{\tau_{21}} \right) \frac{\tau_{21}\,\tau_{31}}{(\tau_{21} + \tau_{31})} \: H(\omega)
\end{equation}

\begin{equation}
{\bf D}^{-1}_{41} \approx 0
\end{equation}
\begin{equation}
{\bf D}^{-1}_{42} = - \tau_{p}\,\left[ j\omega\,\left( 1 + \frac{\tau_{in}}{\tau_{3}} \right) + \left( \frac{1}{\tau_{31}} + \frac{1}{\tau_{32}} \right) \right]\frac{\tau_{21}\,\tau_{31}}{(\tau_{21} + \tau_{31})} \: H(\omega)
\end{equation} 
\begin{equation}
{\bf D}^{-1}_{43} = \tau_{p}\,\left[ j\omega + \left( \frac{1}{\tau_{21}} - \frac{1}{\tau_{32}} \right) \right]\frac{\tau_{21}\,\tau_{31}}{(\tau_{21} + \tau_{31})} \: H(\omega)
\end{equation}
\begin{eqnarray}
{\bf D}^{-1}_{44} & = & \tau_{p}\,\tau_{st}\,\left\{ (j\omega)^{2}\,\left( 1 + \frac{\tau_{in}}{\tau_{3}} \right) + (j\omega)\,\left[ \frac{1}{\tau_{21}}\left( 1 + \frac{\tau_{in}}{\tau_{3}} \right) + \frac{1}{\tau_{31}} + \frac{1}{\tau_{32}} + \frac{1}{\tau_{st}}\left( 2 + \frac{\tau_{in}}{\tau_{3}} \right) \right] \right. \nonumber \\
& & \hspace{2cm} \left. \mbox{} + \left[ \frac{1}{\tau_{st}}\left( \frac{1}{\tau_{21}} + \frac{1}{\tau_{31}} \right) + \frac{1}{\tau_{21}\,\tau_{31}} + \frac{1}{\tau_{21}\,\tau_{32}} \right] \right\} \frac{\tau_{21}\,\tau_{31}}{(\tau_{21} + \tau_{31})} \: H(\omega)
\end{eqnarray}
where $H(\omega)$ is,
\begin{eqnarray}
H(\omega) & = & \frac{1}{\tau_{p}\,\tau_{st}} \left( \frac{1}{\tau_{21}} + \frac{1}{\tau_{31}} \right)\,\left\{ (j\omega)^{3}\,\left(1 + \frac{\tau_{in}}{\tau_{3}} \right) \right. \nonumber \\
& & \left. \mbox{} + (j\omega)^{2}\,\left[ \frac{1}{\tau_{21}}\left( 1 + \frac{\tau_{in}}{\tau_{3}} \right) + \frac{1}{\tau_{31}} + \frac{1}{\tau_{32}} + \frac{1}{\tau_{st}}\left( 2 + \frac{\tau_{in}}{\tau_{3}} \right)  \right] \right. \nonumber \\
& & \left. \mbox{} + j\omega \,\left[ \frac{1}{\tau_{st}}\left( \frac{1}{\tau_{21}} + \frac{1}{\tau_{31}} \right) + \frac{1}{\tau_{21}\,\tau_{31}} + \frac{1}{\tau_{21}\,\tau_{32}} + \frac{1}{\tau_{p}\,\tau_{st}}\left( 2 + \frac{\tau_{in}}{\tau_{3}} \right) \right] \right. \nonumber \\
& & \left. \mbox{} + \frac{1}{\tau_{p}\,\tau_{st}}\,\left(\frac{1}{\tau_{21}} + \frac{1}{\tau_{31}} \right) \right\}^{-1} \label{eqHapprox2} \\
& \approx & \frac{\omega_{R}^{2}}{(\omega_{R}^{2} - \omega^{2} + j\omega \gamma)} \: \: \: \: \: \: \: \: \: \: \: \: \: ({\rm for \: small \: \omega }) \label{eqHapprox}
\end{eqnarray}
In the limit $\omega \tau_{in} \ll 1$, and $\{ \tau_{2} \, , \, \tau_{1} \} \rightarrow \infty$ the current modulation response function $H(\omega)$ has only three poles (Equation (\ref{eqHapprox2})). In order to approximate $H(\omega)$ with a function with only two poles, as in Equation (\ref{eqHapprox}), one may neglect the pole which has the largest magnitude. This approximation will be valid provided that the pole which is dropped has a magnitude much larger than the other two poles. In this paper we have simply dropped the term cubic in $\omega$ in the denominator of $H(\omega)$ in order to obtain a two-pole approximation for $H(\omega)$. Numerical simulations show that this is an excellent approximation. In this approximation $\omega_{R}$ and $\gamma$ are,
\begin{equation}
\omega^{2}_{R}= \frac{\displaystyle \frac{1}{\tau_{p}\,\tau_{st}}\,\left(1 + \frac{\tau_{21}}{\tau_{31}} \right)}{\displaystyle \left[ 1 + \frac{\tau_{21}}{\tau_{31}} + \frac{\tau_{21}}{\tau_{32}} + \frac{\tau_{in}}{\tau_{3}} + \frac{\tau_{21}}{\tau_{st}} \, \left( 2 + \frac{\tau_{in}}{\tau_{3}} \right) \right]} 
\end{equation} 
\begin{equation}
\gamma = \frac{\displaystyle \left[ \frac{1}{\tau_{st}}\,\left(1 + \frac{\tau_{21}}{\tau_{31}} \right) + \frac{1}{\tau_{31}} + \frac{1}{\tau_{32}} + \frac{\tau_{21}}{\tau_{p}\,\tau_{st}}\,\left(2 + \frac{\tau_{in}}{\tau_{3}} \right) \right] }{\displaystyle \left[ 1 + \frac{\tau_{21}}{\tau_{31}} + \frac{\tau_{21}}{\tau_{32}} + \frac{\tau_{in}}{\tau_{3}} + \frac{\tau_{21}}{\tau_{st}} \, \left( 2 + \frac{\tau_{in}}{\tau_{3}} \right) \right]} 
\end{equation}

\section{Noise Spectral Densities and Fano Factors} \label{NoiseSpecDen}
The spectral densities $K_{I}(\omega)$ and $K_{P}(\omega)$ of noise power for the current noise and the intensity noise, respectively, can be computed from the equations,
\begin{equation}
K_{I}(\omega) = \int_{-\infty}^{\infty}\,\frac{d\,\omega'}{2\pi}\:\langle \delta I^{\ast}(\omega)\,\delta I(\omega - \omega') \rangle \label{eqKi}
\end{equation}
\begin{equation}
K_{P}(\omega) = \int_{-\infty}^{\infty}\,\frac{d\,\omega'}{2\pi}\:\langle \delta P_{out}^{\ast}(\omega)\, \delta P_{out}(\omega - \omega') \rangle \label{eqKp}
\end{equation}
Equations (\ref{eqKi}) and (\ref{eqKp}) can be used with Equations (\ref{eqdeltaI}) and (\ref{eqdeltaPo1}) to compute the noise spectral densities. Since all the Langevin noise sources are delta correlated in time domain, they will also be delta correlated in frequency domain, and therefore the fluctuations $\delta I$ and $\delta P_{out}$ in the current and the output power, respectively, will be also be delta correlated in time and frequency domains.

The Fano Factors $F_{I}(\omega)$ and $F_{P}(\omega)$ for the current noise and the intensity noise, respectively, are defined as the ratios of the actual noise spectral densities to the noise spectral densities of shot noise, and are given by the relations,
\begin{equation}
F_{I}(\omega) = \frac{K_{I}(\omega)}{q\,I} \:\:\:\:\:\:\:\:\:\:\:\:\:\:\:\:\:\: {\rm and} \:\:\:\:\:\:\:\:\:\:\:\:\:\:\:\:\:\: F_{P}(\omega) = \frac{K_{P}(\omega)}{h \nu \,P_{out}} \label{eqF}
\end{equation}

The Relative Intensity Noise (RIN) is defined as,
\begin{equation}
{\rm RIN} = 10\,\log_{10}{\left[ \frac{K_{P}(\omega)}{P_{out}^{2}} \right] }
\end{equation}

\section{Current Noise and the Equivalent Circuit Models} \label{CirMod}
Expression for $\delta I^{j}(\omega)$ in Equation (\ref{eqCirMod}) can be found from Equation (\ref{eqdeltaIext}) and Equations (\ref{eqF1})-(\ref{eqF4}),
\begin{equation}
\delta I^{j}(\omega) = q \,\frac{N}{Z(\omega)}\,\frac{\tau_{in}}{C_{inj}}\,\left[ \frac{f^{j}_{in}(\omega)}{(1 + j\omega \, \tau_{in})} - \sum_{k=1}^{3} \sum_{l=1}^{4} \, \left( \frac{1}{\tau_{in}} \, \frac{C_{inj}}{C_{k}} + \frac{1}{\tau_{k}(1 + j\omega \, \tau_{in})} \right) {\bf D}_{kl}^{-1}(\omega)\,F^{j}_{l}(\omega) \right] \label{eqdeltaIj}
\end{equation}  
The correlation between $\delta I^{j}(\omega)$ and $\delta I^{r}(\omega')$ for $j \neq r$ is,
\begin{eqnarray}
\langle \delta I^{j}(\omega)\:\delta I^{r}(\omega') \rangle & = & \left\vert \frac{q}{Z(\omega)}\,\frac{\tau_{in}}{C_{inj}}\,\sum_{k=1}^{3} \, \left( \frac{1}{\tau_{in}} \, \frac{C_{inj}}{C_{k}} + \frac{1}{\tau_{k}(1 + j\omega \, \tau_{in})} \right) {\bf D}_{k4}^{-1}(\omega) \right\vert^{2} \langle F_{L}(\omega)\:F_{L}(\omega') \rangle \nonumber \\
& = & q\eta_{r}\frac{(I-I_{th})}{N} \; \frac{\displaystyle \left( \frac{\tau_{st}}{\tau_{21}} \theta_{3}- \frac{\tau_{st}}{\tau_{31}} \theta_{2}\right)^{2}}{(1 + \theta_{3}+ \theta_{2}+ \theta_{1})^{2}} \: 2\pi \delta (\omega - \omega') \hspace{0.5 cm} (I > I_{th} \: , \: \omega < \omega_{3\:{\rm dB}} )
\end{eqnarray}
which shows that $\delta I^{j}(\omega)$ and $\delta I^{r}(\omega')$ are positively correlated.

\section{Noise Model for Semiconductor Quantum Well Diode Lasers} \label{intnoise}
\begin{figure}[t]
\begin{center}
   \epsfig{file=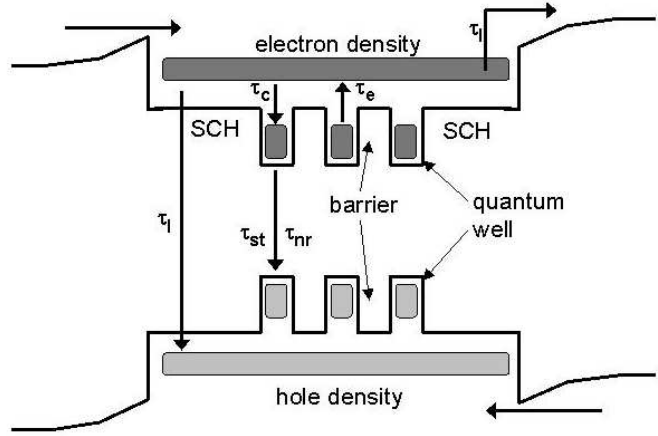,width=3.5in}
\caption{ Active region of a semiconductor quantum well diode laser.} 
\label{FigIntLas}
\end{center}
\end{figure}

A simple model for the current and intensity noise in quantum well interband semiconductor diode lasers is presented~\cite{farhan1,farhan4}. The active region of a quantum well diode laser is shown in Fig.~\ref{FigIntLas}. 
The carriers are injected from the leads into the separate confinement heterostructure (SCH) region either by tunneling or by thermionic emission over the hetero-barrier. The rate equations for the fluctuations $\delta N_{c}$ and $\delta N_{w}$ in the carrier densities (cm$^{-3}$) in the SCH region and the quantum wells, respectively, and the fluctuations $\delta S_{p}$ in the photon density (cm$^{-3}$) are~\cite{farhan1,farhan4},
\begin{equation}
\frac{d\,\delta N_{c}}{d\,t} = \frac{\delta I_{ext}}{qV_{c}} - \delta N_{c} \left( \frac{1}{\tau_{c}} + \frac{1}{\tau_{l}} \right) + \frac{\delta N_{w}}{\tau_{e}}\frac{V_{w}}{V_{c}} - F_{c} - F_{l} + F_{e}\frac{V_{w}}{V_{c}} \label{eqdeltanc}
\end{equation}
\begin{equation}
\frac{d\,\delta N_{w}}{d\,t} =  \frac{\delta N_{c}}{\tau_{c}}\frac{V_{c}}{V_{w}} - \delta N_{w} \left(\frac{1}{\tau_{e}} +  \frac{1}{\tau_{nr}} +  \frac{1}{\tau_{st}} \right) - v_{g} \; g \, \delta S_{p}  + F_{c}\frac{V_{c}}{V_{w}} - F_{e} - F_{nr} - F_{RN} \label{eqN3int}
\end{equation}
\begin{equation}
\frac{d\,\delta S_{p}}{d\,t} = \frac{ \delta N_{w}}{\tau_{st}}\frac{V_{w}}{V_{p}}  +  \left( \Gamma v_{g}\; g - \frac{1}{\tau_{p}} \right) \delta S_{p} + F_{RS} - F_{L} \label{eqSpint}
\end{equation}
and the current and the intensity fluctuations are given as~\cite{farhan1,farhan4},
\begin{equation}
\frac{\delta I_{ext}}{qV_{c}} = \frac{G \: \delta V_{d}}{qV_{c}} - \frac{\delta N_{c}}{\tau_{_{G}}}  +  F_{in}
\end{equation}
\begin{equation}
\delta P_{out} = \eta_{o}\,h \nu \,\frac{V_{p}\:\delta S_{p}}{\tau_{p}} + F_{o}
\end{equation}
It is assumed that the carrier density $N_{c}$ in the SCH region also includes the carriers inside the quantum well barriers and also those in the quantum wells which have energy high enough to not be confined within the quantum wells (Fig.~\ref{FigIntLas}). Only those carriers which are confined within the quantum wells are included in the carrier density $N_{w}$. $V_{c}$ and $V_{w}$ are the volumes of the SCH region and the quantum wells, respectively. $V_{p}$ is the volume of the optical mode. $\tau_{c}$ and $\tau_{e}$ are the capture and emission times for electrons going into and coming out of the quantum wells, respectively. $\tau_{l}$ is the lifetime associated with carrier leakage and recombination in the SCH region. $\tau_{nr}$ is the non-radiative recombination time in the quantum wells. $\tau_{st}$ is the differential lifetime associated with stimulated and spontaneous emission into the lasing mode. $\tau_{p}$ is the photon lifetime inside the laser cavity. $\delta V_{d}$ is the fluctuation in the voltage across the active region. The conductance $G$ relates the increase in the injection current into the SCH region from the leads with the increase in the voltage across the active region at a {\em fixed carrier density}. $\tau_{_{G}}$ relates the decrease in the current injection rate to the increase in the carrier density in the SCH region. $F_{in}$ is the Langevin noise source associated with carrier injection into the SCH region. $F_{l}$, $F_{c}$, and $F_{e}$ model the noise in carrier leakage, carrier capture and carrier emission events. $F_{nr}$ describes the noise in non-radiative recombination in the quantum wells including spontaneous emission into the non-lasing modes. $F_{RN}$ and $F_{RS}$ model the noise associated with photon emission into the lasing mode. $F_{L}$ and $F_{o}$ model the noise in photon loss from the cavity. All the non-zero correlations of the Langevin noise sources can be obtained from the methods described in Ref.~\cite{coldren},
\begin{equation}
V_{c} \, \langle F_{c}(t)\,F_{c}(t') \rangle = \frac{N_{c}}{\tau_{c}} \: \delta (t-t')
\end{equation} 
\begin{equation}
V_{c} \, \langle F_{l}(t)\,F_{l}(t') \rangle = \frac{N_{c}}{\tau_{l}} \: \delta (t-t')
\end{equation} 
\begin{equation}
V_{w} \, \langle F_{e}(t)\,F_{e}(t') \rangle = \frac{N_{w}}{\tau_{e}} \: \delta (t-t')
\end{equation} 
\begin{equation}
V_{w} \, \langle F_{nr}(t)\,F_{nr}(t') \rangle = \frac{N_{w}}{\tau_{nr}} \: \delta (t-t')
\end{equation} 
\begin{equation}
V_{w} \, \langle F_{RN}(t)\,F_{RN}(t') \rangle = v_{g}\,g\,\left[ \left( 2n_{sp} - 1 \right) S_{p} + \frac{n_{sp}}{V_{p}} \right]   \: \delta (t-t')
\end{equation} 
\begin{equation}
V_{p} \, \langle F_{RS}(t)\,F_{RS}(t') \rangle = \frac{V_{w}}{V_{p}} \, v_{g}\,g\,\left[ \left( 2n_{sp} - 1 \right) S_{p} + \frac{n_{sp}}{V_{p}} \right]   \: \delta (t-t')
\end{equation} 
\begin{equation}
V_{p} \, \langle F_{RN}(t)\,F_{RS}(t') \rangle = v_{g}\,g\,\left[ \left( 2n_{sp} - 1 \right) S_{p} + \frac{n_{sp}}{V_{p}} \right]   \: \delta (t-t')
\end{equation} 
\begin{equation}
V_{p} \, \langle F_{L}(t)\,F_{L}(t') \rangle = \frac{S_{p}}{\tau_{p}} \: \delta (t-t')
\end{equation} 
\begin{equation}
\langle F_{o}(t)\,F_{o}(t') \rangle = \eta_{o} \, \left( h \nu \right)^{2} \frac{V_{p}\,S_{p}}{\tau_{p}} \: \delta (t-t')
\end{equation} 
\begin{equation}
V_{p} \, \langle F_{o}(t)\,F_{L}(t') \rangle = \eta_{o} \, \left( h \nu \right) \, \frac{V_{p}\,S_{p}}{\tau_{p}} \: \delta (t-t')  
\end{equation} 
$F_{in}$ has the approximate correlation~\cite{yama4},
\begin{equation}
V_{c}^{2}\,\langle F_{in}(t) \, F_{in}(t') \rangle \approx \left( \frac{I}{q} + 2\,\frac{N_{c}V_{c}}{\tau_{_{G}}} \right) \, \delta (t - t')
\end{equation}
The inclusion of the rate equation for fluctuations in the carrier density in the SCH region is necessary to accurately model the current noise. Carrier leakage in the SCH region results in a less than unity efficiency $\eta_{i}$ for current injection into the quantum wells,
\begin{equation}
\eta_{i} = \frac{\tau_{l}}{\left( \tau_{c} + \tau_{l} \right) }
\end{equation} 
and above threshold the expression for the output power can be written as,
\begin{equation}
P_{out} = \eta_{o} \eta_{i} \frac{h \nu}{q} \, (I-I_{th})
\end{equation}
where $\eta_{o}$ is the output coupling efficiency~\cite{coldren}. 

\subsection{Modulation Response}
The current modulation response of diode lasers follows from the rate equations, and for frequencies less than the inverse of the carrier capture time $\tau_{c}$, it can be put in the form~\cite{coldren},
\begin{equation}
\frac{\delta P_{out}(\omega)}{\delta I_{ext}(\omega)} = \eta_{o}\,\eta_{i}\,\frac{h \nu}{q} \, H(\omega) = \eta_{o}\,\eta_{i}\,\frac{h \nu}{q} \, \frac{\omega_{R}^{2}}{(\omega_{R}^{2} - \omega^{2} + j\omega \gamma)} \label{eqmodres3int}
\end{equation}
The relaxation oscillation frequency $\omega_{R}$ and the damping constant $\gamma$ are,
\begin{eqnarray}
\omega_{R}^{2} & = & \frac{\displaystyle \frac{1}{\tau_{st} \, \tau_{p}} }{\displaystyle 1 + \eta_{i}\,\frac{\tau_{c}}{\tau_{e}}}   \\
\gamma & = & \frac{\displaystyle \left( \frac{1}{\tau_{nr}} + \frac{1-\eta_{i}}{\tau_{e}} + \frac{1}{\tau_{st}}  \right)}{\displaystyle 1 + \eta_{i}\,\frac{\tau_{c}}{\tau_{e}}} = K\,\omega^{2}_{R} + \gamma_{o}
\end{eqnarray}
where,
\begin{equation}
K = \tau_{p} \hspace{1 cm} {\rm and} \hspace{1 cm} \gamma_{o} = \frac{ \displaystyle \left( \frac{1}{\tau_{nr}} + \frac{1-\eta_{i}}{\tau_{e}} \right)}{\displaystyle 1 + \eta_{i}\,\frac{\tau_{c}}{\tau_{e}}} \label{eqKg2} 
\end{equation}
In diode lasers $\gamma$ is small at threshold, and the 3 dB frequency $\omega_{3\:{\rm dB}}$ increases with the bias current until $\gamma/\sqrt{2}$ equals $\omega_{R}$. As the bias current is increased beyond this point the modulation response becomes over-damped, and $\omega_{3\:{\rm dB}}$ starts to decrease. Thus, in diode lasers the value for $\omega_{\rm 3\: dB\:|\: max}$ can be found by equating $\gamma/\sqrt{2}$ with $\omega_{R}$, and it comes out to be,
\begin{equation}
\omega_{\rm 3\: dB\:|\: max} \approx \frac{\sqrt{2}}{\tau_{p}} \label{eqwr3dbmaxint}
\end{equation} 

\subsection{Differential Resistance}
The differential resistance of the laser diode below and above threshold can be derived from the rate equations by removing all the noise sources,
\begin{equation}
R_{d} = \left\{
\begin{array}{lll}
{\displaystyle \frac{1}{G} \, \left( 1 + \theta' \right) \hspace{1 cm} (I < I_{th})  } & \\ & \\
{\displaystyle \frac{1}{G} \, \left( 1+ \theta \right) \hspace{1 cm} (I>I_{th}) }
\end{array} \right. \label{rdint}
\end{equation}
$\theta'$ and $\theta$ are,
\begin{equation}
\theta' =  \frac{1}{\tau_{_{G}}} \, \frac{\tau_{l} \, \tau_{c} }{\left(  \tau_{l}\left( 1 - \eta_{e} \right)  + \tau_{c} \right) } \hspace{0.25 cm} , \hspace{0.25 cm} \theta = \frac{1}{\tau_{_{G}}} \, \frac{\tau_{l} \, \tau_{c} }{\left(  \tau_{l}  + \tau_{c} \right) }
\end{equation}
where the emission efficiency $\eta_{e}$ is $\tau_{nr}/(\tau_{e} + \tau_{nr})$. Values of the time constants $\tau_{c}$, $\tau_{e}$, $\tau_{g}$, $\tau_{l}$ and $\tau_{nr}$ are typically 10 ps, 40 ps, 50 ps, 60 ps, and 1 ns, respectively~\cite{zory}. It follows that $\theta'$ and $\theta$ have the values 0.97 and 0.17, respectively.

The discontinuity $\Delta R_{d}$ in the differential resistance at threshold becomes,
\begin{equation}
\Delta R_{d} = \frac{1}{G}\left( \theta' - \theta \right) = \eta_{i} \, \eta_{e} \frac{\theta'}{G} = \eta_{i} \, \eta_{e} \frac{\theta'}{1 + \theta'} \, R_{d} \Big\vert_{ I \le I_{th} \: {\rm evaluated \: at} \: I=I_{th}}
\end{equation} 
Below threshold the current-voltage characteristics of a laser diode resemble that of an ideal PN junction~\cite{aggarwal},
\begin{equation}
I = I_{o} \left[ \exp{\left( \frac{qV_{d}}{mK_{_{B}}T} \right) } - 1 \right]
\end{equation}
where $m$ is the diode ideality factor with values usually between 1.5 and 2. Therefore,
\begin{equation}
R_{d} \Big\vert_{I \le I_{th}} = m \frac{K_{_{B}}T}{qI}
\end{equation}
and $\Delta R_{d}$ becomes,
\begin{equation}
\Delta R_{d} = \eta_{i} \, \eta_{e} \frac{\theta'}{1 + \theta'} \, m \frac{K_{_{B}}T}{qI_{th}}
\end{equation}
The above equation shows that the discontinuity in the differential resistance at threshold is $K_{_{B}}T/qI_{th}$ times a factor which is close to unity.

\subsection{Differential Impedance}
The differential Impedance $Z(\omega)$ of a diode laser above threshold can be expressed in terms of the modulation response $H(\omega)$,
\begin{equation}
Z(\omega) = \frac{1}{G}\,\left[ 1 + \theta \left( 1 + j\omega \, \eta_{i} \, \frac{\tau_{p}\,\tau_{st}}{\tau_{e}} \, H(\omega) \right) \right]
\end{equation}
The differential resistance $R_{d}$ above threshold, given by Equation (\ref{rdint}), equals $Z(\omega = 0)$.

\subsection{Current Noise}
\begin{figure}[t]
\begin{center}
\epsfig{file=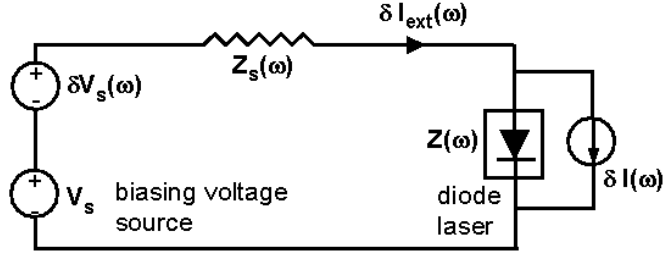,angle=0,width=3.5in}
\caption{\small Circuit model for the current fluctutions in semiconductor diode lasers} 
\label{FigCirModInt}
\end{center}
\end{figure}

As in the case of QCLs, the fluctuations $\delta I(\omega)$ produced by the current noise source that sits in parallel with the laser diode (Fig.~\ref{FigCirModInt}) can be found by looking at the current noise in the external circuit when the voltage fluctuation $\delta V_{d}$ across the diode is zero (because all external sources and impedances are assumed to be shorted). 
This implies that, 
\begin{equation}
\frac{ \delta I(\omega) }{q} =  V_{c} F_{in}(\omega) - \frac{ \delta N_{c}(\omega) V_{c}}{ \tau_{_{G}} } \label{eqdeltaIint0}
\end{equation}
Below threshold, and for frequencies less than $\omega_{3\:{\rm dB}}$, $\delta I(\omega)$ is,
\begin{equation}
\frac{ \delta I(\omega) }{q} = \frac{  
V_{c} F_{in}(\omega) + 
\theta' \left( 1 - \eta_{e} \right) \left[ V_{c}F_{c}(\omega) - V_{w}F_{e}(\omega) \right] + 
\theta' \, V_{c} F_{l}(\omega) + 
\theta' \, \eta_{e} \, V_{w}F_{nr}(\omega) 
}
{ (1 + \theta' ) } \label{eqdeltaIint1}
\end{equation}
Above threshold $\delta I(\omega)$ is,
\begin{equation}
\frac{ \delta I(\omega) }{q} = \frac{  
V_{c} F_{in}(\omega) + 
\theta \left[ V_{c}F_{c}(\omega) - V_{w}F_{e}(\omega) + V_{c} F_{l}(\omega) \right] + 
\theta \left( \frac{\displaystyle \tau_{st}}{\displaystyle \tau_{e}} \right) \left[ V_{p} F_{RS}(\omega) - V_{p}F_{L}(\omega) \right]
}
{ (1 + \theta ) }
\label{eqdeltaIint2}
\end{equation}
Equations (\ref{eqdeltaIint1}) and (\ref{eqdeltaIint2}) show that above threshold the noise associated with carrier injection into the active region is not suppressed (since $\theta \ll 1$). Above threshold, the carrier density in the quantum wells is strongly damped, and only the carrier density in the SCH region provides negative feedback to suppress the noise associated with carrier injection. Below threshold, the carrier injection noise is suppressed (since $\theta' \approx 1$). Below threshold, the carrier densities in both the SCH region and the quantum wells provide feedback to suppress the carrier injection noise. The spectral density $K_{I}(\omega)$ of the current noise $\delta I(\omega)$ follows from Equations  (\ref{eqdeltaIint1}) and (\ref{eqdeltaIint2}),
\begin{equation}
K_{I}(\omega) \Big\vert_{\displaystyle \omega < \omega_{3\:{\rm dB}}} = \left\{ 
\begin{array}{ll}
{\displaystyle
q\,I \hspace{1 cm} (I<I_{th}) } & \\
{\displaystyle
q\,I + 
2q\; I_{th} \;\frac{ \left( \theta' - \theta \right) }{\left( 1 + \theta \right)} +               
\; 2qn_{sp} \; \eta_{i} \; (I-I_{th})\,\frac{\left( \theta \, \frac{\displaystyle \tau_{st}}{\displaystyle \tau_{e}}    \right)^{2}}{\left( 1 + \theta \right)^{2}} \hspace{1 cm} (I>I_{th}) }
\end{array} \right.
\end{equation} 
In the limit $\omega \rightarrow \infty$, the current noise is just the noise associated with carrier injection into the active region and has the spectral density,
\begin{equation}
K_{I}(\omega) \Big\vert_{\displaystyle \omega \rightarrow \infty} = \left\{ 
\begin{array}{ll}
{\displaystyle
q\,I\,\left( 1 + 2\,\theta' \right) \hspace{1 cm} (I<I_{th}) } & \\
{\displaystyle
q\,I\,\left( 1 + 2\,\theta \right) + 2q\;I_{th}\,\left( \theta' - \theta \right) \hspace{1 cm} (I>I_{th}) }
\end{array} \right.
\end{equation} 

\subsection{Suppression of the Current Noise by Large External Impedance}
The current noise $\delta I_{ext}(\omega)$ in the external circuit in the presence of an external impedance $Z_{s}(\omega)$ and an external voltage noise source $\delta V_{s}(\omega)$ is (Fig.~\ref{FigCirModInt}),
\begin{equation}
\delta I_{ext}(\omega) = \frac{\delta V_{s}(\omega)}{(Z(\omega) + Z_{s}(\omega))} + \frac{Z(\omega)}{(Z(\omega) + Z_{s}(\omega))} \, \delta I(\omega)
\end{equation}
where $Z(\omega)$ is the differential impedance of the active region and $Z(\omega = 0) = R_{d}$. The external impedance $Z_{s}(\omega)$ is the Thevenin equivalent of the external circuit impedance and the impedance associated with the laser parasitics (ohmic contact resistance, depletion layer capacitance e.t.c.). Assuming that $\delta V_{s}(\omega)$ represents only the thermal noise originating in $Z_{s}(\omega)$, the spectral density $K_{I_{ext}}(\omega)$ of the current noise in the external circuit becomes,
\begin{equation}
K_{I_{ext}}(\omega) = \frac{K_{V_{s}}(\omega)}{\left\vert Z(\omega) + Z_{s}(\omega) \right\vert^{2}} + \left\vert \frac{Z(\omega)}{Z(\omega) + Z_{s}(\omega)} \right\vert^{2} K_{I}(\omega) = \frac{2K_{B}T\,{\rm Real} \{ Z_{s}(\omega) \}}{\left\vert Z(\omega) + Z_{s}(\omega) \right\vert^{2}} + \left\vert \frac{Z(\omega)}{Z(\omega) + Z_{s}(\omega)} \right\vert^{2} K_{I}(\omega)
\end{equation}
When $Z_{s}(\omega)$ is much larger than the differential impedance $Z(\omega)$ of the active region then the current noise in the external circuit is just the thermal noise originating in the impedance $Z_{s}(\omega)$. When $Z_{s}(\omega)$ is much smaller than $Z(\omega)$ then the current noise in the external circuit is the noise originating inside the active region. By making the impedance $Z_{s}(\omega)$ very large the current noise in the external circuit can be suppressed well below the shot noise value.

\subsection{Intensity Noise}
Above threshold, and for frequencies less than $\omega_{3\:{\rm dB}}$, $\delta P_{out}(\omega)$ is,
\begin{eqnarray}
\delta P_{out}(\omega) & = & \eta_{o}\,\eta_{i}\,h \nu \frac{I_{ext}(\omega)}{q} +  \eta_{o} \,h \nu \bigg\{ \left( 1 - \eta_{i} \right) \left[V_{c}F_{c}(\omega) - V_{w}F_{e}(\omega) \right] - \eta_{i}V_{c}F_{l}(\omega) \nonumber \\
& & \mbox{} - V_{w}F_{nr}(\omega) - V_{w}F_{RN}(\omega) + \gamma \tau_{st} \left[ V_{p}F_{RS}(\omega) - V_{p}F_{L}(\omega) \right] \: \bigg\} + F_{o}(\omega)
\end{eqnarray}

High impedance suppression of the current noise $\delta I_{ext}(\omega)$ in the external circuit can have a profound effect on the laser intensity noise through the first term on the right hand side of the above Equation. If $\delta I_{ext}(\omega)$ is suppressed then the spectral density $K_{P}(\omega)$ of the intensity noise, for frequencies less than $\omega_{3\:{\rm dB}}$, is,
\begin{eqnarray}
K_{P}(\omega) \Big\vert_{\displaystyle \omega < \omega_{3\:{\rm dB}}} & = & h \nu \, P_{out} \, \left[  1 - \eta_{o} + 2\eta_{o}\,n_{sp}\,\left( \frac{1}{\tau_{nr}} + \frac{1-\eta_{i}}{\tau_{e}}  \right)^{2}\,\tau_{st}^{2} \right]  \nonumber \\ 
& & \mbox{} + (\eta_{o} \, h \nu)^{2}\left[ \eta_{i}(1 - \eta_{i}) \frac{I}{q} + \eta_{i}\frac{I_{th}}{q} \right] \label{eqKp5}
\end{eqnarray} 
The low frequency Fano factor of the intensity noise at large bias currents becomes,
\begin{equation}
F_{P}(\omega) \Big\vert_{\displaystyle \left( \omega < \omega_{3\:{\rm dB}} \: , \: I \gg I_{th} \right) } = 1 - \eta_{o}\,\eta_{i} 
\end{equation}
In diode lasers both $\eta_{o}$ and $\eta_{i}$ have typical values around 0.85, and therefore high impedance suppression of the current noise in the external circuit can result in more than 5 dB suppression of the laser intensity noise below the shot noise value. On the other hand if the external impedance $Z_{s}(\omega)$ is much smaller than the impedance $Z(\omega)$ of the active region then it can be shown that at large bias currents the Fano factor of the laser intensity noise approaches unity,
\begin{equation}
F_{P}(\omega) \Big\vert_{\displaystyle \left( \omega < \omega_{3\:{\rm dB}} \: , \: I \gg I_{th} \right) } = 1 
\end{equation}
In practice it is difficult to make $Z_{s}(\omega)$ very small. One way of obtaining a small external impedance $Z_{s}(\omega)$ is by using the circuit B shown in Fig.~\ref{figBiasCir} and shorting the RF port of the bias T. At frequencies of interest $Z_{s}(\omega)$ would then just be the parasitic impedance associated with the laser device and would be dominated by the resistance of the device ohmic contacts.

\newpage

\begin{table}[htb]
\caption{Device Parameters Used in Numerical Simulations (From Ref.~\cite{faistHP})}
\begin{center}
{\tt
\label{table1}
\begin{tabular}{|l|r|}
\hline 
{\bf Parameter} & {\bf Value}  \\ 
\hline
Lasing wavelength $\lambda$ & 5.0 $\mu$m  \\
\hline
Operating temperature  & 20 K  \\
\hline
Number of gain stages $N$ & 25 (unless stated otherwise) \\
\hline 
Total confinement factor ${ \sum_{j=1}^{N}} \Gamma^{j}$ & $erf(0.019\,N) \approx 0.02\,N$ \\
\hline
Cavity width $W$ & 11.7 $\mu$m \\ 
\hline
Cavity length $L$ & 3 mm  \\
\hline 
Facet reflectivities $r_{1}$, $r_{2}$ & 0.27  \\
\hline
Cavity internal loss $\alpha_{i}$ & 11 cm$^{-1}$  \\
\hline
Mode effective index $n_{eff}$ & 3.29  \\
\hline
Mode group index $n_{g}$ & 3.4  \\
\hline
Differential gain $a$ & $\sim 4.0 \times 10^{-9}$ cm  \\
\hline
$\tau_{in}$, $\tau_{out}$, $\tau_{3}$ & 1.0 ps  \\
\hline
$\tau_{2}$, $\tau_{1}$ & $\infty$ \\
\hline
$\tau_{32}$ & 2.1 ps \\
\hline
$\tau_{31}$ & 3.4 ps \\
\hline
$\tau_{21}$ & 0.5 ps \\
\hline
$C_{inj}$ & 0.31 $\mu$F/cm$^{2}$ \\
\hline
$C_{3}$, $C_{2}$ & 0.56 $\mu$F/cm$^{2}$ \\
\hline
$C_{1}$ & 0.81 $\mu$F/cm$^{2}$ \\
\hline
$\chi_{in}$, $\chi_{out}$ & $\sim 1$ \\
\hline
\end{tabular}
}
\end{center}
\end{table}


\newpage

\begin{thebibliography}{100}
\bibitem{buttiker} Y. M. Blanter, M. Buttiker, ``Transition from sub-Poissonian to super-Poissonian shot noise in resonant quantum wells,'' Phys. Rev. B {\bf 59}, 10217-10226 (1999). 

\bibitem{davis} J. H. Davies, P. Hyldgaard, Selman Hershfield, J. W. Williams, ``Classical theory for shot noise in resonant tunneling,'' Phys. Rev. B {\bf 46}, 9620-9633 (1992).

\bibitem{pelleg} G. Iannaccone, M Macucci, B. Pellegrini, ``Shot noise in resonant-tunneling structures,'' Phys. Rev B {\bf 55}, 4539-4550 (1997).

\bibitem{yama1} Y. Yamamoto, S. Machida, ``High-impedance suppression of pump fluctuation and amplitude squeezing in semiconductor lasers,'' Phys. Rev. A {\bf 35}, 5114-5130 (1987).

\bibitem{farhan1} F. Rana, R. Ram, ``Photon noise and correlations in semiconductor cascade lasers,'' Appl. Phys. Lett. {\bf 76}, 1083-1085 (2000). 

\bibitem{farhan2} F. Rana, S. G. Patterson, R. Ram, ``Noise in semiconductor cascade lasers,'' Proceedings Integrated Photonics Research'1999 (1999). 

\bibitem{faistvtQCL} J. Faist, F. Capasso, C. Sirtori, D. L. Sivco, A. L. Hutchinson, A. Y. Cho, ``Vertical transition quantum cascade laser with Bragg confined excited state,'' Appl. Phys. Lett. {\bf 66}, 538-540 (1995).

\bibitem{faistCW} J. Faist, F. Capasso, C. Sirtori, D. L. Sivco, A. L. Hutchinson, A. Y. Cho, ``Continuous wave operation of a vertical transition quantum cascade laser above T=80 K,'' Appl. Phys. Lett.  {\bf 67}, 3057-3059 (1995).

\bibitem{carloCW} C. Sirtori, J. Faist, F. Capasso, D. L. Sivco, A. L. Hutchinson, S. N. G. Chu, A. Y. Cho, ``Continuous wave operation of midinfrared (7.4-8.6 $\mu$m) quantum cascade lasers up to 110 K temperature,'' Appl. Phys. Lett. {\bf 68}, 1745-1747 (1996). 

\bibitem{carloPandC} C. Sirtori, J. Faist, F. Capasso, D. L. Sivco, A. L. Hutchinson, A. Y. Cho, {\em }, ``Pulsed and continuous-wave operation of long wavelength infrared ($\lambda \approx 9.3$ $\mu$m) quantum cascade lasers,'' IEEE J. Quantum Electron. {\bf 33}, 89-93 (1997). 

\bibitem{faistHP} J. Faist, A. Tredicucci, F. Capasso, C. Sirtori, D. L. Sivco, J. N. Baillargeon, A. L. Hutchinson, ``High-power continuous-wave quantum cascade lasers,'' IEEE J. Quantum Electron. {\bf 34}, 336-343 (1998).

\bibitem{carloGaAs} C. Sirtori, P. Kruck, S. Barbieri, P. Collot, J. Nagle, ``GaAs/Al$_{x}$Ga$_{1-x}$As quantum cascade lasers,'' Appl. Phys. Lett. {\bf 73}, 3486-3488 (1998).

\bibitem{cgdiskQCL} C. Gmachl, J. Faist, F. Capasso, C. Sirtori, D. L. Sivco, A. Y. Cho, ``Long-wavelength ($9.5$-$11.5$ $\mu$m) microdisk quantum-cascade lasers,'' IEEE J. Quantum Electron. {\bf 33}, 1567-1573 (1997).

\bibitem{cgsingle} C. Gmachl, F. Capasso, A. Tredicucci, D. L. Sivco, ``Noncascaded intersubband injection lasers at $\lambda \approx 7.7$ $\mu$m,'' Appl. Phys. Lett. {\bf 73}, 3830-3832 (1998). 

\bibitem{cgdfb} C. Gmachl, F. Capasso, J. Faist, A. L. Hutchinson, A. Tredicucci, D. L. Sivco, J. N. Baillargeon, S. N. G. Chu, A. Y. Cho, ``Continuous-wave and high-power pulsed operation of index-coupled distributed feedback quantum cascade laser at $\lambda \approx 8.5$ $\mu$m,'' Appl. Phys. Lett. {\bf 72}, 1430-1432 (1998). 

\bibitem{atsuperQCL1} A. Tredicucci, F. Capasso, C. Gmachl, D. L. Sivco, A. L. Hutchinson, A. Y. Cho, J. Faist, G. Scamarcio, ``High-power inter-miniband lasing in intrinsic superlattices,'' Appl. Phys. Lett. {\bf 72}, 2388-2390 (1998).

\bibitem{atsuperQCL2} A. Tredicucci, F. Capasso, C. Gmachl, D. L. Sivco, A. L. Hutchinson, A. Y. Cho, ``High performance interminiband quantum cascade lasers with graded superlattices,'' Appl. Phys. Lett. {\bf 73}, 2101-2103 (1998).

\bibitem{atsuperQCL3} A. Tredicucci, C. Gmachl, F. Capasso, D. L. Sivco, A. L. Hutchinson, A. Y. Cho, J. Faist, G. Scamarcio, ``Long wavelength superlattice quantum cascade lasers at $\lambda \approx 17$ $\mu$m,'' Appl. Phys. Lett. {\bf 74}, 638-640 (1999).

\bibitem{gssupQCL} G. Strasser, S. Gianordoli, L. Hvozdara, W. Schrenk, K. Unterrainer, E. Gornik, ``GaAs/AlGaAs superlattice quantum cascade lasers at $\lambda \approx 13$ $\mu$m,'' Appl. Phys. Lett. {\bf 75}, 1345-1347 (1999).

\bibitem{coldren} L. A. Coldren, S. Corzine, ``Diode Lasers and Photonic Integrated Circuits,'' John Wiley and Sons, New York (1995).

\bibitem{capasso} J. Faist, F. Capasso, C. Sirtori, D. L. Sivco, D. L. Hutchinson, M. S. Hybertsen, A. Y. Cho, ``Quantum cascade lasers without intersubband population inversion,'' Phys. Rev. Lett. {\bf 76}, 411-414 (1996).

\bibitem{APK} A. Wacker, A-P. Jauho, ``Microscopic modelling of perpendicular electronic transport in doped multiple quantum wells,'' Physica Scripta {\bf T69}, 321-324 (1997). 

\bibitem{gardiner} C. W. Gardiner, ``Handbook of Stochastic Methods,'' Springer Verlag, New York (1996).

\bibitem{farhan3} F. Rana, ``Current noise, photon number fluctuations and squeezing in quantum cascade lasers,'' Proceedings IEEE LEOS Annual Meeting'1999 (1999).

\bibitem{ramoshock} P. D. Yoder, K. Gartner, W. Fichtner, ``A generalized Ramo-Shockley theorem for classical to quantum transport at arbitrary frequencies,'' J. Appl. Phys. {\bf 79}, 1951-1954 (1996).

\bibitem{shore1} N. Mustafa, L. Pesquera, L. Cheung, K. A. Shore, ``Terahertz bandwidth prediction for amplitude modulation response of unipolar intersubband semiconductor lasers,'' IEEE Photonics Tech. Lett. {\bf 11}, 527-529 (1999).

\bibitem{interbandsq} S. Machida, Y. Yamamoto, ``Observation of amplitude squeezing from semiconductor lasers by balanced direct detectors with a delay line,'' Optics Lett. {\bf 14}, 1045-1047 (1989).

\bibitem{cgscale} C. Gmachl, F. Capasso, A. Tredicucci, D. L. Sivco, R. Kohler, A. L. Hutchinson, A. Y. Cho, ``Dependence of the device performance on the number of stages in quantum-cascade lasers,''  IEEE J. Selected Topics Quantum Electronics {\bf 5}, 808-816 (1999).  

\bibitem{yama3} S. Inoue, H. Ohzu, S. Machida, Y. Yamamoto, ``Quantum correlation between longitudinal-mode intensities in a multimode squeezed semiconductor laser,'' Phys. Rev. A {\bf 46}, 2757-2765 (1992).

\bibitem{farhan4} F. Rana, R. Ram, ``Current Noise in Semiconductor Lasers,'' Proceedings IEEE LEOS Summer Topical'2001 (2001).

\bibitem{yama4} J. Kim, Y. Yamamoto, ``Theory of noise in p-n junction light emitters,'' Phys. Rev. B {\bf 55}, 9949-9959 (1997).

\bibitem{aggarwal} G. P. Agrawal, N. K. Datta, ``Long Wavelength Semiconductor Lasers,'' Van Nostrand Reinhold, New York (1993).

\bibitem{zory} K. Y. Lau in ``Quantum Well Lasers,'' P. S. Zory (Editor), Academic Press, New York (1993). 
\end{thebibliography}
\end{document}